\documentclass[12pt]{article}
\usepackage{jheppub}
\usepackage[colorinlistoftodos]{todonotes}
\usepackage{tikz}
\usepackage{url}
\usepackage{comment}

\newcommand{\bC}{\mathbb{C}}
\newcommand{\bR}{\mathbb{R}}
\newcommand{\cI}{{\cal I}}
\newcommand{\cO}{{\cal O}}
\newcommand{\dd}{\mathrm{d}}

\renewcommand{\d}{\text{d}}
\newcommand{\bu}{\bullet}
\newcommand{\dbar}{{\bar \partial}}
\newcommand{\C}{\mathbb{C}}
\newcommand{\del}{\partial}
\DeclareMathOperator{\Tr}{Tr}

\newcommand{\lie}{\mathfrak}
\DeclareMathOperator{\PV}{PV}
\DeclareMathOperator{\Sym}{Sym}
\newcommand{\cN}{\mathcal{N}}
\newcommand{\ep}{\varepsilon}

\newcommand{\p}{\partial}

\newcommand{\cV}{\mathcal{V}}
\newcommand{\bbC}{\mathbb{C}}
\newcommand{\bbZ}{\mathbb{Z}}
\newcommand{\bbR}{\mathbb{R}}
\newcommand{\Spin}{\mathrm{Spin}}
\newcommand{\calO}{\mathcal{O}}

\newcommand{\calH}{\mathcal{H}}
\newcommand{\calI}{\mathcal{I}}

\newcommand{\calL}{\mathcal{L}}
\newcommand{\calM}{\mathcal{M}}
\newcommand{\calN}{\mathcal{N}}
\newcommand{\calR}{\mathcal{R}}
\newcommand{\calS}{\mathcal{S}}

\newcommand{\calZ}{\mathcal{Z}}

\usepackage{color}
\usepackage{soul}

\usepackage{tikz,tikz-3dplot}
\usepackage{pgfplots}
\pgfplotsset{compat=1.17}
\usetikzlibrary{shapes,arrows,cd,chains,decorations.markings,decorations.pathmorphing,calc}
\tikzset{
	->-/.style args={#1rotate#2}{decoration={markings, mark=at position #1 with {\arrow[scale=1.5,rotate = #2 ]{stealth}}}, postaction={decorate}}
}
\tikzset{
	->>-/.style args={#1and#2rotate#3}{decoration={markings, mark=at position #1 with {\arrow[scale=1.5,rotate = #3 ]{stealth}}, mark=at position #2 with {\arrow[scale=1.5,rotate = #3 ]{stealth}}}, postaction={decorate}}
}
\tikzset{
	-s-/.style args={#1rotate#2}{decoration={markings, mark=at position #1 with {\arrow[scale=1.5,rotate = #2]{to}}}, postaction={decorate}}
}
\tikzset{
    marrow/.style={decoration={markings,mark=at position 0.575 with {\arrow{#1}}}, postaction=decorate}
}
\tikzset{
    mmrrow/.style={decoration={markings,mark=at position 0.50 with {\arrow{#1}}}, postaction=decorate}
}
\tikzset{
    farrow/.style={decoration={markings,mark=at position 0.55 with {\arrow[rotate = 180]{#1}}}, postaction=decorate}
}
\tikzstyle{GraphNode}=[circle, draw=black, fill=black, inner sep=2pt, minimum size=5pt]
\tikzstyle{GraphEdge}=[black]
\pgfmathsetmacro{\gS}{1}

\usepackage{tikz} 
\usepackage{tikz-cd} 
\usetikzlibrary{matrix,calc} 
\usetikzlibrary{decorations.markings,shapes.geometric,shapes.misc} 
\usetikzlibrary{decorations.pathreplacing,positioning} 
\usetikzlibrary{arrows}
\usepackage{mathrsfs}

\usepackage{quiver}
\newcommand{\pic}[2]{\vcenter{\hbox{\includegraphics[height=#1]{#2}}}}
 
\graphicspath{ {./images/} }
\newcommand{\pick}[2]{\vcenter{\hbox{\includegraphics[width=#1]{#2}}}}

\title{Semi-Chiral Operators in 4d ${\cal N}=1$ Gauge Theories}

\author[\pick{1.5ex}{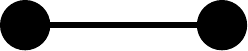},\pic{1.2ex}{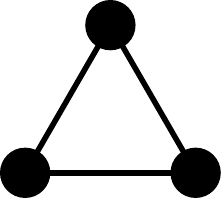}]{Kasia Budzik,}
\author[\pick{1.5ex}{segment}]{Davide Gaiotto,}
\author[\pick{1.5ex}{segment},\pic{1.2ex}{triangle}]{Justin Kulp,}
\author[\pic{1ex}{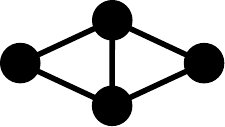}]{Brian R. Williams,}
\author[\pic{1ex}{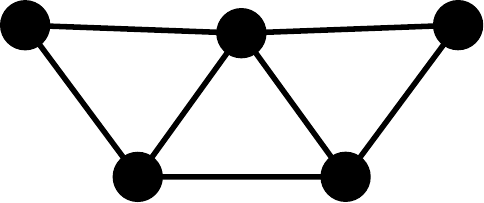}]{Jingxiang Wu,}
\author[\pick{1.5ex}{segment},\pic{1.2ex}{triangle}]{and Matthew Yu}

\affiliation[\pick{1.5ex}{segment}]{Perimeter Institute for Theoretical Physics, Waterloo, ON N2L 2Y5, Canada}
\affiliation[\pic{1.2ex}{triangle}]{Department of Physics \& Astronomy, University of Waterloo, Waterloo, ON N2L 3G1, Canada}
\affiliation[\pic{1ex}{bitriangle}]{Department of Mathematics and Statistics, Boston University,  665 Commonwealth Ave, Boston, MA 02215}
\affiliation[\pic{1ex}{tritriangle}]{Mathematical Institute, University of Oxford, Andrew-Wiles Building, Woodstock Road, Oxford, OX2 6GG, UK}

\emailAdd{kbudzik@perimeterinstitute.ca}
\emailAdd{dgaiotto@perimeterinstitute.ca}
\emailAdd{jkulp@perimeterinstitute.ca}
\emailAdd{bwill22@bu.edu}
\emailAdd{Jingxiang.Wu@maths.ox.ac.uk}
\emailAdd{myu@perimeterinstitute.ca}

\abstract{We discuss the properties of quarter-BPS local operators in four-dimensional ${\cal N}=1$ supersymmetric Yang-Mills theory using the formalism of holomorphic twists. We study loop corrections both to the space of local operators and to algebraic operations which endow the twisted theory with an infinite symmetry algebra. We classify all single-trace quarter-BPS operators in the planar approximation for $SU(N)$ gauge theory and propose a holographic dual description for the twisted theory. We classify perturbative quarter-BPS operators in $SU(2)$ and $SU(3)$ gauge theories with sufficiently small quantum numbers and discuss possible non-perturbative corrections to the answer. We set up analogous calculations for some theories with matter. }


\arxivnumber{2306.01039}

\graphicspath{ {./images/} }

\begin{document}
\maketitle
\section{Introduction}
Four-dimensional ${\cal N}=1$ supersymmetric quantum field theories have been the subject of intense investigation (see \cite{Intriligator:1995au} for a review, and \cite{Tachikawa:2018sae, Razamat:2022gpm} for a review of some modern aspects of the subject). They display phenomena of great physical interest, such as confinement and chiral symmetry breaking, which have a rich interplay with quantities that can be exactly computed or strongly constrained via supersymmetry \cite{Intriligator:1995au}. 

An important tool is the study of F-terms and chiral operators, which are protected by two of the four supercharges. These constrain the space of supersymmetric vacua as well as important classes of deformations of the theories \cite{Seiberg:1993vc,Seiberg:1994bz, Seiberg:1994bp, Intriligator:1994sm, Seiberg:1994pq}. They are also the first quantities one would typically match across dualities such as Seiberg duality and its generalizations \cite{Seiberg:1994pq,Kutasov:1995ss,Kutasov:1995ve,Kutasov:1995np}.

Another important protected quantity which is available for ${\cal N}=1$ superconformal field theories is the superconformal index \cite{Kinney:2005ej, Romelsberger:2005eg, Romelsberger:2007ec, Dolan:2008qi, Rastelli:2016tbz, Gadde:2022pgw}. 
The superconformal index is a power series whose coefficients are Witten indices of the space of local operators with a given set of charges. 
It is a robust quantity which can be computed semi-classically, disregarding interactions. 

Essentially by definition, the coefficients of the superconformal index also count, with signs, the cohomology of the spaces of local operators with respect to one of the four supercharges, denoted simply as $Q$ in the following. 
The cohomology of $Q$ thus provides a natural ``categorification'' of the 
data in the superconformal index \cite{Grant:2008sk,Chang:2013fba}. 

Concrete calculations in these references demonstrated surprising simplifications which can be rigorously proven and extended with the help of the {\it holomorphic twist} of the ${\cal N}=1$ SCFTs \cite{Eager:2018oww}. This is a procedure which maps the full physical theory to a simplified theory whose gauge-invariant local operators are the $Q$-cohomology of the original theory.\footnote{Essentially, one adds $Q$ to the original BRST differential of the physical theory. This enlarges the gauge symmetry of the problem and allows further simplifications to be carried on systematically.}

The notion of $Q$-cohomology of the space of local operators appears to be well-defined even for theories which are not superconformal. It should describe local operators which are protected by a single supercharge. In analogy with the chiral ring, we dub these operators the {\it semi-chiral ring} of the theory. In the main text of the paper, we will discuss why the term ``ring'' is reasonable here. Again, one can explore this structure by the {\it holomorphic twist} of the theory, which is also well-defined in the absence of conformal symmetry. 

The twisting procedure has recently been developed in a mathematically rigorous perturbative setting \cite{CostelloHol, ESW}, allowing one to simplify the twisted SQFTs to holomorphic quantum field theories defined on ${\mathbb C}^2$. The simplified theories are endowed with extra symmetries and structures which are a four-dimensional analogue of a two-dimensional chiral algebra \cite{LMNS, NekThesis,Saberi:2019fkq, Gaiotto:2024gii,Gwilliam:2018lpo}. 
The implications of these structures on the structure of the $Q$-cohomology and the associated constraints on the original physical theory are mostly unexplored.

It is important to observe that the $Q$-cohomology is much less ``robust'' than the Witten index. 
In particular, interactions are expected to modify the naive free field theory answers. 
In the literature, $Q$-cohomology calculations have mostly been done at tree level, invoking or conjecturing non-renormalization theorems to justify disregarding perturbative or non-perturbative corrections. 
For example, this approach has been rather successful for planar ${\cal N}=4$ supersymmetric Yang-Mills theory \cite{Chang:2013fba}. However, the presence of quantum corrections to the chiral ring, such as the generalized Konishi anomaly \cite{Cachazo:2002ry}, suggests the existence of quantum corrections to the $Q$-cohomology of general ${\cal N}=1$ gauge theories, which we will make explicit in this paper.


The main purpose of this work is to explore the effect of interactions on the semi-chiral ring of pure ${\cal N}=1$ gauge theory. We will work with the assumption that the space of local operators can be studied systematically in the far UV where the theory is asymptotically free. Concretely, that means we will start from the semi-chiral ring of the free theory and study how interactions deform the action of $Q$ order-by-order in perturbation theory,
gradually reducing the size of the cohomology by pairing up operators which were in cohomology at the previous loop order.\footnote{This procedure is somewhat analogous to a spectral sequence in mathematics and is formalized by the notion of homotopy transfer. See appendix \ref{sec:HomotopyTransfer}.}

Our main assumption will be that non-perturbative corrections may at most further reduce the size of the perturbative semi-chiral ring. Instanton effects violate the anomalous $R$-symmetry selection rules which constrain the action of $Q$ in perturbation theory, allowing for further cancellation between perturbative cohomology classes.\footnote{See \cite{Dorey:2002ik} for a review of semi-classical instanton calculations. It would be nice to do similar calculations in twisted theories.}

Crucially, all calculations can be done in the twisted theory.\footnote{This statement can typically be promoted to a theorem in perturbation theory, but may require some caution non-perturbatively. In principle, the choice of path integration contour/solution of Ward identities for the simplified theories might not be unique, leading to the higher-dimensional analogue of conformal blocks. We will not explore this issue in this paper, but it would be interesting to do so.} Classically, this reproduces and generalizes the simplifications which were found in direct calculations in ${\cal N}=4$ SYM \cite{Grant:2008sk,Chang:2013fba}. Quantum mechanically, this greatly reduces the complexity of the relevant Feynman diagrams. 
The one-loop corrections are semi-chiral analogues of the generalized Konishi anomaly equations \cite{Konishi:1983hf, Konishi:1985tu, Cachazo:2002ry}. Our first observation is that the perturbative corrections cannot stop at 1-loop, as the resulting differential is not nilpotent. Indeed, for pure $SU(2)$ gauge theory we can even ``bootstrap'' the necessary two-loop corrections, essentially bypassing an actual calculation of the Feynman diagrams. 

Companion papers \cite{Budzik:2022mpd, Gaiotto:2024gii} outline the general definition and computation of the Feynman diagrams which contribute to the differential and to other algebraic structures in twisted gauge theories with a general matter content and interactions. 

In order to explore the structure of the cohomology, we conduct a systematic calculation of perturbative cohomology classes in pure $SU(2)$ gauge theory. We are able to include in the calculation operators with up to fourteen derivatives. The cohomology at one loop is already sufficiently sparse to forbid further perturbative corrections. 

Remarkably, the quantum numbers of the non-zero perturbative cohomology representatives appear compatible with the hypothesis that instanton effects will ultimately lift all operators in the semi-chiral ring which are not in the chiral ring. Recall that the chiral ring of the theory has a single generator, the gaugino bilinear $C_0$.\footnote{The glueball field is usally denoted $S$ in the SUSY literature.} It is expected to satisfy the relation $C_0^2 \sim \Lambda^6$ (in the chiral ring of $SU(2)$) and thus acquire a vev which distinguishes the two gapped vacua of the theory.\footnote{More generally, we expect $C_0^h \sim \Lambda^{2h}$ for a gauge group with dual Coxeter element $h$.}

If the hypothesis is correct, the semi-chiral ring computed in the UV will fully match the semi-chiral ring of the theory in the far IR, which is trivial up to the order parameter distinguishing the two vacua. 
It would be interesting to 
test further the hypothesis that the semi-chiral ring in the UV and IR should coincide for general twisted theories, perhaps by looking at gauge theories which enjoy Seiberg-like dualities. See \cite{Berkooz:1995cb} for an earlier attempt to work outside the chiral ring.

We also present systematic calculations in $SU(3)$ gauge theory up to five derivatives as well as derive some abstract results valid for any gauge group. Our main analytic result is a surprising observation: one-loop corrections to the differential in pure ${\cal N}=1$ gauge theory make the theory topological rather than holomorphic. We are sorely tempted to take this result as a UV manifestation of the IR confinement of the physical theory: a gapped theory is indeed topological at large distances. 

Concretely, this result places strong constraints on the possible IR behaviour of the theory: the holomorphic twist of the IR degrees of freedom must also be secretly topological. Further investigation (and more examples) of such {\it holomorphic confinement} constraints is warranted. 

Another analytic result we present is a full calculation of the semi-chiral ring in the planar approximation for $SU(N)$. The structure of the calculation leads us to propose a holographic dual description for the large $N$ twisted theory, involving the B-model on a certain backreacted version of $\bC^3$. 

Many of our methods extend to more general gauge theories. We anticipate some of these generalizations by a uniform discussion of the large $N$ cohomology and potential twisted holographic duals for a variety of theories, including SQCD, ${\cal N}=4$ SYM and quivers associated to Calabi-Yau cones \cite{Eager:2018oww}. 

\subsection{Structure of the Paper}
Section \ref{sec:symmetries} discusses in detail the expected properties of holomorphic twists of ${\cal N}=1$ supersymmetric QFTs. 
We review a useful superspace technology in section \ref{subsec:chiralSuperspace}, discuss infinite symmetry enhancement in section \ref{subsec:symmetries}, and the twisted version of the stress tensor multiplet in section \ref{subsec:stresstensor}.

Section \ref{sec:twistedAsHolo} discusses the holomorphic twist of pure ${\cal N}=1$ gauge theory and the presentation as a 4d $bc$ system. In section \ref{subsec:localops} we discuss the space of local operators in the free theory and at tree level. In section \ref{subsec:loopCorrections} we discuss loop corrections to the BRST differential acting on local operators. 

In section \ref{sec:addingMatter} we extend the discussion to general Lagrangian gauge theories. 
In section \ref{sec:infiniteN} we perform cohomology calculations in the planar approximation. We discuss a general strategy, apply it in full to pure gauge theory, and implement it schematically for several other examples.  

In section \ref{sec:holography} we use the large $N$ analysis to motivate a holographic proposal for twisted pure gauge theory and discuss twisted versions of standard holographic dualities for certain quiver gauge theories. 

Section \ref{sec:indices} contains some auxiliary computations of characters of spaces of local operators. 
Section \ref{sec:finiteN} contains the results of our concrete calculations of local operator cohomology for pure $SU(2)$ and $SU(3)$ guage theories. 

We conclude with several appendices.
Appendix \ref{appendix:lambdabrackets} reviews some properties of $\lambda$-brackets. Appendix \ref{app:treelevelbracket}
discusses the calculation of these brackets at tree level and appendix \ref{appendix:loopbracket} discusses the calculation of brackets at one loop. 
Appendix \ref{sec:HomotopyTransfer} reviews the notion of homotopy transfer and the homotopy perturbation lemma. 
Appendix \ref{app:MC} reviews an origin of $L_\infty$ algebras in QFT.
Finally, appendix 
\ref{sec:RigidSUSY} contains a careful comparison of different perspectives on twisted supersymmetric QFTs.

\section{Symmetries and Kinematics}
\label{sec:symmetries}
Our starting point is the four-dimensional ${\cal N}=1$ supersymmetry algebra in Euclidean signature. Using the isomorphism $\mathrm{Spin}(4) \cong SU(2) \times SU(2)$, we have an isomorphism between the vector representation of $\Spin(4)$ and (the product of) the spinor representations of the two $SU(2)$'s. Thus we denote our supercharges with the two-component spinors\footnote{We have switched the standard two-component spinor notation for dotted/undotted indices as dotted indices will soon drop out \cite{Wess:1992cp}.} $Q_{\dot \alpha}$ and $\bar Q_{\alpha}$, subject to anticommutation relations:
\begin{align}
    \{ Q_{\dot \alpha}, \bar{Q}_\beta \} 
        &= P_{\dot \alpha \beta}\,, \\
    \{ Q_{\dot \alpha}, Q_{\dot\beta} \} 
        &= \{ \bar{Q}_\alpha, \bar{Q}_\beta \} 
        = 0\,,
\end{align}
where $P_{\dot \alpha \beta}$ are the translation generators. Rotation generators in the two $SU(2)$ subgroups will be denoted $M_{\dot \alpha  \dot \beta}$ and $\bar M_{\alpha \beta}$ respectively.

Without loss of generality, we choose $Q \equiv Q_{\dot -}$ as a twisting supercharge.\footnote{Different choices of $Q$ will simply define a different complex structure on $\bbC^2$. The moduli space of twisting supercharges forms the nilpotence variety of the supersymmetry algebra \cite{Eager:2018dsx, ESW} (see also 2.1.1  of \cite{Garner:2022its} for an introduction).} This choice equips Euclidean spacetime with a complex structure such that $z^\alpha \equiv x^{\dot + \alpha}$ are holomorphic coordinates and $\bar z^\alpha \equiv x^{\dot - \alpha}$ are anti-holomorphic.\footnote{Complex conjugation acts as $(z_\alpha)^* = \epsilon_{\alpha \beta} \bar z^\beta$.}
We immediately see that the anti-holomorphic translations are $Q$-exact:
\begin{equation}
\{Q, \bar Q_{\alpha} \} = \partial_{\dot - \alpha} = \partial_{\bar z^\alpha} \,.
\end{equation}
As a result, the twisted theory is holomorphic in a cohomological sense \cite{CG2,BWhol}.

The choice of twisting supercharge breaks the four-dimensional rotation group to the $SU(2)$ subgroup generated by the $\bar M_{\alpha \beta}$ in the twisted theory, which acts trivially on dotted spinors. These are holomorphic symplectic rotations of the $z^\alpha$, preserving the holomorphic volume form $\d^2 z = \d z^1 \wedge \d z^2$.

In the presence of an unbroken $U(1)_R$ symmetry, we can combine $M_{\dot + \dot -}$ with the $R$-symmetry generator to define an extra twisted rotation generator\footnote{In conventions where $[M_{\dot{+}\dot{-}},Q_{\dot{-}}]=-Q_{\dot{-}}$ and $[R,Q_{\dot{-}}]=-Q_{\dot{-}}$. Charges and gradings are summarized in table \ref{tab:charges}.} $M_R\equiv M_{\dot + \dot -}-R$, which extends the $SU(2)$ above to the entire $U(2)$ group of holomorphic rotations of the $z^\alpha$. The gauge theories we consider in this paper have a $U(1)_R$ symmetry which is anomalous at one loop. The $U(2)$ rotation symmetry formally survives as a symmetry in perturbative calculations, but is broken down to $SU(2)$ by familiar  instanton effects.

For cohomology computations, it is convenient to define a $\mathbb{Z}$-grading on the twisted theory such that $Q$ increases the cohomological degree by $1$. We can combine the Cartan generator $M_{\dot + \dot -}$ with the ghost number symmetry of the physical theory to play the role of the cohomological degree $C\equiv \text{gh}-M_{\dot + \dot -}$. This choice is a bit unusual, but is very convenient in the absence of $U(1)_R$ symmetries.

In perturbation theory, interactions will lift some part of the cohomology but will not add new cohomology classes. Thus we expect the cohomology to still be supported in non-positive cohomological degree for $\mathcal{N}=1$ SQFTs expanded perturbatively around the free point. It would be interesting to know if this property holds non-perturbatively as well, or in generic ${\cal N}=1$ SQFTs. 

\subsection{Semi-Chiral Operators and Semi-Short Superconformal Multiplets}
In a superconformal theory the $Q$-cohomology is somewhat constrained. Operators can be grouped into superconformal multiplets, generated from primary operators which are annihilated by superconformal and special conformal generators. Each primary operator transforms in some representation of $\mathrm{Spin}(4)$, has some charge $R$ under $U(1)_R$ and some scaling dimension $\Delta$. Other ``descendant'' operators are obtained by acting on primaries with supercharges and derivatives. 

For generic $\Delta$, supercharges and derivatives act freely. In particular, an operator in such a supermultiplet is $Q$-closed only if it is $Q$-exact. However, when $\Delta$ achieves some minimum values (aka saturates BPS bounds), some descendant operators are missing and the multiplets become short. An obvious example is the identity operator, which has no descendants. Obviously, it is a non-trivial $Q$-cohomology class. Generally, short multiplets for ${\cal N}=1$ superconformal symmetry are classified, see e.g. \cite{Cordova:2016emh}, with separate shortening conditions for $Q$'s and $\bar Q$'s. While we don't perform the computations explicitly here, one can work out exactly the $Q$-cohomology of such multiplets. 

It is instructive to look at some simple subalgebras of the superconformal algebra. For example, consider the $\mathfrak{su}(1|1)$ algebra generated by our supercharge $Q_{\dot -}$ and its Hermitian conjugate $S_{\dot +}$. Their anticommutator is
\begin{equation}
    \{Q_{\dot -},S_{\dot +}\} = \Delta + \frac32 R - M_{\dot + \dot -}\,.
\end{equation}
Generic multiplets consist of two operators, which cancel 
in $Q$-cohomology. Short multiplets appear when 
$\Delta = M_{\dot + \dot -}-\frac32 R$ and consist of a single $Q$-cohomology class. 

The $\mathfrak{su}(2|1)$ algebra generated by $Q_{\dot \alpha}$ and $S_{\dot \beta}$ is also very useful. It closes on 
\begin{equation}
    \{Q_{\dot \alpha},S_{\dot \beta}\} = (\Delta + \frac32 R)\epsilon_{\dot \alpha \dot \beta} - M_{\dot \alpha \dot \beta}\,.
\end{equation}
Generic multiplets consist of quadruplets of $\mathfrak{su}(2)$ representations: a primary of spin $j$, two $Q$-descendants with spins $j\pm\frac12$, and a $Q^2$ descendant of spin $j$.
For $j=0$, the $Q$-descendants have spin $\frac12$ only. 

For general $j$, the only short multiplet (``$A_1$'') is missing the  $j-\frac12$ $Q$-descendants and the $Q^2$ descendants. It is easy to see that there is a single $Q$-cohomology class, the $Q_{\dot +}$ descendant of the primary operator with the most positive $M_{\dot \alpha \dot \beta}$ charge. It has cohomological degree $-2j-1$.
Being a $Q_{\dot +}$ descendant, it is not in the chiral ring. 

For $j=0$, there are two possible short multiplets. One (``$A_2$'') is missing the $Q^2$ descendant. The $Q_{\dot +}$ descendant gives again a $Q$-cohomology class, of cohomological degree $-1$, which is not in the chiral ring. 

The second type of $j=0$ short multiplet (``$B_1$'') consists of the primary only, which is thus a $Q$-cohomology class of cohomological degree $0$ and also a chiral ring element. 

In conclusion, $Q$-cohomology elements in superconformal theories are always annihilated by the $M_{\dot + \dot +}$ rotation generator and have non-positive cohomological degree. The cohomology of degree $0$ consists of the familiar {\it chiral ring} operators.

\subsection{Chiral Superspace and Dolbeault Forms}\label{subsec:chiralSuperspace}
Four-dimensional ${\cal N}=1$ SQFTs are often formulated in the language of superspace \cite{Wess:1992cp, Weinberg:2000cr}. Spacetime is promoted to a supermanifold with extra odd coordinates $\theta^{\dot \alpha}$ and $\bar \theta^\alpha$, and supersymmetric multiplets of operators are collected into ``superfields''\footnote{We notice the unfortunate convention of using the term ``superfield''  to denote operators on superspace, even when they are not elementary fields.} which depend on the superspace coordinates. 

In our context, it is natural to employ a ``chiral'' superspace which only employs $\bar \theta^\alpha$ odd coordinates, with the property that $\bar Q_{\alpha} = -\partial_{\bar \theta^\alpha}$ when acting on superfields.\footnote{In full superspace, such a definition would be combined with $D_{\dot \alpha} = \partial_{\theta^{\dot \alpha}}$. Given a standard ${\cal N}=1$ superfield we can Taylor expand in $\theta^{\dot \alpha}$ to restrict to the chiral superspace.} The identification $\bar \theta^\alpha \leftrightarrow \d \bar z^\alpha$ allows us to present superfields as  Dolbeault forms of type $(0,\bu)$. The Dolbeault operator can be written as
\begin{equation}
    \bar \partial \equiv \bar \theta^\alpha \partial_{\bar z^\alpha} \,.
\end{equation}

Given such a superfield/form $\calO$, we denote its $n$-th form component as $\calO^{(n)}$. The whole superfield can be reconstructed by a repeated application of $\bar \theta^\alpha \bar Q_{\alpha}$ on the $\calO^{(0)}$ component:
\begin{equation}
    \calO[\bar\theta]= e^{\bar \theta^\alpha \bar Q_\alpha} \calO^{(0)} = \calO^{(0)} + \calO^{(1)} + \calO^{(2)} \,. 
\end{equation}
The combination $Q + \bar \partial$ coincides with the superspace derivative $D_{\dot -}$ in the physical theory  and acts naturally on superfields: 
\begin{equation}
\{ Q + \bar \partial, \bar Q_{\alpha} \} =0 \,.
\end{equation}
As a result, we have the descent relation:
\begin{equation}
    (Q+\bar\partial)\calO[\bar\theta] = (Q\calO)[\bar\theta]\,,
\end{equation}
or, in components,
\begin{equation}
    Q\mathcal{O}^{(k)} + \bar{\partial} \calO^{(k-1)} = (Q\calO)^{(k)} \,.
\end{equation}

We call a superfield $\calO$ {\it semi-chiral}\,\footnote{In analogy to {\it (anti)chiral} superfields, which satisfy $D_{\dot +}\calO=D_{\dot -}\calO=0$. 
} if it satisfies 
\begin{equation}
    D_{\dot-}\calO = (Q + \bar \partial) \calO =0 \,.
\end{equation}
From the descent relations, we see that a superfield $\calO$ is semi-chiral if and only if its 0-form component $\calO^{(0)}$ is $Q$-closed. For semi-chiral superfields, the descent relations become:
\begin{equation}
    Q \calO^{(1)}+ \bar \partial \calO^{(0)}=0\,, \qquad Q \calO^{(2)}+ \bar \partial \calO^{(1)}=0 \,.
\end{equation}
Analogously, if $\calO^{(0)}$ is $Q$-exact, i.e. $\calO^{(0)} = Q \mathcal{U}^{(0)}$ then 
\begin{equation}
\calO = (Q + \bar \partial) \mathcal{U} \,.
\end{equation}

In short, we can identify the cohomology of $Q$ with the space of semi-chiral superfields modulo the image of $(Q + \bar \partial)$, simply by a translation in the superspace directions. We will often do so implicitly. 

Finally, we can comment on the OPE of semi-chiral superfields. As anti-holomorphic translations are $Q$-exact, non-holomorphic terms in the OPE 
must be $Q$-exact as well. Restricting to the $Q$-cohomology, the OPE will thus be meromorphic. Hartog's theorem forbids the existence of meromorphic functions on $\mathbb{C}^2$ singular at one point. As a consequence, the OPE must be non-singular when restricted to the $Q$-cohomology. This justifies the claim that semi-chiral operators can be organized in a \textit{semi-chiral ring}. Next, we will discuss more refined algebraic structures hidden in the OPE of semi-chiral superfields. 

\subsection{Infinite Dimensional Symmetries}\label{subsec:symmetries}
The descent relations satisfied by the components of a semi-chiral superfield can be interpreted as conservation laws which hold in the twisted theory. Each semi-chiral superfield $\calO$ actually gives rise to infinitely many conservation laws: given any $\bar \partial$-closed $(2,\bu)$ form $\rho$ the combination $\rho\, \calO$ is also conserved. 

This is completely analogous to what happens for holomorphic operators in 2d CFTs \cite{Saberi:2019fkq, SWchar}. Recall that a 2d CFT always includes a holomorphic stress tensor $T(z)$, and may include other holomorphic operators. Every holomorphic local operator in 2d is automatically a conserved current, being invariant under $\bar{\partial}$. Moreover, every holomorphic local operator remains a conserved current when multiplied by a holomorphic function, giving rise to various infinite symmetry enhancements. For example, the infinite Virasoro generators arise from the conservation of the holomorphic stress tensor $T(z)$.

More generally, every holomorphic operator $\calO(z)$ in a 2d CFT gives rise to a tower of symmetry generators $\hat{\calO}_n$, $n \in \mathbb{Z}$, acting on the space of local operators near the origin by\footnote{We are using a mathematical indexing convention here, omitting a shift by the conformal dimension of $\calO(z)$ which is standard in the physics literature. With these conventions, $\hat{\calO}_{-1}$ represents a regularized multiplication by $\calO(0)$.}
\begin{equation}
    (\hat{\calO}_n A)(0)= \oint_{S^1} \frac{\d z}{2\pi i} z^{n} \calO(z) A(0)\,.
\end{equation}
For non-negative $n$, these modes capture the singular part of the OPE of $\calO(z)$ with other operators. In the mathematical literature, the non-negative modes are sometimes collected in a generating function called the $\lambda$-{\it bracket} \cite{kac1998vertex}:
\begin{equation}
    \{\calO\, {}_{\lambda}\, A\}(0)= \oint_{S^1} \frac{\d z}{2\pi i} e^{\lambda z} \calO(z) A(0)\,.
\end{equation}
Conversely, the negative modes $\hat{\calO}_{-n-1}$ define the regularized products $(\partial^{n} \calO \, A)$.

The contour-integral definition of the negative modes involves the singular forms $z^{-n-1} \frac{\d z}{2\pi i}$. In order to set up an analogy to problems in higher dimensions, it is useful to identify these singular forms as derivatives of the 2d Bochner-Martinelli kernel $\omega_{\mathrm{BM}} = \frac{1}{2\pi i}\frac{1}{z}$, i.e. the Green's function for the $\bar \partial$ operator.

Finally, the collection of holomorphic 1-forms $\frac{\d z}{2\pi i} z^{n}$ which appear in the definition of the modes spans the Dolbeault cohomology $H^{1,0}(\mathbb{C}^\times) \cong \bbC[z^{\pm 1}] \, \d z$. It is also useful to give a Serre-dual description as the basis of Laurent monomials in $H^{0,0}(\bbC^\times) \cong \bbC[z^{\pm 1}]$.\footnote{The duality pairing is given by integration on $S^1$, with the explicit dual of the Laurent monomial $z^n$ given by the one-form $z^{-n-1} \frac{\d z}{2\pi i}$.} This is just the statement that the modes $\hat{\calO}_n$ are the coefficients of a Laurent expansion of $\calO(z)$ around $z=0$.

We are now ready to generalize these notions to $n$ complex dimensional holomorphic theories. The symmetries 
\begin{equation} \label{eq:Orho}
    \hat \calO_\rho = \oint_{S^{2n-1}} \rho\, \calO \, ,
\end{equation}
of the space of local operators associated to the holomorphic superfield $\calO$ are now labelled by forms $\rho$ living in $H^{n,\bu}(\mathbb{C}^n-0)$ \cite{Gwilliam:2018lpo,Saberi:2019fkq}. Indeed, such an operator varies by $Q$-exact amounts if we deform the $S^{2n-1}$ integration contour and if we vary $\rho$ by a $\bar \partial$-exact form. 

A standard calculation reveals that the Dolbeault cohomology $H^{n,\bu}(\mathbb{C}^n-0)$ is concentrated entirely in degrees 0 and $n-1$. The degree $n-1$ part is dual, under integration on $S^{2n-1}$, to polynomials $\mathbb{C}[z_1,\dots,z_n]$.  It consists of the $n$-dimensional Bochner-Martinelli kernel $\omega_{BM}$ and its holomorphic derivatives.\footnote{The Bochner-Martinelli kernel $\omega_{\mathrm{BM}}$ is the $(0,n-1)$-form defined to be the fundamental class of $S^{2n-1}$ when wedged with $\d^nz$ (some conventions define $\omega_{BM}$ to be this $(n,n-1)$-form instead). So, for any continuously differentiable function $f$ on the closure of a domain $D$,
\begin{equation}
    f(0) = \oint_{\partial D} f(z) \, \omega_{\mathrm{BM}} \,\d^nz - \int_{D} \bar{\partial} f(z) \, \omega_{\mathrm{BM}} \,\d^nz\,.
\end{equation}
When $f$ is holomorphic, the second term disappears. When $f$ is cohomologically holomorphic, the second term is $Q$-exact, being proportional to the first descendant of $f$.} It is analogous to the negative modes in 2d and captures the regularized products $(\partial_1^{k_1} \cdots \partial_n^{k_n} \calO \, A)$. We can denote the corresponding modes as $\hat \calO_{-k_1-1,\cdots,-k_n-1}$.

The degree $0$ part consists of polynomial holomorphic top forms $\mathbb{C}[z_1,\dots,z_n] \d^n z$.\footnote{The dual $(0,n-1)$ forms are best described by \v{C}ech representatives of the form $\mathbb{C}[z_1^{-1},\dots,z_n^{-1}]\frac{1}{z_1\cdots z_n}$ at the intersection of the $n$ patches in $\mathbb{C}^n$ where one coordinate is non-zero.} The integral \eqref{eq:Orho} thus employs the $(n-1)$-th descendant in $\calO$. These modes are analogous to the non-negative modes in 2d and can also be combined into a $\lambda$-bracket \cite{Oh:2019mcg}:
\begin{equation}\label{eq:lambdabracket}
    \{ \calO_1 \,{}_\lambda \calO_2\} =  \oint_{S^{2n-1}} e^{\lambda \cdot z} \d^n z \, \calO_1(z) \calO_2(0) \,,
\end{equation}
which generalizes the secondary products in TFTs \cite{Beem:2018fng}.\footnote{In topologically twisted theories, the modes would instead be characterized by the de Rham cohomology $H^\bu(\bbR^d-0)$. This is supported entirely in degrees $0$ and $d-1$, each with dimension 1. In this case, the ``mode'' extracted by the ``degree 0 class'' is just the $(d-1)$th descendant $\calO^{(d-1)}$ itself. Instead of a $\lambda$-bracket, we just have a single secondary product \cite{Beem:2018fng}. Alternatively, we can think of the $\lambda$-bracket as a generating function for the many secondary products in holomorphically twisted theories \cite{Oh:2019mcg}.} In analogy with the vertex algebra situation, we will say that fields $\calO_1$ and $\calO_2$ have {\em regular} OPE if such $\lambda$-brackets vanish \cite{Bakalov_2003} (see also \cite{Oh:2019mcg, Garner:2022its, Costello:2020ndc}). See appendix \ref{appendix:lambdabrackets} for some algebraic properties of this bracket. 

The $\lambda$-bracket and the regularized product in $n$-dimensional holomorphic theories  satisfy associativity axioms which generalize the axioms of a vertex operator algebra. However, these axioms are further complicated in this derived setting: the operations can be defined on local operators rather than on their $Q$-cohomology, so that extra higher brackets appear in the associativity relations, much as $L_\infty$ or $A_\infty$ algebras appear in two-dimensional TFTs \cite{Garner:2022its}. We leave a further discussion to a companion paper \cite{Gaiotto:2024gii}.

\subsection{Twisting the Stress Tensor Supermultiplet}
\label{subsec:stresstensor}
As a warm-up, consider a theory with 4d ${\cal N}=1$ supersymmetry equipped with a conserved flavour current. 
The conserved current is part of a real {\it linear} supermultiplet ${\cal J}[\theta, \bar \theta]$ such that
\begin{equation}
    D^2 {\cal J} =0\,, \qquad \qquad \bar D^2 {\cal J} =0\,.
\end{equation}
We immediately see that $J[\bar\theta] \equiv D_{\dot +} {\cal J}|_{\theta=0}$ is a semi-chiral superfield, which plays a role analogous to that of a Kac-Moody current in 2d CFT. 

The zero mode $\hat J_{0,0}$ implements global flavour rotations, but 
the remaining positive modes $\hat J_{n,m}$ give an action of a much larger symmetry algebra: holomorphic position-dependent flavour rotations. The negative modes give extra Grassmann-odd symmetry generators which essentially multiply an operator by derivatives of $J$ 
\cite{Gwilliam:2018lpo, faonte2019higher, SWchar}.

A similar analysis, but slightly more involved, is also possible for the stress tensor multiplet. The most general ``${\cal S}$''-supermultiplet was derived in \cite{Dumitrescu:2011iu}. It includes a superfield $ S_{\dot \alpha \beta}$.  The component
\begin{equation}
S_{\alpha} \equiv D_{\dot +} S_{\dot + \alpha}|_{\theta=0}
\end{equation}
satisfies $(Q + \bar \partial) S_{\alpha} = \bar Y_{\dot + \alpha}$, where $ \bar Y_{\dot + \alpha}$ is a component of a closed form $\bar Y_\mu\dd x^\mu$. In particular,
\begin{equation}
    \partial_{\dot{+}\alpha} \bar Y_{\dot{+}}^{\alpha} = 0 \,.
\end{equation}
Therefore, $\partial_{z^\alpha} S^{\alpha}$ is semi-chiral. The operator $\bar Y_\mu$ vanishes for theories endowed with a conserved $R$-symmetry current, in which case $S_{\alpha}$ itself is semi-chiral. 

The one-form component $S^{(1)}_{\alpha}$ which enters the definition of $\widehat{(\partial_\alpha S^\alpha)}_{n,m}$ or $\hat S^\alpha_{n,m}$  includes the holomorphic part of the physical stress tensor. The $\widehat{(\partial_\alpha S^\alpha)}_{n,m}$ positive modes generate 
Hamiltonian holomorphic symplectomorphisms of spacetime: 
\begin{equation}
    \int_{S^3} f(z) \partial_\alpha S^\alpha(z) \calO(w) = - \int_{S^3} (\partial_\alpha f(z)) S^\alpha(z) \calO(w) \, ,
\end{equation}
which are diffeomorphisms generated by the Hamiltonian vectorfield $\partial_\alpha f$. In the case $S_\alpha$ itself is semi-chiral, we expect the positive modes $\hat S^\alpha_{n,m}$ to generate holomorphic diffeomorphisms of spacetime. The negative modes give extra Grassmann-odd symmetry generators which essentially multiply an operator by derivatives of $\partial_\alpha S^\alpha$ or $S_\alpha$  respectively \cite{BWthesis,Saberi:2019fkq,faonte2019higher}.

If a quantum field theory is classically scale invariant but has a non-zero beta-function at one-loop, we will thus find that $(Q + \bar \partial) S_\alpha$ vanishes classically but not at one-loop, while $(Q + \bar \partial)  \partial_{\alpha} S^\alpha$ vanishes exactly. This is a simple example of 1-loop correction to the action of $Q$.

In pure gauge theory, we will find a startling result: \textit{$\partial_{\alpha} S^\alpha$ is $(Q + \bar \partial)$-exact at one-loop} -- so, the theory actually becomes topological at first-order in perturbation theory. We dub this phenomenon {\it holomorphic confinement}, as we expect it to remain true at all orders in perturbation theory and even non-perturbatively. If this expectation holds, it implies remarkable constraints on the IR behaviour of the theory: the low energy effective field theory must also become topological upon holomorphic twist. 

\subsection{A Simple Example: the Chiral Multiplet}
\label{section:chiralmultiplet}
The free (anti)chiral multiplet is described by an (anti)chiral superfield $\Phi$, i.e. a superfield annihilated by both superspace derivatives $D_{\dot \pm}$,  which also satisfies 
an equation of motion $\bar D^2 \Phi=0$. 

Clearly 
\begin{equation}
\gamma \equiv \Phi
\end{equation}
itself is a semi-chiral superfield.\footnote{We do not even need to set $\theta=0$, as $\Phi$ is independent of $\theta$.} 
The complex conjugate $\bar \Phi$ also contains a semi-chiral superfield: 
\begin{equation}
\beta \equiv D_{\dot +} \bar \Phi \, .
\end{equation}
Together they form a four-dimensional analogue of the two-dimensional chiral $\beta \gamma$ system. 
The $U(1)$ current superfield $|\Phi|^2$ gives us $J = \beta \gamma$. A slightly more involved calculation produces the holomorphic stress tensor 
\begin{equation}
    S_\alpha = \beta \partial_\alpha \gamma - \Delta \, \partial_\alpha (\beta \gamma) \, .
\end{equation} 
Much as in a 2d $\beta \gamma$ system, the value of the coefficient $\Delta$ in the second term determines the holomorphic scaling dimension of $\beta$ and $\gamma$ and is proportional to the R-charge of $\Phi$. The dependence drops out for theories with no conserved $R$-symmetry:
\begin{equation}
   \partial_\alpha S^\alpha = \partial_\alpha \beta \partial^\alpha \gamma \, .
\end{equation}

Brackets can be readily calculated using the Green's function pairing $\beta$ and $\gamma$, which is the Bochner-Martinelli kernel. E.g. $\beta$ generates holomorphic shifts of $\gamma$, etc.

A small variation of this system is a chiral multiplet with a superpotential, with an equation of motion $\bar D^2 \Phi= \partial_\Phi {\cal W}(\Phi)$.
Now $\beta$ is not semi-chiral. Instead, we have 
\begin{equation}
    (Q + \bar \partial) \beta = \partial_\gamma {\cal W}(\gamma) \, .
\end{equation}
Notice that 
\begin{equation}
    (Q + \bar \partial) \beta \partial_\alpha \gamma = \partial_\alpha \left[{\cal W} - \Delta \,\gamma \partial_\gamma {\cal W}\right]\,,
\end{equation}
so $S_\alpha=\beta \partial_\alpha \gamma$ is not a semi-chiral superfield unless ${\cal W}$ is quasi-homogeneous with weight $1$, but $\partial_\alpha S^\alpha$ always is. This example is also sufficiently rich to admit 1-loop corrections to some of the brackets \cite{Gaiotto:2024gii}.

\section{Twisted Gauge Theory as a Holomorphic BF Theory}\label{sec:twistedAsHolo}
The holomorphic twist of four-dimensional ${\cal N}=1$ pure gauge theory has been identified in the BV formalism with a holomorphic version of the BF theory \cite{CostelloYangian,ESW,SWchar}.
The identification is non-trivial and involves simplifications which modify the field content \emph{without changing the $Q$-cohomology}. 
The twisted theory exists (classically) on any complex surface $Y$, but at the quantum level there is an anomaly which requires that $Y$ be Calabi--Yau.\footnote{This is a consequence of the Grothendieck--Riemann--Roch theorem. 
On curved space there is an anomaly to the one-loop quantization of holomorphic BF theory which is proportional to the characteristic class $c_1(T_Y)$ times the quadratic Casimir of the gauge group $G$ in the adjoint representation. For semi-simple groups, the only way this anomaly vanishes is if $c_1(T_Y) = 0$.} In the language of the previous section \ref{sec:symmetries}, the holomorphic stress tensor $S_\alpha$ is not closed at 1-loop, so only complex symplectomorphisms are symmetries of the theory. In this paper, we will focus on the properties of local operators, so we can 
take $Y = \bC^2$ with its standard holomorphic volume form $\dd^2z \equiv \dd z^1 \wedge \dd z^2$.

The elementary fields are collected in a bosonic superfield\footnote{If we do not explicitly fix a holomorphic volume form it is more natural to view $b^{(i)}$ as a $(2,i)$ Dolbeault form. 
In this paper we will always fix the volume form $\d^2 z$ and view $b^{(i)}$ as a $(0,i)$ form.} $b$ and a fermionic superfield $c$:
\begin{align}
    b=b^{(0)}+b^{(1)}+b^{(2)} \in \Omega^{0,\bu}(\bC^2,\lie{g}^\vee), \quad c=c^{(0)}+c^{(1)}+c^{(2)} \in \Omega^{0,\bu}(\bC^2,\lie{g})[1] \, ,
\end{align}
valued in the adjoint representation of the gauge group.\footnote{Strictly speaking $c$ is valued in the adjoint representation and $b$ in the coadjoint representation. We will use a nondegenerate invariant pairing on the Lie algebra to identify the two representations.} 

The $c^{(0)}$ field is a ghost and has cohomological degree $1$. The descendant $c^{(1)}$ has cohomological degree $0$ and can be identified with the anti-holomorphic part of the gauge connection. Accordingly, the 2-form part of the combination $\bar \partial c - \frac12 [c,c]$ 
is the anti-holomorphic field strength. 

For matrix Lie groups $G$, the action of holomorphic BF theory is written as 
\begin{equation} \label{eq:BFaction}
\int \d^2 z \, \Tr b \left(\bar \partial c - \frac12 [c,c] \right) .
\end{equation}
Expanding out the action, we find the BF term: 
\begin{equation}
\int \d^2 z \, \Tr b^{(0)} \left(\bar \partial c^{(1)} - \frac12 [c^{(1)},c^{(1)}] \right) .
\end{equation}
The cohomological degree of $b^{(0)}$ is $-2$. The holomorphic part of the connection is absent in this description.

Field redefinitions notwithstanding, we can compare the field content of the BF theory to the $Q$-cohomology of the space of fields of a free ${\cal N}=1$ pure gauge theory: 
\begin{itemize}
    \item the $(2,0)$ part of the field strength $F_{\dot + \dot +}$ corresponds to $\d^2 z \, b^{(0)}$,
    \item the two gaugino components $\bar{\lambda}_\alpha$, $\alpha=1,2$ which satisfy the Dirac equation $\epsilon_{\alpha \beta}\p^{\alpha} \bar{\lambda}^\beta=\partial_\alpha \bar{\lambda}^\alpha =0$ are identified with the two holomorphic derivatives $\partial_{z^\alpha}  c^{(0)}$. 
In fact, $\bar{\lambda}_\alpha$ is the leading component of an anti-chiral superfield $\overline {\cal W}_\alpha$, which can be identified with the superfield $\partial_{z^\alpha}  c$. 
\end{itemize}

The basic action above actually corresponds to a self-dual limit of the twisted theory. The full twisted action also includes 
the twisted version of the usual kinetic F-term $\int d^2 \bar \theta \, \tau \, \overline {\cal W}_\alpha \,\overline {\cal W}^\alpha$:
\begin{equation}
   \int \d^2 z \, \tau \Tr \partial_\alpha c \, \partial^\alpha c \,.
\end{equation}
This is a total derivative in the BF theory, though, so it only affects instanton calculations. It is also possible to make the coupling $\tau(z)$ position dependent and holomorphic. Mathematically, this corresponds to the statement that the space of translation invariant quantizations of holomorphic BF theory on $\bC^2$ is a torsor for the abelian group $H^4(\lie{g}[[z_1,z_2]])$.

\subsection{Local Operators}
\label{subsec:localops}

In the BV/BRST formalism, the $c^{(0)}$ field plays the role of a ghost for the gauge symmetry of the BF theory and thus $c^{(0)}$ itself is forbidden from appearing in local operators: only its derivatives are allowed. 
Indeed, Faddeev–Popov ghosts are only suitable to gauge-fix the non-zero modes of the gauge group. 
The correct way to deal with constant gauge transformations is to only consider local operators which are strictly invariant under them.\footnote{The rule is different when one studies interactions, which can include the field $c^{(0)}$. Furthermore, interactions are defined up to total derivatives, so they are identified as the cohomology of the combination $Q + \partial$. See our companion paper \cite{Gaiotto:2024gii} for a more detailed review.} This is consistent with the form of semi-chiral operators in the physical free gauge theory, which involves the gauginos $\bar{\lambda}_\alpha \sim \partial_\alpha c^{(0)}$ and their derivatives, up to the equation of motion $\partial_\alpha \bar{\lambda}^\alpha=0$.

\subsubsection{Local Operators in the Free Theory}
The $Q$-cohomology of the space of local operators in the free limit of the holomorphic BF theory consists of (graded symmetric) polynomials in $b^{(0)}$ and its holomorphic derivatives and holomorphic derivatives of $c^{(0)}$ which are invariant under constant gauge transformations. 
Thus, as a graded vector space, the space of local operators is the $G$-invariant set
\begin{equation}\label{eq:freeQcoh}
    {\cal V} \equiv \bC\left[\partial^{n}_{z^1} \,\partial^{m}_{z^2}\,c^{(0)} , \partial^k_{z^1} \,\partial^l_{z^2}\, b^{(0)}, \,n + m > 0, \,k+l \geq 0 \right]^G \,.
\end{equation}
This is also consistent with the field content of semi-chiral operators in the free physical theory. Operators in the free physical theory that 
are built from the holomorphic field strength $F_{\dot +\dot +}$ are mapped to $b^{(0)}$, and the two modes of the gaugino $\bar{\lambda}_\alpha$ modulo the equation of motion are mapped to $\partial_\alpha c^{(0)}$. 

Notice that the kinetic term in \eqref{eq:BFaction} pairs $b^{(0)}$ and $c^{(1)}$ as well as $b^{(1)}$ and $c^{(0)}$. As a consequence $b^{(0)}$ and $c^{(0)}$ have a regular OPE and the above composite operators have no renormalization ambiguities in the free theory. 
In the BV formalism, the quantum corrections to $Q$ in the free theory involve the {\it BV Laplacian} $\Delta_{\text{BV}}$, which roughly implements a single Wick contraction on the fields in the operator.\footnote{To avoid UV divergences, this operator must be suitably regularized.} The absence of $b^{(1)}$ and $c^{(1)}$ in the above cohomology representatives 
thus also means that quantum corrections\footnote{We emphasize that we are studying the free theory and quantum corrections thereof in this subsection.} do not change the free $Q$-cohomology ${\cal V}$. 

The operators in the $Q$-cohomology for the free BF theory can be promoted to semi-chiral superfields which are polynomials in 
$b$ and its holomorphic derivatives as well as holomorphic derivatives of $c$. The resulting semi-chiral operators also have no renormalization ambiguities.

\subsubsection{Local Operators in the Classical Interacting Theory}
Classically, the BRST transformations are modified in the interacting theory to 
\begin{equation}\label{eqn:ct1}
Q c = -\dbar c + \frac12 [c ,c] \, ,\qquad \qquad Q b = -\dbar b + [c, b] \, .
\end{equation}
Observe that these BRST transformations act sensibly on the space of $G$-invariant local operators which do not contain 
$c$'s but only their derivatives: the BRST transformation of any operator 
equals an infinitesimal $G$ rotation with parameter $c$ plus corrections proportional to derivatives of $c$. For example, 
\begin{align} 
(Q + \bar \partial) \partial_\alpha c &= [ c, \partial_\alpha c] \,, \nonumber \\
(Q + \bar \partial) \partial_\alpha \partial_\beta c &= [ c, \partial_\alpha \partial_\beta c] + [ \partial_\alpha c, \partial_\beta c]\,,  \label{eqn:Qint2}  \\ 
(Q + \bar \partial)  \partial_\alpha b &= [ c,\partial_\alpha b] + [\partial_\alpha c, b] \,. \nonumber
\end{align}
Thus we can drop the terms containing $c$ when $Q $ acts on $G$-invariant operators.

Here we encounter the simplest example of {\it homotopy transfer}, or a spectral sequence. The cohomology of the local operators with respect to the full interacting (tree level) differential
\begin{equation}
    Q = Q_{\mathrm{free}} + Q_{\mathrm{int}}
\end{equation}
can be computed in stages by first computing the $Q_{\mathrm{free}} = -\dbar$ cohomology, which we denoted $\cV$, defined in \eqref{eq:freeQcoh}, and then computing the cohomology of an appropriate version of $Q_{\mathrm{int}}$ acting on $\cV$.\footnote{
Homotopy transfer is a method used to recognize the full cohomology as the cohomology of a particular differential acting on $\cV$.
Said differently, there is a spectral sequence whose first page is the free cohomology $\cV$.
The differential computing the next page is given by the restricted action of $Q_{\mathrm{int}}$ on $\cV$. 
In general, there may be further terms in the spectral sequence, but for the example at hand the spectral sequence collapses at this stage.}

Omitting the ${}^{(0)}$ superscripts for legibility, the restricted action of $Q_{\rm int}$ is just 
\begin{equation}\label{eqn:ct0}
Q_{\mathrm{int}} c = \frac12 [c, c]\,, \qquad \qquad Q_{\mathrm{int}} b = [ c, b] \, .
\end{equation}
The situation is particularly simple here because $Q_{\mathrm{int}}$ does not change the overall form degree or number of anti-holomorphic derivatives in an operator. In particular, it maps ${\cal V}$ back to itself.   

This action matches naturally the interacting SUSY transformations in the twisted theory,
as formulated in past work on the categorification of the superconformal index \cite{Grant:2008sk, Chang:2013fba, Chang:2022mjp}. In these works, the role of $\cV$ is played by the space of $G$-invariant words in a collection of BPS letters and their holomorphic covariant derivatives, which precisely matches the semi-chiral superfields in the twisted theory and their holomorphic derivatives, with $c$ omitted. The leading correction to the action of the supercharge comes from the presence of the holomorphic connection in the covariant derivatives, so that $Q_{\dot -}$ maps $D_{z^\alpha} \to [\bar{\lambda}_\alpha,\cdot\,]$. This reproduces the corrections proportional to $\partial_\alpha c$ in \eqref{eqn:Qint2}. 

The holomorphic twist explains this simplification and extends it to the quantum theory, allowing one to compute the further loop corrections to the cohomology which will be discussed momentarily. 

\subsubsection{Lie Algebra Cohomology}\label{sub:LieAlgCoho}

A concise way to describe the action of $Q_{\mathrm{int}}$ on $\cV$ is in terms of relative Lie algebra cohomology. 
The Lie algebra cohomology of a Lie algebra $\tilde{\lie{g}}$ is computed by the Chevalley--Eilenberg complex $C^\bu(\tilde{\lie{g}})$ which is the space of graded polynomials on the vector space $\tilde{\lie{g}}[1]$ equipped with the Chevalley--Eilenberg differential.
If $G$ is a group whose Lie algebra $\lie{g}$ is a Lie subalgebra of $\tilde{\lie{g}}$, then
the relative Lie algebra cohomology of $\tilde{\lie{g}}$ relative to $\lie{g}$ is the cohomology of a subcomplex $C^\bu(\tilde{\lie{g}}|\lie{g})$ of $C^\bu(\tilde{\lie{g}})$ which as a graded vector space is $G$-invariants of the graded symmetric algebra on the graded vector space $(\tilde{\lie{g}} / \lie{g})[1]$.
The differential is the usual Chevalley--Eilenberg differential. 
There is a similar construction for relative Lie algebra cohomology with values in a module $M$ which we denote by $C^\bu(\tilde{\lie{g}} | \lie{g}; M)$ \cite{ChevEil}. 
For appearances of relative Lie algebra cohomology in the context of supersymmetric gauge theory see \cite{Chang:2013fba, Chang:2022mjp}.

In the example of holomorphic BF theory, the tree-level cohomology of local operators is equivalent to the Lie algebra cohomology of $\tilde{\lie{g}}=\lie{g}[[z_1,z_2]]$ relative to the subalgebra $\lie{g}$ (consisting of constant polynomials) with coefficients in the module $M = {\rm Sym}(\lie{g}[[z_1,z_2]]^*) =  \bC\left[\partial^k_{z^1} \, \partial^l_{z^2}\, b,\,k,l \in \mathbb{Z}_{\geq 0}\right]$. Following the notation above, we denote this by
\begin{equation}\label{eqn:relative1}
C^\bu\left(\lie{g}[[z_1,z_2]] \; | \; \lie{g} ; \,{\rm Sym}(\lie{g}[[z_1,z_2]]^*) \right) .
\end{equation}
The generators of $\lie{g}[[z_1,z_2]]$ play the role of derivatives $\partial_1^n \partial_2^m c^{(0)}$ of the ghost field, while subalgebra $\lie{g} \subset \lie{g}[[z_1,z_2]]$ represents the Lie algebra of constant gauge transformations, which in the notation above was given by the ghost $c^{(0)}$ and is removed from the calculation. 

Unfortunately, the computation of most Lie algebra cohomologies is notoriously difficult. Remarkable simplifications occur in the large $N$ limit for $U(N)$ gauge theories. In section \ref{sec:infiniteN} we will discuss the large $N$ limit and in section \ref{sec:finiteN} some explicit calculations in $SU(2)$
and $SU(3)$ gauge theories. 

We can present a collection of cohomology classes which exist for all $G$.
Consider any $G$-invariant polynomial $A_a \equiv P_a(b)$. As it does not contain derivatives, this is trivially $Q_{\text{int}}$-closed. As it does not contain $c$ fields, it cannot be $Q_{\text{int}}$-exact. In $SU(N)$ gauge theory, such operators can be written as polynomials in 
\begin{equation}
    A_n = \Tr b^n \, , \qquad \qquad n\geq 1 \, .
\end{equation}
As $Q_{\text{int}}$ commutes with derivatives $\partial_\alpha$ and satisfies a Leibniz rule, polynomials in the $A_a$ and their derivatives will also be $Q_{\text{int}}$-closed and give a large collection of cohomology classes. 

We can do better. We can introduce an auxiliary fermionic variable $\varepsilon$ and combine the elementary fields as $C(\varepsilon) \equiv c + \varepsilon b$. The BRST transformations become 
\begin{equation}\label{eqn:Ct0}
Q_{\text{int}}C(\varepsilon) = \frac12 [C(\varepsilon), C(\varepsilon)]\,.
\end{equation}
First of all, this allows us to recast the problem as a Lie algebra cohomology problem:
\begin{equation}\label{eqn:relative2}
    C^\bu\left(\lie{g}[[z_1,z_2,\varepsilon]] \; | \; \lie{g} \right) .
\end{equation}
Second, as the commutator is local in $\varepsilon$, $Q_{\text{int}}$ must commute with all vectorfields in $z_\alpha$ and $\varepsilon$. This includes vectorfields
\begin{align}
    \xi &= \partial_\varepsilon \,\,\,\;:\, c \to b \cr
    \tilde \xi_\alpha &= \varepsilon \partial_\alpha \,:\, b \to \partial_\alpha c \, ,
\end{align}
which anticommute to translations $\partial_\alpha$.

We can define 
\begin{align}
    B_{a-1,\alpha} &\equiv \tilde \xi_\alpha A_a \cr
    C_{a-2} &\equiv \tilde \xi_1 \tilde \xi_2 A_a \, .
\end{align}
These are manifestly $Q_{\text{int}}$-closed. As $\xi A_a=0$, we have
\begin{align}
    \xi B_{a-1,\alpha} &= \partial_\alpha A_a \cr
    \xi C_{a-2} &= 2\partial_\alpha B_{a-1}^\alpha \, .
\end{align}
The first relation implies that $B_{a-1,\alpha}$ cannot be $Q_{\text{int}}$-exact, 
as $\partial_\alpha A_a$ is not. In order to draw the same conclusion for $C_{a-2}$, we need to verify that $\partial_\alpha B_{a-1}^\alpha$ is not $Q_{\text{int}}$-exact. That follows from the relation 
\begin{equation}
    \partial_\alpha B_{a-1}^\alpha = \tilde \xi^\alpha \partial_\alpha A_a \, .
\end{equation}
We thus obtained three collections of tree-level cohomology classes. For $SU(N)$ gauge group we can denote them as symmetrized traces:
\begin{alignat}{3}
    A_n &= \Tr b^n, &&\quad n\geq 2 \nonumber \\
    B_{n,\alpha} &= \mathrm{STr}\, b^n \partial_\alpha c, &&\quad n\geq 1,\; \alpha=1,2 \label{eq:ABC} \\
    C_n &= \mathrm{STr}\, b^n \partial_\alpha c \, \partial^\alpha c, &&\quad n\geq 0 \, . \nonumber
\end{alignat}

The first element in the $B$ tower is the stress tensor superfield:  
\begin{equation}
S_\alpha = \Tr b \partial_\alpha c \,.
\end{equation}
This is perfectly analogous to the stess tensor in a two-dimensional $bc$ system.
The pure gauge theory is scale-invariant and has a $U(1)_R$ $R$-symmetry at tree level, and indeed $S_\alpha$ is a semi-chiral superfield at tree-level, as anticipated in section \ref{subsec:stresstensor}.

The auxiliary symmetries we introduced are also the zero-modes of certain currents. The symmetry $\xi$ is generated by $A_2 = \Tr b^2$. The symmetries
$\tilde \xi_\alpha$ are generated by the primitive $\Tr c \partial_\alpha c$ of $C_0$. 
Higher modes of these currents, together with those of the ghost number current $\Tr b c$ and the stress tensor $\Tr b \partial_\alpha c$, give the action of more general vector-fields in $z_\alpha$ and $\epsilon$. 

We list our conventions for the Lie algebra of $SU(N)$ and its Chevalley--Eilenberg complex that we will use in the remainder of the paper.
We will use capital letters $A,B, \dots$ for the adjoint gauge indices. The Killing form is taken to be $\delta^{AB}$ so that adjoint indices can be raised and lowered freely: 
\begin{equation}
    \Tr t^A t^B = -\frac12 \delta^{AB}, 
    \quad \Tr t^A[t^B, t^C] =  -\frac{1}{2} f^{ABC} \, ,
\end{equation}
where the trace is taken in the defining representation and $f^{ABC}$ is the structure constant $[t^A,t^B] = f^{ABC} t^C$ such that
\begin{equation}
    \sum_{A,B} f^{ABC} f^{ABD} = N \delta^{CD} \, .
\end{equation}
In particular, for $SU(2)$, we take the generators to be $t^A = -\frac{i}{2}\sigma^A$ and $f^{ABC} = \epsilon^{ABC}$.

In this notation, the tree level differential (the Chevalley--Eilenberg differential) $Qb =[c,b]$, $Qc = \frac12 [c,c]$ reads
\begin{equation}
    Qb^A = \epsilon^{ABC}c^B b^C, \quad Qc^A = \frac12\epsilon^{ABC}c^B c^C .
\end{equation}
We will also use the relation
\begin{equation}
    b^2 = b_Ab^A = -2 \Tr b^2 \, .
\end{equation}

\subsection{Loop Corrections} \label{subsec:loopCorrections}

The classical answer for the cohomology of local operators in the interacting theory must receive perturbative quantum corrections. As an example, consider the holomorphic stress tensor.  Famously, the pure gauge theory is {\it not} scale invariant quantum-mechanically. Both scale-invariance and the $R$-symmetry are broken at one-loop. The latter anomaly is one-loop exact and is computed by a famous triangle diagram and is proportional to $\Tr F \wedge F$. It is supersymmetrized to the Konishi anomaly equation \cite{Konishi:1983hf,Konishi:1985tu,Cachazo:2002ry}, which involves the chiral superfield $\Tr \overline {\cal W}_\alpha \overline {\cal W}^\alpha$. (See \cite{Yonekura:2010mc} for a discussion of the Konishi anomaly in terms of the stress tensor supermultiplets from \cite{Dumitrescu:2011iu}.)

These anomalies must have a counterpart in the holomorphic theory. A direct translation of \cite{Yonekura:2010mc} gives e.g. 
\begin{equation}
(Q + \bar \partial) S_\alpha \sim \partial_\alpha \Tr \left(\partial_\beta c \, \partial^\beta c \right) \, .
\end{equation}
We will momentarily recover this by a direct calculation of a one-loop triangle diagram in the holomorphic BF theory.

As we discussed in the previous section, this relation controls the quantum-mechanical failure of the theory to be invariant under general holomorphic coordinate transformations: 
an infinitesimal coordinate transformation generated by a holomorphic vector field $v^\alpha$ shifts the action by an amount proportional to 
\begin{equation}
    \int \d^2 z \, v^\alpha \, \partial_\alpha \Tr \left(\partial_\beta c \, \partial^\beta c \right) \, .
\end{equation}
In turns, this shifts the holomorphic gauge coupling $\tau$ by a multiple of $\partial_\alpha v^\alpha$. This is the holomorphic generalization of the scale anomaly of the physical theory.  

If $v^\alpha$ is divergence-free with respect to the standard holomorphic volume form on $\bC^2$, then this anomaly vanishes as the density above is a total derivative. Thus complex symplectomorphisms of spacetime remain a symmetry of the quantum theory.

A reader may be surprised by the statement that the action of a supercharge may receive quantum corrections. Indeed, statements of the opposite flavour are sometimes made in the literature.\footnote{A more shocking statement is actually true: even the action of translations receives quantum corrections in QFTs. Consider a current $J^\mu$ which is conserved classically but 
broken at 1-loop. Then $P_\mu J^\mu \equiv \partial_\mu J^\mu$ vanishes classically but not quantum-mechanically.}

The origin of this phenomenon is, of course, operator renormalization. As we regulate a composite  operator we introduce extra terms in the action of symmetries, which may then survive as we send the regulator to $0$. In our companion paper \cite{Gaiotto:2024gii}, we derive in such a manner all the loop corrections to the action of $Q$. The origin of these corrections is manifest in the fact that they do not act ``a letter at the time'' on the operator, but they simultaneously change multiple fields.\footnote{The action of $Q$ still satisfies the Leibniz rule, but only for the loop corrected product $O\circ O' \equiv \widehat O_{-1,-1} O'$. One may try to get rid of the quantum corrections by using the loop-corrected product to multiply elementary letters in a composite operator, but such an effort is thwarted by the fact that the regularized product is not associative. 

For example, we could try to define the stress tensor as $\Tr (b \circ \partial_\alpha c)$. 
This would actually coincide with $\Tr b \partial_\alpha c$. The Leibniz rule for $Q_{\text{int}}$ would produce 
\begin{equation}
\Tr ([c,b] \circ \partial_\alpha c) + \Tr (b \circ [c,\partial_\alpha c]) \, ,
\end{equation}
but that fails to vanish due to a 1-loop correction to the regularized product and reproduces the expected anomaly. }

\subsubsection{Loop Corrections and Higher Brackets}\label{subsub:looppresecroption}

More formally, the full quantum corrected differential in the BV formalism is defined on the space of operators:
\begin{equation}
\dbar + \hbar \Delta_{\text{BV}} + \{\mathcal{I},\cdot\}_{\text{BV}} \, ,
\end{equation}
where the BV Laplacian $\Delta_{\text{BV}}$ accounts for the quantum corrections of the free theory and the last term encodes the effect of the interactions. A proper definition of the differential requires a careful renormalization, addressing both UV and IR divergences. 

Our general expectation/conjecture is that the cohomology of the full quantum differential is computed as the cohomology of a certain differential ${\bf Q}$ acting on the cohomology ${\cal V}$ of the underlying free and classical theory, at least in perturbation theory. The differential ${\bf Q}$ should be interpreted as a quantum-corrected version of the classical interacting differential $Q_{\mathrm{int}}$. 

Operationally, the action of ${\bf Q}$ on some operator $\calO$ is computed in two steps.  First regularize the combination of the local operator and the exponential of the integrated interaction Lagrangian. Then act with $Q_{\mathrm{free}}$ and rewrite the answer as the regularization of the combination of some new local operator ${\bf Q} \calO$, and the exponential of the integrated interaction Lagrangian. The resulting ${\bf Q}$ operator will be scheme-dependent. In order for the answer to be scheme-independent, we expect that the ${\bf Q}$ obtained in different schemes should be related by quasi-isomorphisms. We refer to our companion paper for a more careful discussion and a convenient renormalization scheme \cite{Gaiotto:2024gii}.

The result is a perturbative expansion for ${\bf Q}$
\begin{equation}\label{eq:fullQ}
{\bf Q} = Q_0+  Q_1 + \cdots ,
\end{equation}  
where $Q_n$ is computed by certain $n$-loop Feynman diagrams in the holomorphic theory.\footnote{One may wonder how to take the cohomology of an operator defined by a formal power series. A natural strategy is to observe that $Q_0^2=0$ and use homotopy transfer to 
map $Q_1 + \cdots$ to the $Q_0$ cohomology. One then takes the cohomology of the new leading term, etc. See appendix \ref{sec:HomotopyTransfer} for details.}

In order to develop some intuition, we can show that the tree-level differential $Q_0$ is the expected BRST operator $Q_{\mathrm{int}}$ arising from the classical interaction.
Denote the interacting part of the action \eqref{eq:BFaction} as
\begin{equation}
    \int_{\mathbb{R}^4} {\cal I}^{(2)}(z, \bar z) \d^2 z\,.
\end{equation}
i.e. $\mathcal{I} = -\frac12\Tr b[c,c]$. This integrated operator is naively BRST invariant, as its variation is a total derivative. 
In the presence of a local operator $\calO \in {\cal V}$ placed at the origin, though, renormalization may lead to a BRST anomaly. If we simply turn off the interaction in a small ball of radius $\epsilon$ surrounding the operator, the BRST anomaly will be  
\begin{equation}
    Q_0 \calO  = Q_{\mathrm{free}}\left[\int_{|z|>\epsilon} {\cal I}(z,\bar z) \d^2z \ \calO(0,0)\right]=
    \int_{|z|=\epsilon}  \, {\cal I}(z, \bar z) \d^2 z \, \calO(0,0) \, .
\end{equation}
Here we have used the fact that $\calO,\mathcal{I} \in \cal V$ are both semi-chiral and thus are annihilated by $Q_{\mathrm{free}} = -\bar \partial$. Integration by parts gives an integral on a small 3-sphere surrounding the origin. 

By definition \eqref{eq:Orho}, this is the zero-mode of ${\cal I}$ acting on $\calO$ in the free theory, which we can also write as 
\begin{equation}
    Q_0 \calO = \{ {\cal I},\calO\} \, ,
\end{equation}
where the bracket is the $\lambda=0$ specialization of the $\lambda-$bracket \eqref{eq:lambdabracket}.  When seen as an element of the $Q_{\mathrm{free}}$ cohomology, it is independent of $\epsilon$. 
The integral greatly simplifies in the $\epsilon \to 0$ limit. Indeed:
\begin{itemize}
    \item In the absence of Wick contractions, the integral goes to $0$ as $\epsilon^3$.
    \item Each Wick contraction gives a propagator $P(z, \bar z)$ which is a $(0,1)$ form as explained below. So for the dimensional reason, we can have at most one Wick contraction, involving the single descendant in ${\cal I}^{(1)}$. Denoting the remaining collection of uncontracted fields as $F(z,\bar{z})$, we have 
    \begin{equation}
        \int_{|z|=\epsilon} \d^2 z\, F(z,\bar z) P(z,\bar z) \to F(0,0) + O(\epsilon)
    \end{equation}
    as $\epsilon \to 0$. In other words, the fields in ${\cal I}$ which survive the Wick contractions can be placed 
    at the origin. As a consequence, removing the regulator projects $Q_0 \calO$ back to ${\cal V}$.
\end{itemize}

The general form of $Q_n$ is that of a multi-linear ``higher bracket''\footnote{In a companion paper \cite{Gaiotto:2024gii}, we also propose an explicit description for the higher brackets in a general class of twisted theories.} with $n+2$ slots, see appendix \ref{appendix:lambdabrackets}, with all but one of the slots taken by the interaction:
\begin{equation}
   Q_n \calO = \frac{1}{(n+1)!} \left\{{\cal I}, {\cal I}, \cdots, {\cal I},\calO \right\} \, . \label{eq:QnintermsofBracket}
\end{equation}

One can argue that these higher brackets satisfy certain quadratic axioms and that the resulting ${\bf Q}$ is nilpotent order-by-order in perturbation theory thanks to these axioms and an anomaly-cancellation Maurer-Cartan condition on the interaction Lagrangian:
\begin{equation}
   \frac12  \{{\cal I},{\cal I}\} + \frac16  \{{\cal I},{\cal I},{\cal I}\} + \cdots = 0 \, .
\end{equation}
See \cite{Gaiotto:2024gii} for more details. The basic procedure for evaluating the higher brackets can be described in the following steps:
\begin{enumerate}
    \item Take the basic integral 
\begin{equation}
   \int_{\mathbb{R}^{4n}} \prod_{i=1}^n {\cal I}(z_i, \bar z_i) \d^2 z_i \, \calO(0,0) \, .
\end{equation}
\item Do a maximal number of Wick contractions using regularized propagators.
\item Act with $Q_{\mathrm{free}}$ and integrate the resulting $\bar \partial$ by parts to act on the product of propagators. 
\item Set to $0$ all descendants and project all surviving fields to the origin, by setting $\bar z_i=0$ and Taylor-expanding in $z_i$ to get an element of ${\cal V}$. 
\item Do the resulting Feynman integrals, which turn out to be finite and independent of the regulator. 
\end{enumerate}

The main feature of the Feynman diagram calculations is that they involve superspace propagators which coincide with the Bochner-Martinelli kernel: one-forms $P(x,\bar x,\dd\bar x)$ of Dolbeault type $(0,1)$ (and Grassman odd):
\begin{equation}
	P(x, \bar x, \dd \bar x) \equiv \frac{1}{4\pi^2} \frac{\bar x^2 \dd \bar x^1 - \bar x^1 \dd \bar x^2}{|x|^4} \,,
\end{equation}
where $|x|^2 = x^1 \,\bar{x}^1 + x^2 \,\bar{x}^2 $. The propagator is the Green's function for $\bar{\p} = \dd\bar{x}^1 \frac{\p}{\p \bar{x}^1}+\dd\bar{x}^2 \frac{\p}{\p \bar{x}^2}$, giving the relation
\begin{equation}
   \bar{ \p}	P(x, \bar x, \dd \bar x) = \dd\bar{x}^1 \dd\bar{x}^2  \delta^4(x)\,.
\end{equation}
This puts an immediate upper bound on the number of propagators allowed at any loop order: $Q_n$ involves an integral over $2n+2$ anti-holomorphic variables and thus $2n+1$ propagators. In what follows we will avoid writing the dependence on the antiholomorphic coordinate and write propagators as $P(x)$.
Each propagator removes a $b$ and a $c$ superfield:
\begin{equation}
     \begin{tikzpicture}
        [
    	baseline={(current bounding box.center)},
    	line join=round
    	]
    	\coordinate (pd1) at (1.*\gS,0.*\gS);
    	\coordinate (pd2) at (-1.*\gS,0.*\gS);
    	\draw (pd1) node[GraphNode] {} ;
    	\draw (pd2) node[GraphNode] {};
        \draw[GraphEdge] (pd1) -- (pd2);
         \draw[] (-1.*\gS-.2,0.*\gS-.5) node {$c(x)$};
 \draw[](1.*\gS+.3,0.*\gS-.4) node{$b(y)$};
 \draw[](-1.*\gS+1,0.*\gS+.5) node{$P(x-y)$};
    \end{tikzpicture}
\end{equation}
and each interaction $\mathcal{I}(x) = -\tfrac{1}{2}\Tr b[c,c](x)$,
adds two $c$ superfields and one $b$ superfield.
The net number of $c$ fields thus increases by $1$, while the net number of $b$ fields decreases by $n$.
A simple scaling argument also shows that $Q_n$ will increase the number of each type of derivative by $n$. 

Notice that the action of the differential is well-defined and can in principle be computed on all polynomials in the elementary fields and their holomorphic derivatives. For a gauge theory with compact gauge group, we then restrict the action of the general differential to operators which involve derivatives of $c$ but not $c$ itself and are $G$-invariant. 

\subsubsection{Tree Level}
The only tree-level diagram has a single internal line, contracting a single field in $\calO(x,\bar x)$ with a field in $\mathcal{I}(y,\bar y)$. In particular, it satisfies the Leibniz rule: other uncontracted fields in $\calO(x,\bar x)$ go along for the ride. It also commutes with $\partial_\alpha$. 

If we take $\mathcal{O} = b$, under the action of the tree level $Q_0$ we have
\begin{equation}
\begin{split}
Q_0 \, b(x) =  \{\cI,b \}(x) &=
     \int_{|y-x|>\epsilon}\ \dd^2 y \ \,\bar{\p}_y P(y-x)  \, [c(y),b(y)]   \\
     &= [c(x),b(x)] - \bar{\p}_x \int_{|y-x|>\epsilon} \dd^2 y  \, P(y-x)\, [c(y),b(y)]\, ,
\end{split}
\end{equation}
where we performed a single Wick contraction between $b$ in $\calO$ and $c$ in $\cal I$ in two ways. We then integrated by parts the free differential to get to the second line. The result is just $[c(x),b(x)]$ up to a term exact in the free cohomology $\mathcal{V}$. Similarly, we find $Q_0\, c = \{\cI,c \} = \frac12 [c,c]$, reproducing the action of $Q_{\mathrm{int}}$ in eq. \eqref{eqn:ct0}. It is easy to see that $Q_0 \calO$ coincides with $Q_{\mathrm{int}} \calO$ in general, as they both satisfy the Leibniz rule and commute with derivatives.

\subsubsection{One Loop and Beyond}\label{subsection:1loopandbeyond}

At one-loop, $Q_1 \calO =\frac12 \{\mathcal{I},\mathcal{I},\calO\}$ only receives contributions from triangle diagrams. The resulting Wick contractions remove two fields in $\calO$ and two fields in each of the two interaction vertices, as shown below:
\begin{center}
\begin{tikzpicture}
    [
    baseline={(current bounding box.center)},
    line join=round
    ]
    \coordinate (pd1) at (-0.866*\gS,-0.5*\gS);
    \coordinate (pd2) at (0.866*\gS,-0.5*\gS);
    \coordinate (pd3) at (0.*\gS,1.*\gS);
    \draw (pd1) node[GraphNode] {};
    \draw (pd2) node[GraphNode] {};
    \draw (pd3) node[GraphNode] {};
    \draw[GraphEdge] (pd1) -- (pd2);
    \draw[GraphEdge] (pd1) -- (pd3);
    \draw[GraphEdge] (pd2) -- (pd3);
    \draw[] (-0.866*\gS-.2,-0.5*\gS-.5) node{$\mathcal{I}(x)$};
    \draw[] (0.866*\gS+.2,-0.5*\gS-.5) node{$\mathcal{I}(y)$};
    \draw[] (0.*\gS+.2,1.*\gS+.3) node{$\calO(0)$};
\end{tikzpicture}\,.
\end{center}
Without explicitly evaluating the integral, we see the resulting $Q_1$ acts on {\it pairs} of fields within an operator. It also has to respect the gauge symmetry and the perturbative spacetime symmetry $U(2)$. For example, it will transform one $b$ field into a $c$ field and add two holomorphic derivatives, with spacetime indices contracted. For more details involving  computing this bracket for specific operators $\calO$ see appendix \ref{appendix:loopbracket}. An important computation \eqref{eq:Qbb} involves acting with $Q_1$ on $\Tr b^2$, i.e. 
\begin{equation} \label{eq:kon}
Q_1 \Tr b^2 \propto \partial_\alpha \left(\Tr b \partial^\alpha c \right)\,.
\end{equation}
This is $\partial_\alpha S^\alpha \equiv \partial_\alpha \left(\Tr b \partial^\alpha c \right) $, i.e. the derivative of the stress tensor. Even though $\partial_\alpha S^\alpha$  belongs to a nontrivial cohomology class at tree level, it becomes $Q$-exact at one loop! As a consequence of the filtration discussed above and more explicitly in \eqref{eq:QnkillsshortOps}, higher order corrections all act trivially on $\Tr b^2$, so \eqref{eq:kon} does not receive further perturbative corrections.

As we discussed before and review in detail in the appendices \ref{appendix:hamiltoniansyp} and \ref{appendix:loopbracket}, $\partial_\alpha S^\alpha$ generates Hamiltonian symplectomorphisms. The exactness of $\partial_\alpha S^\alpha$ implies that the image of all $Q$-closed operators under Hamiltonian symplectomorphisms must be $Q$-exact as well. 

As a consequence, the holomorphic derivatives of all local operators vanish in cohomology. Effectively, the holomorphic theory becomes topological at one loop. We dub this peculiar phenomenon {\it Holomorphic Confinement}, as it is compatible with the expectation that the physical gauge theory should become gapped, i.e. topological, in the IR. Conversely, whatever low energy effective description applies to the physical theory, it should become topological upon holomorphic twist. 

Although the general triangle diagram is straightforward to evaluate, and we do so in the companion paper \cite{Budzik:2022mpd}, it turns out that the general structure we just described together with the constraint $\{ Q_0,Q_1 \}=0$ we derive momentarily, is sufficient to ``bootstrap'' the detailed form of $Q_1$ up to two unknown coefficients, which can be determined by direct integration.

Experimentally, the analogous statement seems to apply to all $Q_n$'s. The crucial requirement is that the relation 
\begin{equation}
    \mathbf{Q}^2 \calO= 0, \quad \forall \, \calO \in \mathcal{V}
    \label{eq:Qfull2O}
\end{equation}
with $\textbf{Q}$ as in \eqref{eq:fullQ}, should hold order-by-order in perturbation theory. The first non-trivial equation is $\{ Q_0,Q_1 \}=0$. At higher order, one has 
\begin{equation}
    \sum_{n_1+n_2 = n}Q_{n_1}Q_{n_2} \calO = 0, \quad \forall \, \calO \in \mathcal{V}, \, \forall \, n\geq 0\,. \label{eq:Q2pert}
\end{equation}

Denote the space of the monomials by $V(N_b,N_c,J_1,J_2)$ where $b$ or its derivatives appear $N_b$ times, $c$ or its derivatives appear $N_c$ times, and the total number of derivatives with respect to $z_i$ is $J_i$. We can make the following observation based on the structure of Feynman diagrams:
\begin{itemize}
    \item The action of $Q_{n}$ is linear. 
    \item $Q_n$ annihilates monomials of length shorter than $n+1$, since there are not enough fields in the monomial to perform the Wick contractions needed:
    \begin{equation}
     Q_n V(N_b,N_c,J_1,J_2) = 0, \quad \text{if } N_b+N_c <n+1\,. \label{eq:QnkillsshortOps}
    \end{equation}
    \item $Q_n$ takes a monomial of length $n+1$ and outputs monomial of length $2$. More precisely, it takes a monomial with $(n+1)$ $b$ fields and outputs a monomial of length two that contains a single $b$ field and a single $c$ field.  It also takes a monomial with $n$ $b$ fields and one $c$ field to a monomial with two $c$ fields:
    \begin{align}
        Q_n: V(n+1,0,J_1,J_2) &\rightarrow V(1,1, J_1+n,J_2+n),\\
        Q_n: V(n,1,J_1,J_2) &\rightarrow V(0,2, J_1+n,J_2+n) \, .
    \end{align}
    \item $Q_n$ acts as derivation if the length is longer than $n+1$.
\end{itemize}

Another experimental observation is that $Q_1^2 \neq 0$, 
which implies $Q_2 \neq 0$. In the companion paper \cite{Budzik:2022mpd} we indeed compute the only Feynman diagram which contributes to the two-loop differential and find it non-vanishing. 

 Using the contents of the third bullet above, one can define an auxiliary classical symmetry $U(1)_b$ under which $b$ has charge $1$ and $c$ has charge $0$, commuting with derivatives. Then, the tree level $Q_0$ preserves the symmetry, while higher $Q_n$ reduce the $U(1)_b$ charge. 
 
 The existence of this filtration makes the action of ${\bf Q}$ well-defined on any given monomial in the fields. It also helps ensure that one can define a good cohomology problem in perturbation theory via homotopy transfer, as a spectral sequence.

\section{Adding Matter} \label{sec:addingMatter}

In this section, we discuss briefly how the holomorphic theory is modified for gauge theories with matter. As explained in section \ref{section:chiralmultiplet},
chiral multiplets transforming in some representation $R$ of the gauge group give rise to bosonic superfields $\gamma$ transforming in $R$ as well as fermionic superfields $\beta$ transforming in $\bar R$; after complexification we identify this with the dual representation $R^*$.

\subsection{Twisted SQCD}
\label{s:sqcd}
The action for the matter field takes the form of a ``gauged $\beta \gamma$ system''\footnote{Analogous gauged $\beta \gamma$ systems occur in 2d, e.g. in the chiral algebra subsector of 4d ${\cal N}=2$ gauge theories \cite{Beem:2013sza}.} 
\begin{equation} \label{eq:BFactionmatter}
\int \d^2 z \, \beta \left(\bar \partial \gamma - [c,\gamma] \right) + W(\gamma) \, ,
\end{equation}
where we included a superpotential $W(\gamma)$, see \cite{Saberi:2019fkq}.
The BRST transformation of $c$ is unchanged from \eqref{eqn:ct1}. The remaining fields transform at tree-level as
\begin{equation}\label{eqn:ct2}
Q_0 b = -\dbar b + [c, b] +  \gamma\beta \,, \quad \quad Q_0 \gamma = -\dbar \gamma +[c ,\gamma]\,, \quad \quad Q_0 \beta = -\dbar \beta + [c, \beta] + \partial_\gamma W \, .
\end{equation}
Local operators are built as before, including $\beta$, $\gamma$ and their derivatives. 
From the point of view of the free theory, the $b c$ and $\beta \gamma$ systems are essentially identical, so the computation of the free brackets involves literally the same Feynman diagrams. The main difference is in the form of the interactions. 

An immediate consequence of the modified BRST transformations is that $G$-invariant polynomials in $b$ are not $Q$-closed anymore. On the other hand, chiral operators of the physical theory map to $G$-invariant polynomials in $\gamma$, modulo polynomials multiple of $\partial_\gamma W$.

In the absence of superpotentials, $\beta$'s and $\gamma$'s appear on the same footing in the twisted theory, potentially leading to interesting symmetry enlargements which persist beyond tree level. 

For example, a natural extension of supersymmetric Yang-Mills theory for $SU(N)$ is supersymmetric QCD (SQCD). 
At the level of the twist this gives rise to matter fields
$(\gamma^i, \tilde \beta^a)$ transforming in the fundamental (respectively anti-fundamental) representation and 
$(\tilde \gamma_a, \beta_i)$ anti-fundamental (respectively fundamental) representation with no superpotential. 
In particular, the partners $\tilde \beta^a$ of the anti-fundamental chiral multiplets and the fundamental chiral multiplets $\gamma^i$ have the same gauge quantum numbers, 
extending the naive $SU(N_f)_L \times SU(N_f)_R \times U(1)_B$ flavour symmetry to $SU(N_f|N_f)$.\footnote{For $SU(2)$ gauge group, there is an extension of the flavor symmetries to the strange Lie supergroup $P(2N_f - 1)$ which contains $SU(2N_f)$ as its even part. It would be interesting to compare these symmetry enhancements across dualities such as Seiberg duality. They would provide a non-trivial test of the conjecture that the simplification of twisted gauge theories to gauged $\beta\gamma$ systems holds beyond perturbation theory.}

Concretely, we can build a holomorphic stress tensor of the form 
\begin{equation}
    S_\alpha = \mathrm{Tr}\, b \partial_\alpha c + c_1 \tilde \Gamma^A \partial\, \Gamma_A+ c_2 \partial \,\tilde \Gamma^A \Gamma_A
\end{equation}
for appropriate constants $c_1$, $c_2$. The $\Tr b^2$ operator is not in tree-level cohomology anymore, and neither the stress tensor or its derivatives are exact anymore, at least in perturbation theory. 

Another important example is that of ${\cal N}=4$ SYM. In that example, the holomorphic twist consists of a holomorphic $bc$ system together with three $\beta \gamma$ systems each transforming in the adjoint representation. 
The superpotential is $\Tr [\gamma^1, \gamma^2]\gamma^3$.
Following \cite{Chang:2013fba} we can collect all the fields together into a single adjoint superfield 
\begin{equation}
\label{eqn:N=4sym}
    C(\varepsilon_1, \varepsilon_2, \varepsilon_3) \equiv c + \varepsilon_i \gamma^i - \frac12 \epsilon^{ijk} \varepsilon_i \varepsilon_j \beta_k - \varepsilon_1 \varepsilon_2 \varepsilon_3 b \, ,
\end{equation}
which transforms at  tree-level as
\begin{equation}
\label{eqn:N=4Q0C}
    Q_0 C(\varepsilon_i) = \frac12 [C(\varepsilon_i),C(\varepsilon_i)]\,.
\end{equation}
This makes manifest an enhanced group of symmetries, consisting of holomorphic vector-fields in the $\bC^{2|3}$
parameterized by $z_\alpha$ and $\varepsilon_i$ which preserve the volume element $\d^2 z \d^3 \ep$.

The tree-level BRST cohomology can thus be presented in terms of relative Lie algebra cohomology as 
\begin{equation}\label{eqn:relative4tree}
    C^\bu\left(\lie{gl}_N \otimes \bC[[z_1,z_2,\varepsilon_i]] \; | \; \lie{gl}_N \right) .
\end{equation}
In the next section, this will allow us to make contact with the large $N$ analysis of \cite{Chang:2013fba} to provide an explicit description of the large $N$ BRST cohomology.

\subsection{Twisted Theories from Twisted Branes}
In the context of holography, another very rich class of examples include certain quiver gauge theories with superpotentials, which arise from D3 branes located at the tip of a three-dimensional Calabi-Yau cone $X$.
The holomorphic twist of such quivers gives, somewhat obviously, a gauged $\beta \gamma$ system based on the same quiver with superpotential.

Less obviously, but rather naturally, the twisted theory takes the form of an open string field theory for the corresponding D-branes in the B-twisted Calabi-Yau sigma model with target $X$ \cite{Eager:2018oww}. 
This follows from the relation between twists of D-brane world-volume theories and twisted supergravity \cite{Costello:2016mgj, Costello:2012cy, Costello:2018zrm}, which here should be a B-model on $\bC^2 \times X$.
From this perspective, the description of the twisted worldvolume theory on $\bC^2$ becomes very explicit.
If $B$ denotes the algebra of functions on a non-singular Calabi--Yau manifold $X$, then the complex of fields of the worldvolume theory on a stack of $N$ twisted D3 branes located at a point $p \in X$ is
\begin{equation}
    \Omega^{0,\bu}(\bC^2) \otimes \text{Ext}_B(\cO^{\oplus{N}}_p,\cO^{\oplus{N}}_p) [1] \, ,
\end{equation}
where $\cO_p$ is the skyscraper sheaf at $p$.
Moreover, in this situation the Ext-algebra $\text{Ext}_B(\cO_p,\cO_p)$ is equipped with a cyclic structure which determines the BV structure on this complex of fields.

When $X$ is a Calabi--Yau cone then one should replace the algebra of functions by a non-commutative resolution $B$.
In terms of the quiver determining the Calabi--Yau cone this is the non-commutative Jacobi algebra associated to the superpotential.
In any case, the resulting complex of fields of the holomorphic twist of the worldvolume theory probing the singular point $p$ of $X$ is completely analogous to the smooth case; it can be written as
\begin{equation}
    \Omega^{0,\bu}(\bC^2) \otimes \lie{gl}_N [A] [1]
\end{equation}
where we have introduced the Koszul dual algebra $A = \text{Ext}_B (\cO_p,\cO_p)$.
Notice that $A$ will, in general, admit a model as an $A_\infty$-algebra.

We will unpack this description.
The $N \times N$ matrix valued fields $C_i$ of the holomorphic twist of the four-dimensional gauge theory can be put in correspondence  with the boundary local operators $a^i$ for the B-branes wrapping $\bC^2$ at the tip of the cone $X$, and the tree-level differential takes the form 
\begin{equation}
    Q C_i = f_i^{jk} C_j C_k + f_i^{jkt}C_j C_k C_t + \cdots
\end{equation}
where $f$ are the structure constants for the $A_\infty$ algebra $A$ of boundary local operators in the B-twisted sigma model.\footnote{Their transformations should follow from the open string field theory action 
\begin{equation}
    \int \d^2 z \, \eta^{ij} C_i \bar \partial C_j + \eta^{ijk} C_i C_j C_k + \cdots.
\end{equation}
where $\eta^{\cdots}$ are disk correlation functions in the B-twisted sigma model.}

A single D3 brane usually decomposes into several fractional D-branes, each of which gives rise to a separate gauge group. Elements in a certain subalgebra $A_0$ of $A$ will be dual to the $c$ ghosts for the quiver gauge fields, others will be dual to the $b$ fields and the rest to the $\beta$ and $\gamma$ fields from the chiral multiplets. 

In a compact form, we can write 
\begin{equation}
    C =\sum_i C_i a^i
\end{equation}
and the differential as a Maurer-Cartan equation for deformations of the B-branes:
\begin{equation}
    Q C = \frac12(C,C) + \frac13 (C,C,C) + \cdots
\end{equation}
where the parentheses indicate the operations of the $A_\infty$ algebra.

This is the standard string theory dictionary: the boundary couplings $C_i$ for the world-volume theory of the string can be promoted to fields on the D-brane world-volume. 
In the BV formalism, the Maurer--Cartan equation simultaneously encodes gauge invariance of a deformed boundary and determines the BRST differential for the world-volume theory. 
This is a version of the well-known relation between boundary beta functions and world-volume equations of motion.\footnote{It is also completely analogous to the standard presentation of open string field theory---the field theory encoding the world-volume theory of D-branes \cite{Witten:1985cc}. See also \cite{Herbst:2004jp} and many references therein. 
String field theory is usually not a local field theory, but it is for the B-model \cite{Bershadsky:1993cx}, essentially due to the absence of worldsheet instantons.}

The tree-level BRST cohomology can be expressed in terms of Lie algebra cohomology as
\begin{equation}\label{eqn:relative2tree}
    C^\bu\left(\lie{gl}_N\left[A\right] [[z_1,z_2]] \; | \; \lie{gl}_N\left[A_0\right] \right) .
\end{equation}
where $\lie{gl}_N\left[A_0\right]$ is the gauge theory Lie algebra, which is a direct sum of $\lie{gl}_{k_a N}$ subalgebras. 
This form will allow us to give a very streamlined analysis of the large $N$ cohomology, following \cite{Eager:2012hx,Eager:2018oww}. This ``categorifies'' well-known simplifications in the computation of superconformal indices at large $N$ \cite{Kinney:2005ej,Eager:2012hx}. 

We provide a few examples. Perhaps surprisingly, we can also associate the holomorphic twist of pure gauge theory to a brane system. Indeed, the tree-level cohomology can be recovered from $A=\bC[\varepsilon]$, which would formally arise from a point-like brane in $Y = \bC$. 

\subsubsection{\texorpdfstring{$\cN=4$}{N=4} SYM and \texorpdfstring{$\bC^3$}{C^^3}}

The example of $\cN=4$ supersymmetric Yang--Mills theory arises from probing the tip of $X = \bC^3$ (the origin, say) and corresponds to the quiver with one vertex and three edges together with the superpotential $W = xyz - xzy$, shown in figure \ref{fig:quiverN=4}.
\begin{figure}[t]
\centering
\begin{minipage}{0.45\textwidth}
\centering
\begin{tikzpicture}[thick]
    \def\r{0.5};
    \def\thta{-7.5};
    \def\nubb{1.1};
    \draw[mmrrow=>] ({\r*cos(\thta)},{\r*sin(\thta)})
        arc (245:545:0.5);
    \draw[mmrrow=>] ({\r*cos(\thta+120)},{\r*sin(\thta+120)}) 
        arc (245+120:545+120:0.5);
    \draw[mmrrow=>] ({\r*cos(\thta+240)},{\r*sin(\thta+240)}) arc (245+240:545+240:0.5);
    \node[style={circle,draw},fill=white,minimum size=0.7cm](NR) at (0,0){\Large $N$};
    \node (l1) at ({\nubb*cos(\thta-7.5)},
        {\nubb*sin(\thta-7.5)}){$\gamma_1$};
    \node (l2) at ({\nubb*cos(\thta+120-5)},
        {\nubb*sin(\thta+120-5)}){$\gamma_2$};
    \node (l3) at ({\nubb*cos(\thta+240-5)},  
        {\nubb*sin(\thta+240-5)}){$\gamma_3$};
\end{tikzpicture}
\end{minipage}
\begin{minipage}{0.45\textwidth}
    \centering
\begin{tikzpicture}[thick]
    \def\r{0.4};
    \def\thta{-12.5};
    \def\nubb{1.05};
    \draw[mmrrow=>] ({\r*cos(\thta)},{\r*sin(\thta)})
        arc (245:545:0.5);
    \draw[mmrrow=>] ({\r*cos(\thta+120)},{\r*sin(\thta+120)}) 
        arc (245+120:545+120:0.5);
    \draw[mmrrow=>] ({\r*cos(\thta+240)},{\r*sin(\thta+240)}) arc (245+240:545+240:0.5);
    \node[style={circle,draw},fill=white,minimum size=0.7cm](NR) at (0,0){\Large $e$};
    \node (l1) at ({\nubb*cos(\thta-2.5)},
        {\nubb*sin(\thta-2.5)}){$\varepsilon_1$};
    \node (l2) at ({\nubb*cos(\thta+120)},
        {\nubb*sin(\thta+120)}){$\varepsilon_2$};
    \node (l3) at ({\nubb*cos(\thta+240)},  
        {\nubb*sin(\thta+240)}){$\varepsilon_3$};
\end{tikzpicture}
\end{minipage}
\caption{Quiver diagram for $\mathcal{N}=4$ SYM.}
\label{fig:quiverN=4}
\end{figure}
In the holomorphic twist the resulting quiver gauge theory is a $bc$ system valued in $\lie{gl}(N)$ together with three adjoint-valued $\beta\gamma$ systems that we denote by $\beta_i, \gamma^i$ for $i=1,2,3$. The superpotential is $W = \Tr \gamma^1[\gamma^2,\gamma^3]$ and from \eqref{eqn:ct2} we find the action of the tree level BRST charge
\begin{equation}
\begin{aligned}
     Q_0 c &= \frac12[c,c]\\
     Q_0 b & = [c,b] +\sum_{i=1}^3 [\beta_i,\gamma^i]\\
     Q_0 \gamma^i & = [c,\gamma^i]\\
     Q_0 \beta_i & = [c,\beta_i] + \frac12 \sum_{j,k=1}^3 \epsilon_{ijk} [\gamma^j,\gamma^k] \, .
\end{aligned}
\end{equation}
The algebra $A$ is defined as follows. Each path gives rise to a Grassman generator, i.e.
\begin{equation}
    \varepsilon_i \varepsilon_j = -\varepsilon_j \varepsilon_i \, .
\end{equation}
The node gives rise to an identity $e$
\begin{equation}
    ee = e, \quad  e\varepsilon_i = \varepsilon_i e = \varepsilon_i \, ,
\end{equation}
which recovers \eqref{eqn:N=4sym} and \eqref{eqn:N=4Q0C}. This agrees with the description of the holomorphic twist of $\cN=4$ supersymmetric Yang-Mills theory given in \eqref{eqn:N=4sym}, which we cast as the $bc$ system valued in the super Lie algebra $\lie{gl}(N)[\ep_1,\ep_2,\ep_3]$.
Therefore, in this example, we have $A = \bC[\ep_1,\ep_2,\ep_3]$ and $A_0 = \bC$.

\subsubsection{The Conifold}
Next, let's consider a stack of $N$ D3 branes probing the Calabi Yau cone over the Einstein manifold $T^{1,1}$.
This is a quiver with two nodes and four edges $x_1,x_2 \colon 0 \to 1$, $y_1,y_2 \colon 1 \to 0$, shown in Figure \ref{fig:quiverconifold}. 
This quiver is equipped with the Klebanov--Witten superpotential
\begin{equation}
W = x_1 y_1 x_2 y_2 - x_1 y_2 x_2 y_1 ,
\end{equation}
see~\cite{Klebanov:1998hh}.
The field content of the holomorphically twisted gauge theory is
\begin{itemize}
    \item a $bc$ system valued in $\lie{g} = \lie{g}_1 \oplus \lie{g}_2 = \lie{gl}(N) \oplus \lie{gl}(N)$,
    \item a pair of $\beta\gamma$ systems where
    \begin{equation}
        \gamma_i \in \text{Hom}(\bC^N, \bC^N) , \quad i=1,2
    \end{equation}
    transform in the fundamental of $\lie{g}_1$ and the antifundamental of $\lie{g}_2$,
    \item another pair of $\beta\gamma$ systems where
    \begin{equation}
        \tilde{\gamma}_i \in \text{Hom}(\bC^N, \bC^N) , \quad i=1,2
    \end{equation}
    transform in the antifundamental of $\lie{g}_1$ and the fundamental of $\lie{g}_2$.
\end{itemize}

\begin{figure}[t]
\centering
\begin{minipage}{0.45\textwidth}
\centering
\begin{tikzcd}[execute at end picture={
    \node[style={circle,draw},fill=white,minimum size=1.1cm](NL) at (-2,0){ $\mathfrak{g}_1$};
    \node[style={circle,draw},fill=white,minimum size=1.1cm](NR) at (2,0){$\mathfrak{g}_2$};
    }, thick, every label/.append style = {font = \footnotesize}
    ]
    {\begin{array}{c}\bullet\\\bullet\end{array}} &&&& {\begin{array}{c}\bullet\\\bullet\end{array}}
        \arrow[thick, no head, "{\gamma_2}"{description, pos=0.4}, curve={height=-37.5pt}, from=1-1, to=1-5, marrow=>]
        \arrow[thick, no head, "{\gamma_1}"{description, pos=0.4}, curve={height=-12.5pt}, from=1-1, to=1-5, marrow=>]
        \arrow[thick, no head, "{\tilde\gamma_2}"{description, pos=0.4}, curve={height=-12.5pt}, from=1-5, to=1-1, marrow=>]
        \arrow[thick, no head, "{\tilde\gamma_1}"{description, pos=0.4}, curve={height=-37.5pt}, from=1-5, to=1-1, marrow=>]
\end{tikzcd}
\end{minipage}
\begin{minipage}{0.45\textwidth}
\centering
\begin{tikzcd}[execute at end picture={
    \node[style={circle,draw},fill=white,minimum size=1.1cm](NL) at (-2,0){$e_1$};
    \node[style={circle,draw},fill=white,minimum size=1.1cm](NR) at (2,0){$e_2$};
    }, thick, every label/.append style = {font = \footnotesize}
    ]
    {\begin{array}{c}\bullet\\\bullet\end{array}} &&&& {\begin{array}{c}\bullet\\\bullet\end{array}}
        \arrow[thick, no head, "{\varepsilon_2}"{description, pos=0.4}, curve={height=-52.5pt}, from=1-1, to=1-5, marrow=>]
        \arrow[thick, no head, "{\eta_2}"{description, pos=0.4}, curve={height=-37.5pt}, from=1-1, to=1-5, farrow=>]
        \arrow[thick, no head, "{\varepsilon_1}"{description, pos=0.4}, curve={height=-22.5pt}, from=1-1, to=1-5, marrow=>]
        \arrow[thick, no head, "{\eta_1}"{description, pos=0.4}, curve={height=-7.5pt}, from=1-1, to=1-5, farrow=>]
        \arrow[thick, no head, "{\tilde\eta_2}"{description, pos=0.4}, curve={height=-52.5pt}, from=1-5, to=1-1, farrow=>]
        \arrow[thick, no head, "{\tilde\varepsilon_2}"{description, pos=0.4}, curve={height=-37.5pt}, from=1-5, to=1-1, marrow=>]
        \arrow[thick, no head, "{\tilde\eta_1}"{description, pos=0.4}, curve={height=-22.5pt}, from=1-5, to=1-1, farrow=>]
        \arrow[thick, no head, "{\tilde\varepsilon_1}"{description, pos=0.4}, curve={height=-7.5pt}, from=1-5, to=1-1, marrow=>]
\end{tikzcd}
\end{minipage}
\caption{Conifold quiver.}
\label{fig:quiverconifold}
\end{figure}
The superpotential becomes 
\begin{equation}
    W = \text{Tr} \, \left(\gamma_1 \tilde{\gamma}_1 \gamma_2 \tilde{\gamma}_2 - \gamma_1 \tilde{\gamma}_2  \gamma_2 \tilde{\gamma}_1 \right) .
\end{equation}
From \eqref{eqn:ct2} we find the action of the tree level BRST charge
\begin{equation}
\label{eq:conifoldQ0}
\begin{aligned}
     Q_0 c_1 &= \frac12[c_1,c_1], &  Q_0 c_2 &= \frac12[c_2,c_2] \\
     Q_0 \gamma_i & = c_1\gamma_i -  \gamma_i c_2 & Q_0 \tilde{\gamma}_i & = -\tilde{\gamma}_i c_1 + c_2 \tilde{\gamma}_i\\
    Q_0 b_1 &= [c_1,b_1] +\frac12 \sum_{i=1}^2\left(\gamma_i \beta_i -  \tilde{\beta}_i \tilde{\gamma}_i \right)   & 
    Q_0 b_2 &= [c_2,b_2] - \frac12 \sum_{i=1}^2\left( \beta_i\gamma_i - \tilde{\gamma}_i\tilde{\beta}_i\right)  \\
     Q_0 \beta_1 & = \beta_1c_1 + c_2 \beta_1 +\tilde{\gamma}_1 \gamma_2 \tilde{\gamma}_2 -  \tilde{\gamma}_2  \gamma_2 \tilde{\gamma}_1  & &\\
     Q_0 \beta_2 & = \beta_2c_1 + c_2 \beta_2 +\tilde{\gamma}_2 \gamma_1 \tilde{\gamma}_1 -  \tilde{\gamma}_1  \gamma_1 \tilde{\gamma}_2  & &\\
     Q_0 \tilde{\beta}_1 & = c_1\tilde{\beta}_1 + \tilde{\beta}_1 c_2 + {\gamma}_2 \tilde{\gamma}_2 {\gamma}_1 -  {\gamma}_1  \tilde\gamma_2 {\gamma}_2  & & \\
     Q_0 \tilde{\beta}_2 & = c_1\tilde{\beta}_2 +  \tilde{\beta}_2 c_2 + {\gamma}_1 \tilde{\gamma}_1 {\gamma}_2 -  {\gamma}_2  \tilde\gamma_1 {\gamma}_1 \,. & & 
\end{aligned}
\end{equation}

This BRST differential encodes the $A_{\infty}$ algebra $A$ which abstractly is the Koszul dual of the non-commutative Jacobi ring associated to the quiver \ref{fig:quiverconifold} with potential.
Each node defines an generator $e_{v}$ with $v = 1,2$, which are idempotent and pairwise orthogonal, i.e.
\begin{equation}
    e_1 e_1 = e_1, \quad e_2 e_2 =e_2 \quad e_1e_2 = 0 \,.
\end{equation}
It has natural multiplication with the paths $\epsilon_i$ and $\tilde{\epsilon}_{i}$
\begin{equation}
    \epsilon_{x\rightarrow y} e_y = \epsilon_{x\rightarrow y}, \quad e_x\epsilon_{x\rightarrow y} = \epsilon_{x\rightarrow y} \,.
\end{equation}
In particular, it is easy to see $e \equiv \sum e_v = e_1 +e_2$ is a multiplicative identity 
\begin{equation}
    e \epsilon_{v_1\rightarrow v_2} = \epsilon_{v_1\rightarrow v_2} e = \epsilon_{v_1\rightarrow v_2} \,.
\end{equation}
If a path and a node don't compose, their multiplication vanishes, e.g. $e_1 \eta_1 = \eta_1 e_2 = 0$. We also require 
\begin{equation}
    \epsilon_i \tilde{\epsilon}_j = \tilde{\epsilon}_i \epsilon_j = \epsilon_i {\epsilon}_j = \tilde{\epsilon}_i \tilde{\epsilon}_j = 0
\end{equation}
Furthermore there are nontrivial 3-nary $A_\infty$ brackets\footnote{2-nary bracket is simply denoted as the multiplication.}$(-,-,-)\colon A^{\times 3} \to A$, dual to the relations determined by the non-commutative derivatives of the superpotential.
\begin{equation}
    \begin{aligned}
        \eta_1 \equiv (\tilde{\epsilon}_1,\epsilon_2 ,\tilde{\epsilon}_2) &= - (\tilde{\epsilon}_2, \epsilon_2 ,\tilde{\epsilon}_1 ) \\
        \eta_2 \equiv (\tilde{\epsilon}_2, \epsilon_1, \tilde{\epsilon}_1)& =-(\tilde{\epsilon}_1,\epsilon_1 ,\tilde{\epsilon}_2 )\\
        \tilde{\eta}_1 \equiv (\epsilon_2 ,\tilde{\epsilon_2},\epsilon_1) & =-(\epsilon_1 ,\tilde{\epsilon_2}, \epsilon_2)\\
        \tilde{\eta}_2 \equiv (\epsilon_1 ,\tilde{\epsilon_1},\epsilon_2) & =-(\epsilon_2, \tilde{\epsilon_1},\epsilon_1) \,.
    \end{aligned}
\end{equation}
All of the 2-nary multiplications between $\epsilon_i$, $\tilde{\epsilon}_i$, $\eta_i$ and $\tilde{\eta}_i$ vanish except
\begin{equation}
    \tilde{\eta}_i \tilde{\epsilon}_i  = \epsilon_i \eta_{i}, \quad  \eta_{i} \epsilon_i = \tilde{\epsilon}_i \tilde{\eta}_i, \quad 
 \forall i \,.
\end{equation}
$\eta_i$ and $\tilde{\eta}_i$ multiply with the node $e_i$ naturally according to the path, e.g. $\eta_1 e_1 = e_2 \eta_1 = \eta_1$. All other 3-nary brackets vanish. 

Then if we define the superfield to be
\begin{align*}
    C &= c_1 e_1 + c_2 e_2 + \gamma_i \epsilon_i + \tilde{\gamma}_i \tilde{\epsilon}_i  + \beta_1 \eta_1 + \beta_2 \eta_2 + \tilde{\beta}_1 \tilde{\eta}_1+ \tilde{\beta}_2 \tilde{\eta}_2   - b_1    \epsilon_1 \eta_{1}  
    - b_2  \eta_{2} \epsilon_2 \,.
\end{align*}
We find \eqref{eq:conifoldQ0} is simply
\begin{equation}
    Q_0 C =  [C,C] +  (C,C,C) \,,
\end{equation}
where with a slight abuse of notation, $[x,y]$ denotes the appropriate action according to the representations. For example, if both $x$ and $y$ are in adjoint representation, $[-,-]$ is $\frac12$ times the bracket $[x,y]$ for adjoint action; if $x$ is in adjoint representation and $y$ is in the fundamental, then the $[-,-]$ is simply the matrix multiplication. Note that $e_v$, $\eta_i$ and $\tilde{\eta}_i$ commute with all the fields and $\epsilon_i$, $\tilde{\epsilon}_i$ anticommute with the $c_i$, $\beta_i$ and $\tilde{\beta}_i$.

Lastly, in this example $A_0$ is determined by the number of nodes, i.e. $A_0 = \bC\oplus \bC$.

\subsection{\texorpdfstring{${\cal N}=2$}{N=2} Gauge Theories}
For ${\cal N}=2$ gauge theories, the ``algebra'' presentation of the tree-level cohomology problem involves an auxiliary Lie algebra which appeared before in \cite{Costello:2018fnz},
built from the data of a Lie algebra $\lie{g}$ and a symplectic representation $R$ of $\lie{g}$.
The bosonic part of the algebra is $T^* \lie{g}$, with generators $t^a$ and $\tilde t_a$
and non-trivial commutators $[t^a,t^b] = f^{ab}_c t^c$ and $[t^a,\tilde t_b] = f^{ac}_b \tilde t_c$.\footnote{Here we are simplifying the discussion and only working with the totalized $\bbZ/2$ grading. 
Like twists of $\cN=1$ theories (without superpotential) there exists a mixed ghost/parity grading by the integers where the bosonic part is $T^*[2] \lie{g}$, for example.} The fermionic part of the algebra is $\Pi R$, with generators $\rho^i$ and commutators 
$[t^a,r^i] = f^{ai}_j r^j$ and $\{r^i,r^j\} = f^{ija} \tilde t_a$, where we raised an index in the structure constants using the symplectic form. 

Then \begin{equation}
    C^\bu\bigg( \; \left(T^* \lie{g} \oplus \Pi R\right) \otimes \bC[[z_1,z_2,\ep]] \; | \; \lie{g}\; \bigg) ,
\end{equation}
reproduces the tree-level cohomology of the twisted ${\cal N}=2$ gauge theory with gauge group $G$ and matter representation $R$. The chirals in the matter hypermultiplets give $\gamma_i$ and $\beta_i$ fields packaged in $\Pi R[[\epsilon]]$, while the adjoint chirals in the vectormultiplets are associated to the $\tilde t_a$ generators. 

\subsection{Quivers from Orbifold Projections}
Affine ADE ${\cal N}=2$ quivers with $SU(d_a N)$ gauge groups arise from $N$ D3 branes in probing the singular point of $\bC^2/\Gamma \times \bC$, where $\Gamma$ is the corresponding finite subgroup of $SU(2)$. More general ${\cal N}=1$ quivers arise from 
cones of the form $\bC^3/\Gamma$ for a discrete subgroup $\Gamma$ of $SU(3)$\cite{Douglas:1996sw,Lawrence:1998ja}. 

In all of these examples, the matter content of the theory is obtained by ``orbifold projection'' of the fields in an $U(|\Gamma|N)$ ${\cal N}=4$ gauge theory. The idea is to embed $\Gamma$ into the gauge group as a permutation group of $|\Gamma|$ copies of $\bC^N$ as well as in the $SU(3)$ flavour group in the obvious way. One then projects onto $\Gamma$-invariant fields. 

The permutation action of $\Gamma$ decomposes $\bC^{N |\Gamma|}$ into $N d_a$ copies 
of each representation $R_a$ of dimension $d_a$. Then the gauge fields are projected to $U(d_a N)$ gauge fields, while fields which transform in the (anti)fundamental representation $(3)$ of $SU(3)$
form $d_a N\times d_b N$ matrices for every copy of $R_a$ into $(3)\otimes R_b$ (or viceversa). 

We can implement the orbifold projection directly in the twisted theory. We start from the algebra $\lie{gl}_{|\Gamma|}\otimes \bC[[\ep_1,\ep_2,\ep_3]]$ with $\Gamma$ acting as permutations on $\bC^{|\Gamma|}$ and 
as a subgroup of $SU(3)$ on $\ep_i$. The correct algebra $A$ is then the $\Gamma$-invariant part of $\lie{gl}_{|\Gamma|}\otimes \bC[[\ep_1,\ep_2,\ep_3]]$.

\section{The Large \texorpdfstring{$N$}{N} Cohomology}
\label{sec:infiniteN}

We have argued in section \ref{sub:LieAlgCoho} and section \ref{sec:addingMatter} that the BRST complex of local operators at tree level for a variety of twisted $SU(N)$ theories can be expressed as a certain cochain complex
\begin{equation}
    C^\bu\left(\lie{gl}_N\left[A \otimes \bC[[z_1,z_2]]\right] | \lie{gl}_N\left[A_0\right]\right) \,.
\end{equation}
For pure gauge theory, $A = \bC[[\epsilon]]$ for an odd variable $\epsilon$. For 
${\cal N}=4$ supersymmetric Yang--Mills theory, $A = \bC[[\epsilon_i]]$ for three odd variables $\epsilon_i$. For quivers associated to D-branes at Calabi-Yau singularities, $A$ is the algebra of boundary local operators in the associated B-model D-branes. 

If we work at large $N$, these cohomology problems can be solved in a rather uniform way. For ${\cal N}=4$ SYM, a conjectural solution was proposed in \cite{Chang:2013fba}, in a form which can be easily generalized. 
As in the previous section, denote as 
\begin{equation}
    C(z,\ep) = c(z) + \cdots
\end{equation}
the polynomial in $\ep_i$ which collects all the fields in the problem and depends holomorphically on $z_\alpha$. We view this as a linear element in the complex \eqref{eqn:relative1}.
The differential (called the Chevalley--Eilenberg differential) acting on $C(z,\ep)$ is dual to the Lie bracket on $\lie{gl}_N[[z_\alpha,\ep_i]]$ and hence can be schematically written as
\begin{equation}
    Q C(z,\ep) = \frac12[C(z,\ep), C(z,\ep)] \, .
\end{equation}

We formally extend the space of fields to include not only functions in the variables $z_\alpha, \ep_i$, but also differential forms in these variables.
This amounts to replacing $\bC[[z_\alpha,\ep_i]]$, which we view as functions on the formal super disk $\Hat{D}^{2|\cN-1}$, with de Rham forms on the formal super disk.
There is also a formal de Rham differential defined by
\begin{equation}
    \d C(z,\ep) = \d z_\alpha \frac{\partial C}{\partial z_\alpha} + \d \ep_i \frac{\partial C}{\partial \ep_i} \, .
\end{equation}
We extend the de Rham differential as a derivation and 
declare that it commutes with the original differential as in
\begin{equation}
    Q \d C(z,\ep) = [C(z,\ep),\d C(z,\ep)] \, .
\end{equation}
Notice that the right hand side contains no $c(z)$-term. 
So, all coefficients in the expansion of the expression
\begin{equation}\label{eq:QOps}
    \Tr \left[\d C(z,\ep)\right]^n
\end{equation}
to powers in the variables $z$, $\ep$, $\d z$, and $\d \ep$ are valid $Q$-closed operators. 

Some linear combinations of these operators actually vanish, simply because \begin{equation}\label{eq:relations}
    \d \Tr \left[\d C(z,\ep)\right]^n =0 \,.
\end{equation}

The conjecture of \cite{Chang:2013fba} is that the coefficients of the expansion of these operators in $\ep_i$, $\d\ep_i$ and $\d z^\alpha$, modulo the above relation,  generate the single-trace cohomology at large $N$. We will discuss below the possible loop corrections to this statement.  

If we repeat the analysis exactly in the same way with a single $\epsilon$ variable, 
we can formulate a similar conjecture for pure gauge theory. 
Expanding $C(z,\varepsilon) = c(z) + \ep b(z)$, we have 
\begin{equation}
    \d C(z,\varepsilon)= \d\ep \, b(z) + \d z_\alpha \partial_\alpha c(z) + \ep \d z_\alpha \partial_\alpha b(z)
\end{equation}
and thus 
\begin{align}\label{eq:deRhamexpand}
    \Tr [\d C(z,\varepsilon)]^n &= \d\ep^n \Tr b^n + \d z_\alpha \d\ep^{n-1} \mathrm{STr}\,b^{n-1} \partial_\alpha c+ \d z^1 \d z^2 \d\ep^{n-2} \mathrm{STr}\,b^{n-2} \partial_\alpha c\,\partial^\alpha c \cr& \quad + \ep \d z_\alpha \d\ep^{n-1}\mathrm{STr}\,b^{n-1} \partial_\alpha b + 
    \ep \d z^1 \d z^2 d\ep^{n-2}\mathrm{STr}\,b^{n-2} \partial_\alpha b\partial_\alpha c \, .
\end{align}
For general $n$, the first three terms are our familiar $A_n$, $B_{n-1,\alpha}$ and $C_{n-2}$ towers: 
\begin{align}\label{eq:deRhamexpand2}
    \Tr [\d C(z,\varepsilon)]^n &= \d\ep^n A_n + \d z_\alpha \d\ep^{n-1} B_{n-1,\alpha}+ \d z^1 \d z^2 \d\ep^{n-2} C_{n-2} \cr& \quad + \ep \d z_\alpha \d\ep^{n-1}n^{-1} \partial_\alpha A_n + 
    \ep \d z^1 \d z^2 d\ep^{n-2}(n-1)^{-1} \partial_\alpha B_{n-1,\alpha} \, .
\end{align}
The two last terms do not give us new fields, as they can be rewritten as derivatives of other representatives. These are precisely the relations from \eqref{eq:relations}.

For small $n$ the statements have to be adjusted in a minor way and possibly corrected to account for the difference between $U(N)$ and $SU(N)$.

\subsection{Cohomology of the Tree-Level Differential}

We now provide a proof of these expectations and a generalization for all $A$, following \cite{Eager:2018oww,Costello:2018zrm}.

Our method uses a theorem of Loday, Quillen, and Tysgan (LQT) which relates the (non-relative) Lie algebra cohomology to the \textit{cyclic homology} of $\bC[[z_\alpha,\ep_i]]$ \cite{Quillen1984,Tsygan_1983}. 

Applied to our situation, the Hochschild--Serre spectral sequence implies a general relationship between the relative Lie algebra cohomology and the absolute cohomology which takes the form of an isomorphism
\begin{equation}
H^\bu(\lie{gl}_N(A') | \lie{gl}_N(A_0)) \otimes H^\bu(\lie{gl}_N(A_0)) \simeq H^\bu(\lie{gl}_N(A')) .
\end{equation}
Here $A'$ is a general algebra, and $A_0$ is a subalgebra, but we will employ the result for $A' = A\otimes \bC[[z_1,z_2]]$ with $A_0 \subset A$.
We assume that $\lie{gl}_N(A_0)$ is a reductive subalgebra in order for the above isomorphism to hold; this is true in all examples we consider.
Thus, to obtain relative Lie algebra cohomology we will simply remove the contribution coming from $H^\bu(\lie{gl}_N(A_0))$.
With this in mind, we follow the Loday, Quillen, Tsygan prescription for computing the absolute cohomology, and after this we will extract the relative answer.

The theorem of Loday, Quillen, and Tsygan can be explained in an intuitive way. 
We study single-trace operators
\begin{equation}
    \Tr C_{i_1} \cdots C_{i_n} \,.
\end{equation}
The action of the differential closes on single-trace operators
\begin{equation}
   Q \Tr C_{i_1} \cdots C_{i_n} = \sum_{a,k} (-1)^{\cdots} f_{i_a}^{j_1, \cdots, j_k} \Tr C_{i_1} \cdots C_{i_{a-1}} C_{j_1}\cdots C_{j_k} C_{i_{a+1}} 
 \cdots C_{j_n} \,.
\end{equation}
Here the sign $(-1)^{\cdots}$ accounts for various Koszul signs. This differential defines a degree-shifted version of the dual of cyclic homology of $A'$, denoted as $HC_\bu(A')^\vee [-1]$. 

Including multi-trace operators, we arrive to the theorem of Loday, Quillen, and Tsygan \cite{Quillen1984,Tsygan_1983} relating the large $N$ Lie algebra cohomology of $\lie{gl}_N(A')$ to the cyclic homology of the algebra
\[
H^\bu \left(\lie{gl}_\infty(A')\right) \simeq \Sym \left(HC_\bu(A')^\vee [-1]\right) .
\]
Let $a_m$ denote the elements in the $A_\infty$ algebra $A' =A\otimes \mathbb{C}[[z_1,z_2]]$ with $A_\infty$ operations $(-,\dots, -)$. The cyclic homology itself can be described by linear combinations of formal symbols $[a_1 \cdots a_{i_n}]$ defined up to cyclic rotations (with appropriate Koszul signs) dual to the single-trace operators, placed in a degree shifted by $n-1$, with differential 
\begin{align}
   Q [a_1 \cdots a_n] &= \sum_{u<v}
   (-1)^{\cdots} [a_1 \cdots a_{u-1}(a_u, \cdots,a_v) a_{v+1} \cdots a_{n}]+ \cr &+\sum_{u>v}
   (-1)^{\cdots} [ a_{v+1} \cdots a_{n}(a_u, \cdots,a_v) a_1 \cdots a_{u-1}] \,,
\end{align}
where the sum is over all pairs of $u$ and $v$.

Before we continue, we should remind ourselves of the 2d TFT interpretation of cyclic homology. Intuitively, each boundary condition in the TFT gives a ``boundary state'', i.e. a state for the 2d TFT compactified on the circle. If we produce a complicated boundary state by taking $N$ copies of an elementary boundary condition and deforming it by some couplings $C_i$, we should be able to expand the boundary state  in powers of the couplings.

A more precise statement is that the coefficient of $\Tr C_{i_1} \cdots C_{i_n}$ 
in the expansion is mapped to an element 
in the space $Z_{S^1}[S^1]$ of ``$S^1$-equivariant'' states for the 2d TFT compactified on a circle. The 
images of the planar $Q$ are the coefficients of the BRST variation of the boundary state, 
so the map from single-trace operators to 
$Z_{S^1}[S^1]$ commutes with the BRST differential. 

In a string field theory language, the map tells us how the closed string fields couple to local operators in the brane world-volume.

The cyclicity of the trace turns out to be an important complication in the computation of cyclic homology. A very similar differential acting on words which are not cyclically symmetric defines the notion of Hochschild homology $HH_\bu(A')$ of $A'$ (valued in $A'$), which is usually more computable. 

For example, we can think about a commutative  algebra~$A'$ as defining the affine scheme $X = {\rm Spec}(A')$. Then, a theorem of Hochschild, Kostant, and Rosenberg \cite{Hochschild1962DifferentialFO} asserts that the Hochschild homology of $A'$ can be identified with the algebra of de Rham forms on~$X$.\footnote{The grading is opposite to the one usually used for de Rham forms.}

Crucially, the cyclic cohomology can be computed from $HH_\bu(A')$ via a spectral sequence involving the \textit{Connes $B$ operator}, which is an algebraic analogue to the de Rham differential~\cite{Weibel}.

The first page of the spectral sequence is 
\[
HH_\bu(A')\otimes \bC [u^{-1}] .
\]
Here $u$ is a parameter of ghost number $+2$ and $HH_\bu(A')$ is the ordinary Hochschild homology of the algebra $A'$.
The differential at this page in the spectral sequence is $u B$ where $B$ is the Connes $B$-operator acting on the Hochschild homology.
The variable $u$ corresponds to the generator of $H^\bu(B S^1)$ and its presence encodes the fact that the cyclic homology is an $S^1$-equivariant version of ordinary Hochschild homology.\footnote{The Hochschild homology $HH_\bu(A')$ also has a 2d TFT interpretation, involving TFT states without $S^1$ equivariance and boundaries decorated by a boundary local operator. The difference between equivariant and non-equivariant states in a string theory language concerns the inclusion or exclusion of the $c_0 - \bar c_0$ mode of the bulk ghosts associated to circle rotations. }

Order-by-order in $u$, at order $u^0$ we find a copy $HH_\bu(A')$ modulo the image 
of $B$. At the next order $u^{-1}$, we have a copy in $HH_\bu(A') u^{-1}$ of the kernel of $B$ modulo the image of $B$, etcetera. In other words, we get a single copy of $HH_\bu(A')/(B\, HH_\bu(A'))$ and infinitely many copies of the cohomology of $B$. 

For our algebra $A' = \bC[[z_1,z_2,\ep]]$, the Connes $B$ operator is explicit to describe.
We apply the Hochschild, Kostant, Rosenberg theorem to express the Hochschild homology of $A'$ as 
\[
HH_\bu(A') \simeq \bC[[z_\alpha, \d z_\alpha, \ep, \d \ep]] .
\]
Here the cohomology degree of $z_\alpha$ is zero, $\d z_\alpha$ is degree $-1$, $\ep$ is degree $+1$, and $\d \ep$ is degree zero.
With this identification, the Connes $B$ operator is simply the de Rham differential
\[
B = \d z_\alpha \del_{z_\alpha} + \d \ep \del_{\ep} .
\]

The cohomology of $B$ is very simple and consists of $\bC$ only. It gives us a summand of $\bC[u^{-1}]$ which is actually the difference between the relative, and the absolute Lie algebra cohomologies.

The part we are interested in is the quotient of $\bC[[z_\alpha, \d z_\alpha, \ep, \d \ep]]$ by the image of the de Rham differential. At this point, we have already essentially reproduced our conjectures above: the cohomology is generated by the coefficients in $\Tr \left[\d C(z,\ep)\right]^n$ modulo the relation following from 
$\d \Tr \left[\d C(z,\ep)\right]^n =0$.

More concretely, we compute the second page of the spectral sequence converging to $HC_\bu(A')$ by using an auxiliary spectral sequence that splits the differential $B$ into two pieces. A similar computation with a single complex coordinate and two odd variables was done in \cite{Costello:2018zrm}.
The first differential in this auxiliary spectral sequence is the piece of $u B$ which is the differential acting on the $\ep$-coordinate, and does not affect the holomorphic $z_\alpha$-variables.\footnote{This term in the differential can be thought of as the Koszul differential resolving a point inside of $\mathbb{A}^1 = \text{Spec}(\bC[\d \ep])$.} 
The cohomology of $HH_\bu(A')[u^{-1}]$ with respect to $u \d \ep \del_\ep$ is isomorphic to
\begin{equation} \label{eqn:page1}
    \ep \bC[[z_\alpha, \d z_\alpha, \d\ep]] \oplus \bC[[z_\alpha,\d z_\alpha]][u^{-1}] .
\end{equation}

The remaining differential is simply $u \d z_\alpha \del_{z_\alpha}$, which is basically the holomorphic de Rham differential on $\bC^2$.
The next page of the spectral sequence is the cohomology of \eqref{eqn:page1} with respect to this differential.
This cohomology is
\begin{equation}
\label{eqn:4d1a}
\ep \bC[[z_\alpha, \d \ep]] \oplus \left(\oplus_\alpha \ep \d z_\alpha \bC[[z_\alpha, \d \ep]] \right) \\ 
\oplus \bC[[z_\alpha]] \oplus \d z_1 \d z_2 \bC[[z_\alpha]] [1] \oplus u^{-1} \bC[[u^{-1}]] .
\end{equation}
For the summand in degree $-1$ have made the identification of the quotient space of one-forms on $\bC^2$ modulo exact one-forms with two-forms on $\bC^2$ using the Poincar\'e lemma.
There are no further terms in the spectral sequence.

Removing the summand $\bC \subset \bC[[z_\alpha]]$ and $u^{-1} \bC[u^{-1}]$ corresponds to taking the large $N$ limit of the relative, rather than the absolute, Lie algebra cohomology. 
Thus, in summary we see that the relative large $N$ cohomology is $\Sym^\bu(V^\vee[-1])$ where $V$ is the vector space 
\begin{equation}
\label{eqn:4d1b}
\ep \bC[[z_\alpha, \d \ep]] \oplus \left(\oplus_\alpha \ep \d z_\alpha \bC[[z_\alpha, \d \ep]] \right) \\ 
\oplus \bC[[z_\alpha]] \slash \bC \oplus \d z_1 \d z_2 \bC[[z_\alpha]] [1]
\end{equation}
Here, $\bC[[z_\alpha]]/\bC$ is power series modulo constant functions.
We interpret each of these summands as tree-level single trace operators present in the large $N$ limit:

\begin{itemize}
\item Consider the term $\ep \bC[[z_\alpha, \d \ep]]$.
This corresponds to single trace operators which have ghost number zero. 
Tracing through \eqref{eq:deRhamexpand}, which compares with de Rham cohomology, this term is generated by the primary tower of operators
\[
A_n = \Tr{b^n} , \quad n \geq 0 
\]
and their $z_\alpha$-derivatives.
\item Next consider the term $\ep \d z_\alpha \bC[[z_\alpha, \d \ep]]$.
This corresponds to single trace operators which have ghost number $+1$, and by comparing with \eqref{eq:deRhamexpand}, is generated by the primary tower of operators 
\[
B_{n,\alpha} = \Tr b^n \del_{\alpha} c , \quad n \geq 0 , \; \alpha = 1,2
\]
and their $z_\alpha$-derivatives. 
\item Consider the term $\bC[[z_\alpha]] \slash \bC$. 
This corresponds to single trace operators which have ghost number $+1$.
This term is generated by holomorphic descendants of the local operator $\Tr c$ where at least one holomorphic derivative appears.
\item Consider the term $\d z_1 \d z_2 \bC[[z_\alpha, \d \ep]] [1]$. 
This corresponds to single trace operators which are of ghost number $+2$ and is generated by the primary tower of operators 
\[
C_n = \Tr b^n \del_{\alpha} c \,\del^\alpha c , \quad n \geq 0 
\]
and their $z_\alpha$-derivatives.\footnote{Notice that the relative order of the fields in the trace is immaterial, as we can change the relative order by adding $Q$-exact operators with a single insertion of $\partial_\alpha b$: 
\begin{equation} 
Q \partial_\alpha b = [\partial_\alpha c,b] + \cdots 
\end{equation}}
\end{itemize}

Summarizing, we find the following single-trace primaries in the tree-level cohomology:
\begin{alignat}{3}
    A_n &= \Tr b^n, &&\quad n\geq 1 \nonumber \\
    B_{n,\alpha} &= \Tr b^n \partial_\alpha c, &&\quad n\geq 1,\; \alpha=1,2 \\
    C_n &= \Tr b^n \partial_\alpha c \, \partial^\alpha c, &&\quad n\geq 0\,. \nonumber
\end{alignat}
By the computation above, these operators together with their derivatives and the derivatives of $\mathrm{Tr}\,c$ 
form a basis of the large $N$ single-trace tree-level cohomology.  

The general structure of this calculation should generalize to all algebras of the form 
$A'=A \otimes \bC[[z_\alpha]]$. 
Then 
\[
HH_\bu(A') \simeq \bC[[z_\alpha, \d z_\alpha]] \otimes HH_\bu(A).
\]
The Connes B operator for $A'$ still splits as
\[
B = \d z_\alpha \del_{z_{\alpha}} \otimes \text{id} + \text{id} \otimes B_A
\]
where $B_A$ is the Connes operator for the algebra $A$.

The copies of the $B$ cohomology at non-trivial powers of $u^{-1}$ generalizes the $u^{-1} \bC[u^{-1}]$ factor above. It has the correct quantum numbers to represent the difference between the absolute and relative cohomology.
At order $u^{0}$ we can apply a similar approach as above by employing a spectral sequence which first computes the $B_A$ cohomology. 
The first page in this spectral sequence computing the large $N$ relative Lie algebra cohomology is thus the $u^{0}$ part of the cyclic homology of $A$ tensored with $\bC[[z_\alpha, \d z_\alpha]]$.

If $A$ is commutative, we can give a more intuitive explanation of this by employing superfields $C$ valued in $\bC^2 \times \mathrm{Spec}(A)$. Then we could formally define $dC$ using the Connes $B$ as a differential $d$ and build $Q$-closed operators $\Tr (dC)^n$ as before.

\subsection{Cohomology of the Loop-Corrected Differential}
Starting with the above results about the tree level operators at large $N$, we can now discuss the effect of the one-loop correction to the differential.

The first important observation is that the differential now mixes operators with different numbers of traces: a single-trace operator can be mapped to a single trace operator with an extra factor of $N$, if the differential acts on neighbouring letters in the trace, or to 
a product of two traces if it acts on other pairs of letters. 

Conversely, the 1-loop differential acting on the product of two traces could act on each individually or merge them into a single trace.\footnote{This ``problem'' could be avoided
if we define multi-trace operators via regularized products, effectively adding to the operator extra terms with fewer traces. This would introduce a different, possibly worse problem: the regularized product is not associative.}

In order to have a standard large $N$ combinatorics, it is useful to incorporate a factor of $N^{-1}$ in the propagators of the fields and a factor of $N$ in the interaction. One can normalize single-trace operators with an overall power of $N$. Then the planar loop differential maps single-trace to single-trace with no extra factors of $N$. The leading $1-2$ process is suppressed by a power of $N^2$, but the $2-1$ process is of order $1$. 

In the pure gauge theory, we can schematically discuss the action of the one-loop 
differential on single-trace operators. 

The action on $A_n$ is particularly simple: 
\begin{equation}\label{eq:Q1An}
Q_1 A_n \sim n \mathrm{Tr} \,b^{n-2} \partial_\alpha b \partial^\alpha c\,.
\end{equation}
The relative order of the symbols in the trace can be changed by adding tree-level exact terms such as $Q_0 \mathrm{Tr}\, b^{k_1} \partial_\alpha b\, b^{k_2} \partial^\alpha b$.
Then, by using the fact that the action of $Q_0$ on the individual fields gives commutators, and canceling with the operator in \eqref{eq:Q1An}, we can thus rewrite the $Q_1$ action on the tree-level cohomology in terms of $B_{n,\alpha}$:
\begin{equation}
Q_1 A_n = \frac{2 n}{n-1}\partial_\alpha B_{n-1}^\alpha\,.
\end{equation}
In cohomology, therefore, the one-loop differential eliminates the $A_n$ tower for $n>1$ and divergences of the $B_{n,\alpha}$ tower.
The action on $B_{n,\alpha}$ is a bit more complicated. 
Barring magical cancellation, it must produce a multiple  of $\partial_\alpha C_n$ up to tree-level exact terms. 
On the other hand, the action of $Q_1$ on $C_n$ must vanish as there are no single-trace operators with three $c$'s in the tree-level cohomology. 

Summarizing, in the planar limit we see that the one-loop differential eliminates almost every single-trace operator. In the $SU(N)$ theory, only the $C_n$ survive, even though the derivatives of $C_n$ are trivial in one-loop cohomology.
These remain even in the $SU(N)$ theory.\footnote{We do not expect the differential mapping two traces to a single trace to change this conclusion, as acting on $C_n C_m$ it would produce single-trace operators with too many $c$'s.}

Stripping off the singleton, i.e. working with an $SU(N)$ gauge theory, we thus find that the one-loop corrections produce a {\it topological} theory with single-trace observables associated to $C_n$. The quantum numbers of the operators prevent further perturbative corrections to this answer. Indeed, as they have all even ghost number, even non-perturbative corrections are impossible.

\section{Hints of Twisted Holography}
\label{sec:holography}

We will now discuss a holographic realization of holomorphic BF theory in the B-model topological string, leading to conjectural examples of twisted holography for the twists of pure gauge theory and SQCD with a fixed number of flavours. 
At tree level, this setup can be treated in parallel to other twisted holography examples based on the B-model, e.g. the cases of the 2d chiral algebra in ${\cal N}=4$ supersymmetric Yang--Mills theory \cite{Costello:2018zrm} or the holomorphic twist of ${\cal N}=4$ supersymetric Yang--Mills theory \cite{Costello:2016mgj}, and other theories associated to D3 branes at singularites as we touched upon in Section~\ref{sec:addingMatter}.

For the $B$-model with flat target $\C^{d_1+d_2}$ one can consider branes wrapping $d_1$ of the directions.
As we recalled in section \ref{sec:addingMatter}, the topological B-model presents the tree-level local operators in the world-volume theory on $N$ branes in terms of the Lie algebra cohomology of $\lie{gl}_N[[z_\alpha,\ep_i]]$, where $\alpha=1,\ldots,d_1$ label the bosonic holomorphic coordinates and $i=1,\ldots,d_2$ label the fermionic coordinates. 
We can describe it, as before, in terms of the composite superfield $C(z,\ep)$.

In the case at hand we are considering $d_1 = 2, d_2=1$, so the total dimension is $3$. This is a familiar situation, where the two-dimensional B-twisted sigma-model computes top forms on the moduli space of complex structures of a Riemann surface, which are integrated to give topological string amplitudes. The closed string sector of the B-model topological string on a Calabi--Yau threefold admits an elegant description in terms of Kodaira--Spencer theory \cite{Bershadsky:1993cx, Costello:2012cy, Costello:2015xsa}.

Theories associated to D3 branes at singularities, on the other hand, have $d_1 =2, d_2=3$ and involve a more exotic ``supercritical'' topological string theory, defined in dimension greater than $3$. Loop amplitudes in such a theory vanish unless insertions with sufficiently negative ghost number are present on the worldsheet. The result can still be expressed in the language of Kodaira--Spencer theory \cite{Costello:2012cy}. 

Generally speaking, the bulk B-model fields are divergence-free holomorphic poly-vectorfields on $\bC^{d_1+d_2}$, but in the neighbourhood of the brane they can be roughly thought of as divergence-free holomorphic poly-vectorfields on $\bC^{d_1|d_2}$, essentially by identifying $\partial_{\ep_i}$ as bosonic coordinates $\tilde z^i$ and $\ep_i$ as $\partial_{\tilde z^i}$ in the poly-vectorfield. They can thus be paired up naturally with the forms $\Tr (\d C)^n$ on $\bC^{d_1|d_2}$ to describe the coupling of the brane to the bulk fields.

This setup can be employed to study the world-volume theory of the branes, the BF theory, in two related ways:
\begin{itemize}
    \item The bulk closed string fields must admit a non-anomalous coupling to the world-volume theory living on the brane. Anomaly cancellation can be used to constrain the differential and brackets of the worldvolume theory, for any number $N$ of branes.\footnote{In lower-dimensional examples (topological defects of complex dimension one) the constraints can be discussed elegantly in the language of Koszul duality \cite{Costello:2020jbh, Paquette:2021cij, Garner:2022its, Gaiotto:2019wcc}. It is expected that the notion of Koszul duality can be extended to higher dimensional defects and a similar language will apply here.}
    
    \item  Open-closed duality predicts that the open-closed string theory in the presence of a stack of $N$ branes should be equivalent to the closed string theory in a backreacted geometry. The standard dictionary of open-closed duality relates the planar loop expansion of the field theory to classical calculations in the bulk closed string theory, with a precise relation between the number of loops and the power of the backreaction on the two sides. Quantum effects in the bulk govern the $N^{-1}$ expansion in the field theory. 
\end{itemize}

On general grounds, the backreaction of the D-branes is a $d_1$-vector valued in $(0,d_2-1)$ forms:
\begin{equation}
    N \omega_{\mathrm{BM}} \prod_{i=\alpha}^{d_1} \partial_{z_\alpha}
\end{equation}
where $\omega_{\mathrm{BM}}$ is now the Green's function for $\bar \partial$ in the transverse directions. 

The effects of this backreaction can be drastically different depending on the value of $d_2$ and are typically non-geometric in nature. It is not necessarily obvious how to define a decoupling limit to derive a standard holographic correspondence from the open-closed duality statement. The anomaly cancellation results, on the other hand, can be safely used to constrain the properties of the D-brane worldvolume theory. 

\subsection{Holomorphic BF Theory from the B-Model}

The world-volume theory of a stack of $N$ D-branes wrapping $\bC^2$ in $\bC^3$ is precisely holomorphic BF theory with gauge group $U(N)$.\footnote{Holomorphic BF theory is the dimensional reduction of holomorphic Chern-Simons theory, which is the open string field theory of the B-model topological string on a space-filling brane  \cite{Witten:1992fb}.} 
The lowest component of the $b$ field describes transverse fluctuations of the branes while the lowest component of the $c$ field is the ghost for the world-volume gauge fields. 

We will now show more explicitly that the operators $A_n,B_{n,\alpha},C_n$ \eqref{eq:ABC} in BF theory can be naturally coupled to the closed string fields of the B-model in the neighbourhood of the brane.

The closed string fields of the Kodaira--Spencer theory are given by
\begin{align}
    \beta \in \PV^{0,\bu}(\mathbb{C}^3)[2], \quad \mu \in \PV^{1,\bu}(\bC^3)[1], \quad \pi \in \PV^{2,\bu} (\bC^3) .
\end{align}
These are Dolbeault $(0, \bullet )$ forms with coefficients in the zero, first, and second exterior power of the holomorphic tangent bundle of the target.
The ghost number of a field is determined by both the polyvector degree and Dolbeault degree. 
If $\alpha^{(i)}$ denotes the $(0,i)$ Dolbeault component of a field, then $\beta^{(i)}$ has ghost number $i-2$, $\mu^{(i)}$ has ghost number $i-1$, and $\pi^{(i)}$ has ghost number $i$. 
The fields $\mu,\pi$ satisfy the constraint that they are divergence-free, meaning that $\del_{\Omega} \mu = 0$ and $\del_{\Omega} \pi = 0$. 
In local coordinates $\mu = \mu_\alpha \del_{z_\alpha}$, the first constraint is simply $\del_{z_\alpha} \mu_\alpha = 0$ and similarly for the bivector field. 

Let us take coordinates $(z_1,z_2,w)$ on $\bC^3$ with the complex codimension brane wrapping the first two coordinate planes.
The most basic first-order coupling of closed string fields to a stack of $N$ D-branes of complex codimension one is
\begin{equation}
    N \int_{\mathbb{C}^2} \partial^{-1} \beta \, .
\end{equation}
Notice that only the $(0,2)$ Dolbeault component appears in the coupling expression above.
Here, we view $\beta$ as a Dolbeault type $(3,2)$ form on $\bC^3$ via the holomorphic volume form; $\del^{-1} \beta$ is a non-local expression which schematically denotes a $(2,2)$ form $\beta'$ with the property that $\del \beta' = \beta$.
Fluctuations in the transverse direction, whose coordinate is $w$, are controlled by the eigenvalues of the lowest component of the open string $b$-field via:
\begin{equation}
    \Tr \beta^{(0)}(b) = \sum_{n=0}^\infty \frac{1}{n!} \Tr b^n \partial_{w}^n \beta|_{w=0} \, .
\end{equation}
Therefore, the basic coupling leads to a tower of single-trace couplings
\begin{equation}
    \int_{\mathbb{C}^2} N \frac{1}{n!}\Tr b^n \, \partial_{w}^{n-1} \beta|_{w=0} \, \d z_1 \d z_2
\end{equation}
involving the $A_n=\Tr b^n$ local operators in the tree-level cohomology of the 4d holomorphic BF theory.

Notice that in our B-model inspired analysis of the large $N$ tree-level cohomology, $A_n$ was identified as the coefficient of $\dd\varepsilon^n$ in $\Tr (\d C)^n$. The above coupling is consistent with an identification $\dd\varepsilon \to \partial_w$. 

There are analogous couplings between the $B_{n,\alpha}=\Tr b^n\partial_{\alpha}c$ and $C_n=\Tr b^n \partial_\alpha c \partial^\alpha c$ operators involving the transverse derivatives of the components $\mu_{\alpha} \del_{z_{\alpha}}$ of the field $\mu$ and the field $\pi$ given schematically by
\begin{align}
    &\int_{\mathbb{C}^2} \frac{1}{(n-1)!}B_{n,\alpha} \, \partial_{w}^{n-1} \mu_{\alpha}|_{w=0} \, \d z_1 \d z_2 \\
    &\int_{\mathbb{C}^2} \frac{1}{(n-1)!}C_n \, \partial_{w}^{n-1} \pi |_{w=0} \, \d z_1 \d z_2 \, .
\end{align}

All the couplings can be expressed concisely in terms of $\Tr (\d C)^n$. 
Observe that the replacement $\dd\varepsilon \to \partial_w$, $\varepsilon \to \dd w$
maps the differential $B = \d \varepsilon \partial_\varepsilon + \d z^\alpha \partial_\alpha$ to the $\partial$ operator in $\bC^3$. 
Thus $\Tr (\d C)^n$ becomes a $\partial$-closed form in $\bC^3$, which can be 
naturally paired with the closed string fields. 

These linear couplings can give rise to an anomaly which is bilinear in the bulk field, involving the bracket of two local operators on the brane. We expect it to be cancelled by a tree-level bulk Feynman diagram involving a cubic bulk vertex. This is the general principle of Koszul duality: the algebraic structures in the brane world-volume theory should be constrained or even determined by anomaly cancellation in terms of bulk Feynman diagrams. It should represent a target space manifestation of the relations between open- and closed- string field theory. 

The brane backreaction\footnote{The backreaction can be obtained by solving the equation of motion of the Kodaira-Spencer theory with a D-brane source term $N\delta_{w=0}$.} is captured by the following closed string field which is a bivector:\footnote{This field has charge $-1$ under the $U(1)_b$ rotation symmetry in the plane transverse to the branes, so that a perturbative expansion in this back-reaction breaks $U(1)_b$
just in the same way as the quantum corrections in the 4d holomorphic BF theory.}
\begin{equation}
\pi_{br} = N \frac{1}{w} \partial_{z_1} \wedge \partial_{z_2} .
\end{equation}
At leading order, the effect of $\pi_{br}$ on the classical closed string theory is to modify the differential acting on polyvector fields to

\begin{equation}
    \bar \partial \rightarrow \bar \partial + \{\pi_{br}, - \}_S \, ,
\end{equation}
where $\{-,-\}_S$ is the Schouten bracket.\footnote{In local coordinates, for $\beta,\gamma \in PV^{\bu,\bu}(\mathbb{C}^3)=C^{\infty}(\mathbb{C}^3)[\varepsilon_i\equiv \partial_{i},\dd\bar z^i]$, it is computed by
\begin{align}
    \{ \beta , \gamma \}_S = (\partial_{\varepsilon_i}\beta)(\partial_{z^i}\gamma) + (-1)^{|\beta|} (\partial_{z^i}\beta)(\partial_{\varepsilon_i}\gamma) \, .
\end{align}}

This backreaction also produces a BRST anomaly in the couplings to the brane, which we expect to be cancelled by the planar part of the 1-loop BRST transformations of $A_n$, $B_{n,\alpha}$ and $C_n$.
Generally speaking, turning on a non-trivial closed string field which is a bivector has the effect of making spacetime noncommutative.

\subsection{Large \texorpdfstring{$N$}{N} SQCD}

In this section we discuss some preliminary observations about the holomorphic twist of SQCD, see section \ref{s:sqcd}.
We briefly recall that the twist of SQCD can be described as a higher dimensional gauged $\beta\gamma$ system, with $\beta$, $\gamma$ and $\tilde \beta$, $\tilde \gamma$
fields arising respectively from (anti)fundamental chiral multiplets. 

Recal also that the $\beta$ fields from the fundamental chiral multiplets have the same gauge quantum numbers as the $\tilde \gamma$ fields from the anti-fundamental chiral multiplets, and vice versa. 
As a result, the $U(N_f)_L \times U(N_f)_R$ classical global symmetry of physical SQCD is enhanced to $U(N_f|N_f)$.

Remarkably, we can engineer this holomorphic theory in the B-model as well, by adding $(N_f|N_f)$ space-filling branes,
which support an $U(N_f|N_f)$ holomorphic Chern-Simons theory. 
This opens up the possibility of a twisted holography analysis. 

In the large $N$ limit, we expect the (anti-)fundamental matter to contribute mesons $\tilde \Gamma_A b^n \Gamma^B$
to the BRST cohomology of local operators, together with their derivatives. 

There does not seem to be space for any loop corrections at the leading order in $N$.

It thus appears that the holomorphic twist of SQCD, at the leading order in $N$, enjoys a symmetry by the super Lie algebra
$\mathfrak{u}(N_f|N_f)[z_1,z_2,\partial_{w}]$, deformed by the effect of the 
\begin{equation}
\eta = \frac{1}{w} \partial_{z_1} \wedge \partial_{z_2}
\end{equation}
background. 

At low energy, for $N_f \ll N$, SQCD is expected to have a runaway behaviour due to an instanton-generated superpotential with no minimum \cite{Affleck:1983mk,Affleck:1983rr}. It would be interesting to study this phenomenon in the holomorphic twist and holographically. In practice, it should mean that the operator ``$1$'' is $Q$-exact after instanton corrections are included. 

\section{Indices and Characters for Local Operators} \label{sec:indices}

Supersymmetric indices offer a good way to gain some intuition about the cohomology calculations which characterize BPS operators. 
Depending on the specific calculation we are interested in, we can flavour the indices with fugacities for various actual or spurious symmetries. Crucially, the indices can be computed in the free theory and are unaffected by the interacting differentials. 

Index calculations are not useful to study the full non-perturbative pure gauge theory case, which only has an $SU(2)$ rotational symmetry commuting with the super charge: the collection of protected local operators in the free theory with a given $SU(2)$ charge is infinite-dimensional and thus cannot be sensibly counted.\footnote{We do expect the cohomology for each cohomological degree to be typically finite-dimensional, so that a weighed character of the cohomology is well-defined. For example, in the pure free gauge theory the only letters of non-negative cohomological degree are $\partial_\alpha c$, and each can only appear once in an operator. As long as the interactions can be described as turning on a differential on the free cohomology, the result will be finite-dimensional. An exception would be theories with an infinite-dimensional chiral ring which cannot be decomposed into finite-dimensional sectors of given charge under some flavour symmetry.}

In a perturbative setup, we can grade operators by the full $U(2)$ symmetry, as explained at the beginning of section \ref{sec:symmetries}. Both the $b$ field and the holomorphic derivatives have positive charge under the diagonal generator $M_R$ and no fields of the BF theory have negative charge (see table \ref{tab:charges}). The counting problem is thus well-defined. 

Introducing traditional fugacities $p,q$ for the $U(2)$ Cartan generators, the contribution 
to the index of a $b$ field of gauge fugacity $z$ and its derivatives is 
\begin{equation}
    \frac{1}{\prod_{n,m\geq 0}(1-p^{m+1}q^{n+1} z)} ,
\end{equation}
which we can regard as the plethystic exponential of the single particle index~$\frac{pqz}{(1-p)(1-q)}$.
Here $p q z$ is the fugacity for $b$, which is a boson, and each derivative adds a factor of $p$ or $q$. The contribution of a $c$ field, keeping only derivatives, is
\begin{equation}
    \prod_{n,m\geq 0|(n,m) \neq (0,0)}(1-p^{m}q^{n} z) \, .
\end{equation}
As we build the full index, the contributions of $b$ and $c$ fields with the same gauge charge almost cancel out, leaving the products
\begin{align}
    \prod_{m\geq 1} (1-p^mz) \prod_{n\geq 1} (1-q^nz) \, 
\end{align}
for each generator of the gauge group $G$.

We can manipulate the product of $q$-dependent factors over the generators of $G$ with the help of the Kac-Weyl denominator formula:
\begin{equation}
\prod_{m\geq 1}(1- q^m)^r \prod_{\alpha}\prod_{m\geq 1}(1- q^m z_\alpha) = \sum_{\lambda\in \Lambda_{G}^+} (-1)^{\epsilon(\lambda)} q^{(\lambda, \lambda +2 \rho)/(2 h_G)}\chi_\lambda(z) \, .
\end{equation}
Here $(-1)^{\epsilon(\lambda)}$ is a sign which will be immaterial in the following and the sum on the right hand side runs over the positive roots $\Lambda_{G}^+$ of the group. The product on the left hand side runs over roots $\alpha$ for $G$. 

We can expand the product of $p$-dependent factors in the same manner and combine the two sums. The projection over $G$ invariants maps $\chi_\lambda(z) \chi_\mu(z) \to \delta_{\lambda,\mu}$ as the product of two irreps of $G$ contains a (single) $G$-invariant element if and only if the two irreps coincide.

We thus get 
\begin{equation}\label{eq:pertindex}
I^{G}_{\mathrm{pert}}[p,q]  = \sum_{\lambda\in \Lambda_{G}^+} (p q)^{(\lambda, \lambda +2 \rho)/(2 h_G)} \, .
\end{equation}
For example,
\begin{equation}
I^{SU(2)}_{\mathrm{pert}}[p,q]  = \sum_{\ell \geq 0} (pq)^{\frac{\ell(\ell+1)}{2}} \, .
\end{equation}
Notice that the index is a function of $pq$ only: it does not receive contributions from 
operators with non-trivial $SU(2)$ spin. This is compatible with the exactness of $\partial_\alpha S^\alpha$, discussed in section \ref{subsection:1loopandbeyond}, which implies that all operators with non-trivial $SU(2)$ spin
are $Q$-exact at 1-loop.

\subsection{Large \texorpdfstring{$N$}{N} Index}
If we include the fugacity $x$ for the spurious symmetry $U(1)_b$ giving charge $1$ to $b$, as well as the usual $p$, $q$ rotation fugacities employed in calculations of the superconformal index, the single-letter index $f(x;p,q)$ is given by
\begin{equation}
1-f(x;p,q) = \frac{1-x}{(1-p)(1-q)} \, .
\end{equation}
The fugacity $x$ should be set to $p q$ when we introduce loop corrections. Non-perturbative corrections would further constrain $p^N q^N = 1$ as the $U(1)_R$ symmetry is broken to a discrete subgroup by anomalies.  

The large $N$ index for an $U(N)$ theory is computed by a standard formula \cite{Kinney:2005ej}:
\begin{equation}
I^\infty_{\mathrm{pert}}[x;p,q] = \prod_{n=1}^\infty \frac{1}{1-f(x^n;p^n,q^n)} = \prod_{n=1}^\infty \frac{(1-p^n)(1-q^n)}{1-x^n} \, .
\end{equation}
This index should count polynomials in single-trace operators. The single-trace index is the Plethystic logarithm: 
\begin{equation}\label{eq:singindex}
I_{\mathrm{single}}[x;p,q] = \frac{x}{1-x} - \frac{p}{1-p} - \frac{q}{1-q} \, .
\end{equation}
We expect all single trace operators in the tree-level cohomology to be organized into towers of derivatives of a ``primary'' field, 
with the exception of $\partial_\alpha \mathrm{Tr}\,c$, as $\mathrm{Tr}\,c$ is disallowed. Excluding the 
tower of derivatives of $\mathrm{Tr}\,c$, the index for the remaining single-trace primaries is
\begin{equation}
1+(1-p)(1-q)\left(-1+I_{\mathrm{single}}[x;p,q]\right) = \frac{x - p x - q x + p q}{1-x} \, .
\end{equation}
This agrees with our large $N$ cohomology calculations in section \ref{sec:infiniteN}: we see the contribution of the primaries
\begin{alignat}{3}
    A_n &= \Tr b^n, &&\quad n\geq 1 \nonumber \\
    B_{n,\alpha} &= \Tr b^n \partial_\alpha c, &&\quad n\geq 1,\; \alpha=1,2 \\
    C_n &= \Tr b^n \partial_\alpha c\, \partial^\alpha c, &&\quad n\geq 0 \, . \nonumber
\end{alignat}
This confirms the result that these operators together with their derivatives, and the derivatives of $\mathrm{Tr}\,c$,
form a basis of the large $N$ single-trace tree-level cohomology.  

In the case of gauge group $SU(N)$ and infinite $N$, the derivatives of $\Tr c$ vanish and the cohomology is generated by primaries
\begin{align}
    A_n, \, n\geq 2, \qquad  B_{n,\alpha}, \, n\geq 1, \qquad  C_n, \, n\geq 0
\end{align}
and their derivatives.

The 1-loop corrected cohomology for $U(N)$ appears to only include the $C_n$ operators and no derivatives, as well as derivatives of $\Tr c$. 
This reproduces the single-particle index in \eqref{eq:singindex} where $x=pq$,
without any cancellation, i.e. each term in the single-trace index corresponds to one single-trace cohomology class.   

\begin{table}[t]
\centering
\begin{tabular}{c|c|c|c|c|c|}
\cline{2-6}
                                   & $R$ & $M_{\dot+\dot-}$ & $M_R$ & $C_R$ & $C$ \\ \hline
\multicolumn{1}{|c|}{$Q$}          & -1  & -1               & 0     & 1     & 1   \\ \hline
\multicolumn{1}{|c|}{$b$}          & 0   & 2                & 2     & 0     & -2  \\ \hline
\multicolumn{1}{|c|}{$c$}          & -1  & -1               & 0     & 1     & 1   \\ \hline
\multicolumn{1}{|c|}{$\partial_\alpha$} & 0   & 1                & 1     & 0     & -1  \\ \hline
\end{tabular}
\label{tab:charges}
\caption{We collect here the charges of elementary fields under various generators used in the main text. The twisted rotation generator was defined as $M_R= M_{\dot+\dot-}-R$ and the ``perturbative" and ``non-perturbative" cohomological degrees as $C_R=\text{gh}-R$ and $C=C_R-M_R=\text{gh}-M_{\dot+\dot-}$, respectively.}
\end{table}

\section{Numerical Analysis for \texorpdfstring{$N=2,3$}{N=2,3}}
\label{sec:finiteN}

In this section, we study the cohomology of local operators for the holomorphic twist of pure $\mathcal{N}=1$ SYM for $SU(N)$ gauge group with $N=2,3$, formulated as holomorphic BF theory as described in section \ref{sec:twistedAsHolo}. The space of local operators can be organized by the charge $M_R\equiv M-R$, with finite-dimensional eigenspaces. We use symbolic manipulation software to compute the cohomology up to the following values for the charge:
\begin{equation}
\begin{alignedat}{3}
    &M_R= 14 \quad &\text{for} \quad N=2 \, , \\
    &M_R=8 \quad &\text{for} \quad N=3 \, .
\end{alignedat} \label{eq:levels}
\end{equation}
Up to the tree-level stage, our calculation is the pure gauge theory analog of \cite{Chang:2013fba, Chang:2022mjp}, which computed numerically the low energy cohomology of semi-chiral operators in $\mathcal{N}=4$ SYM for $SU(N)$ with $N=2,3,4$. The main bottleneck in numerical calculations of cohomology of local operators appears to be the large number of gauge invariant operators and of monomials which appear in individual gauge invariant operators, especially for $N>2$.

The computational strategy is straightforward:
\begin{enumerate}
    \item We pick a specific basis for the adjoint representation of the gauge group and work with the explicit components for all the fields. In this way, the finite $N$ trace relations are implemented automatically.\footnote{This choice is also the main limitation of our method, as the number of possible monomials in the fields grows dramatically as $M_R$ or $N$ increase.}
    \item We produce a basis for the space of gauge invariant operators 
    \[
    \cV_{\bu} (n_1,n_2) = \oplus_{|b|} \mathcal{V}_{|b|}(n_1,n_2) 
    \]
    built out of the operator $b$ and its derivatives as well as derivatives of $c$. 
    Operators in $\mathcal{V}_{|b|}(n_1,n_2)$ have charges $n_1,n_2$ under the Cartan generators of $U(2)$, which add up to $M_R$, have charge $|b|$ under the auxiliary $U(1)_b$ which commutes with the tree-level differential $Q_0$ and the cohomological grading (which we have left implicit in the expression above).
    Recall that $|b|$ simply counts the number of times $b$ appears in a local operator; all other fields are weighted trivially under $U(1)_b$.
    The perturbative differential commutes with $U(2)$, so that we get a collection of complexes $\mathcal{V}_{\bu}(n_1,n_2)$ indexed by the $n_i$.  
    \item We compute the action of 
    \begin{equation}
        Q_0 \colon \mathcal{V}_{|b|}^i (n_1,n_2)\to \mathcal{V}^{i+1}_{|b|} (n_1,n_2)
    \end{equation} on the basis and find a basis for the representatives of the $Q_0$ cohomology \[H^i(\mathcal{V}_{|b|}(n_1,n_2),Q_0) .\]
    \item We compute the action of the one-loop differential $Q_1$ on the representatives of the $Q_0$ cohomology:
    \begin{equation}
        Q_1 \colon H^i(\mathcal{V}_{|b|}(n_1,n_2),Q_0)\to H^{i+1}(\mathcal{V}_{|b|-1}(n_1,n_2),Q_0) \, .
    \end{equation}
    We then compute the representatives of the $Q_1$ cohomology \[H^\bu (H^{\bu}
    (\cV_{\bu}(n_1,n_2),Q_0),Q_1).\]
\end{enumerate}
This procedure gives the one-loop approximation to the perturbative cohomology. In principle, we should continue and compute the two-loop differential 
 \begin{equation}
        Q_2 \, :\,H^\bu(H^\bu (\cV_{|b|}(n_1,n_2),Q_0),Q_1)\to H^\bu (H^\bu(\cV_{|b|-2}(n_1,n_2),Q_0),Q_1)
    \end{equation}
as in a spectral sequence. Up to the values of $M_R$ we consider, though, we find that the 1-loop cohomology is sufficiently sparse to forbid non-zero higher loop differentials. Our answer 
\begin{equation}
    H^\bu(\cV_{\bu} (n_1,n_2),Q_{\mathrm{pert}}) \equiv \bigoplus_{|b|} H^{\bu+n_1+n_2}(H^\bu (\cV_{|b|}(n_1,n_2),Q_0),Q_1) 
\end{equation}
is thus exact to all orders in perturbation theory.

A potential strategy for non-perturbative calculations would be to start from $H^\bu(\cV_{\bu}(n_1,n_2),Q_{\mathrm{pert}})$ and compute a 1-instanton correction to the differential. 
Such a correction must lower $n_1+n_2$ by multiples of $2N$, the anomaly introduced by a 1-instanton effect, while increasing cohomological degree by $1$:
 \begin{equation}
        Q_{1\text{-}\mathrm{inst}}\, :\,H^i(\cV_{\bu}(n_1,n_2),Q_{\mathrm{pert}})\to H^{i+1} (\cV_{\bu} (n_1-N,n_2-N),Q_{\mathrm{pert}}) \, .
\end{equation}
It would be the beginning of another spectral sequence-like calculation incorporating $n$-instanton effects.

For $N=2$, we will find multiple pairs of perturbative cohomology classes which could in principle be lifted by instanton effects. Indeed, our calculations are compatible with the conjecture that the whole perturbative cohomology should be lifted by instanton effects, with the exception of the identity operator and of the gaugino condensate $C_0 = \partial_\alpha c \partial^\alpha c$, with no non-trivial operator with the charge of $C_0^2$. This is somewhat remarkable, as the perturbative cohomology is otherwise rather sparse. 

Such a result would fully agree with the space of local operators in the far infrared: the IR theory has two gapped vacua distinguished by the vev of $C_0$, which is expected to square to a multiple of the identity operator, proportional to the appropriate power of the strong coupling scale. The far infrared effective theory in each gapped vacuum is not expected to support any other non-trivial local operators. We will comment on this point further at the end of the section.

For $N=3$ one may similarly hope for instanton effects to cancel all perturbative cohomology classes except $C_0$ and $C_0^2$, as the theory is expected to have three gapped vacua in the IR.  Unfortunately, we cannot push our calculation to the order $p^5 q^5$, which would be needed to find a perturbative cohomology class whose 1-instanton image could cancel the other perturbative cohomology class we find below.  

\subsection{Cohomology of the Tree-Level Differential}

By the numerical analysis outlined above, we find that all cohomology classes at tree level (up to the studied values in equation \eqref{eq:levels}) are of the ``same type" as in the infinite $N$ case, i.e. their representatives are polynomials in single trace primaries $A_n,B_{n,\alpha},C_n$ given in \eqref{eq:ABC} and their derivatives. Therefore, there is a surjective map from infinite $N$ tree-level cohomology into $SU(2)$ and $SU(3)$ cohomology, up to the level we checked:\footnote{The map is not injective: e.g. due to trace relations operators $C_n$ vanish for odd $n$ in $SU(2)$.}
\begin{align}
    H_{SU(\infty)}^{i}(\mathcal{V}_{|b|}(n_1,n_2),Q_0) \twoheadrightarrow H_{SU(N)}^{i}(\mathcal{V}_{|b|}(n_1,n_2),Q_0) \, , \qquad N=2,3 \,.
\end{align}
In the analogous setting for ${\cal N}=4$ SYM with $SU(2)$ gauge group, this statement fails at $M_R\simeq 8$, due to the appearance of a ``non-multitrace'' cohomology representative which is not a polynomial in infinite $N$ single-trace cohomology representatives \cite{Chang:2022mjp,Choi:2022caq,Choi:2023znd}. Here we do not find any non-multitrace cohomology classes of the tree level differential up to $M_R=14$. It is possible that such states will appear at higher quantum numbers. We do not have a clear expectation either way. 

The tree-level characters for $N=2,3$ and $N=\infty$ are listed in table \ref{tab:tree_char}.

\begin{table}[t]
\begin{tabular}{|c|l|l|l|}
\hline
$M_R$ & \multicolumn{1}{c|}{$N=2$} & \multicolumn{1}{c|}{$N=3$} & \multicolumn{1}{c|}{$N=\infty$} \\ \hline
2 & $p q u^2$ &  &  \\ \hline
3 & $-p u x-q u x+ p^2 q u^2+p q^2 u^2$ &  &  \\ \hline
4 & \begin{tabular}[c]{@{}l@{}}$x^2-p^2 u x-2 p q u x-q^2 u x$\\ $+ p^3 q u^2+p^2 q^2 u^2+p q^3 u^2$\end{tabular} & $+p q u^2 x+p^2 q^2 u^4$ &  \\ \hline
5 & \begin{tabular}[c]{@{}l@{}}$p^4 q u^2+p^3 q^2 u^2-p^3 u x+p^2 q^3 u^2$\\ $-2 p^2 q u x+p q^4 u^2-2 p q^2 u x+p x^2$\\ $-q^3 u x+q x^2$\end{tabular} & \begin{tabular}[c]{@{}l@{}}$+p^3 q^2 u^4+p^2 q^3 u^4-p^2 q u^3 x$\\ $+p^2 q u^2 x-p q^2 u^3 x+p q^2 u^2 x$\\ $-p u x^2-q u x^2$\end{tabular} &  \\ \hline
6 & \begin{tabular}[c]{@{}l@{}}$p^5 q u^2+p^4 q^2 u^4+p^4 q^2 u^2-p^4 u x$\\ $+p^3 q^3 u^4+p^3 q^3 u^2-p^3 q u^3 x$\\ $-2 p^3 q u x+p^2 q^4 u^4+p^2 q^4 u^2$\\ $-p^2 q^2 u^3 x-2 p^2 q^2 u x+p^2 x^2$\\ $+p q^5 u^2-p q^3 u^3 x-2 p q^3 u x$\\ $+p q u^2 x^2+p q x^2-q^4 u x+q^2 x^2$\end{tabular} & \begin{tabular}[c]{@{}l@{}}$+p^4 q^2 u^4+p^3 q^3 u^4-p^3 q u^3 x$\\ $+p^3 q u^2 x+p^2 q^4 u^4-3 p^2 q^2 u^3 x$\\ $+p^2 q^2 u^2 x-p^2 u x^2-p q^3 u^3 x$\\ $+p q^3 u^2 x+p q u^2 x^2-2 p q u x^2$\\ $-q^2 u x^2+x^3$\end{tabular} & \begin{tabular}[c]{@{}l@{}}$+p^3 q^3 u^6$\\ $+p^2 q^2 u^4 x$\\ $+p q u^2 x^2$\end{tabular} \\ \hline
\end{tabular}
\caption{Tree-level characters counting cohomology classes with charges $p^{n_1} q^{n_2} u^{C_R} x^b $. Each column should be added to the previous one to get the full answer.}
\label{tab:tree_char}
\end{table}

\subsection{Cohomology of the Loop-Corrected Differential}

\begin{table}[t]
\begin{tabular}{|c|l|c|l|}
\hline
$M_R$ & \multicolumn{1}{c|}{$N=2$} & $N=3$                            & \multicolumn{1}{c|}{$N=\infty$}                                                                                                                                               \\ \hline
2     & $C_0$                      & \multicolumn{1}{l|}{$C_0$}       & $C_0$                                                                                                                                                                         \\ \hline
4     & \multicolumn{1}{c|}{-}     & \multicolumn{1}{l|}{$C_1,C_0^2$} & $C_1,C_0^2$                                                                                                                                                                   \\ \hline
6     & $C_2$                      & ?                                & $C_2,C_0C_1,C_0^3$                                                                                                                                                            \\ \hline
8     & \multicolumn{1}{c|}{-}     & ?                                & $C_3,C_2C_0,C_1^2,C_1C_0^2,C_0^3$                                                                                                                                             \\ \hline
10    & $C_4, B_{1,2} \partial_2 C_0 \partial_1^2 C_0$                 & ?                                & $C_4,C_3C_0,C_2C_1,C_2C_0^2,C_1^2C_0,C_1C_0^3,C_0^5$                                                                                                                          \\ \hline
12    & $B_{1,2}\partial_1\partial_2 B_{1,2} \partial_1^2 C_0$                      & ?                                & \begin{tabular}[c]{@{}l@{}} $C_5,C_4C_0,C_3C_1,C_3C_0^2,C_2^2,C_2C_1C_0,C_2C_0^3,$ \\
$C_1^3,C_1^2C_0^2,C_1C_0^3,C_0^6$          \end{tabular}                                                                         \\ \hline
14    & $C_6,B_{3,2} \partial_2 C_0 \partial_1^2 C_0$                  & ?                                & \begin{tabular}[c]{@{}l@{}}$C_6,C_5C_0,C_4C_1,C_4C_0^2,C_3C_2,C_3C_1C_0,C_3C_0^3,$\\
$C_2^2C_0,C_2C_1^2,C_2C_1C_0^2,C_2C_0^4,C_1^3C_0,C_1^2C_0^3,C_1C_0^5,C_0^7$\end{tabular} \\ \hline
\end{tabular}
\caption{Cohomology representatives at 1-loop. For $N=2$ we find cohomology classes which are not of the form of the cohomology classes at infinite $N$.}
\label{tab:1loop_ops}
\end{table}

As predicted, the derivatives of cohomology classes are exact at 1-loop for finite and infinite $N$. More surprisingly, we find several cohomology classes at $N=2$, which are not polynomials in the infinite $N$ single-trace cohomology of the 1-loop differential. These are the ``non-multitrace" cohomology classes with respect to the 1-loop differential.

The 1-loop cohomology classes for $N=2,3$ compared to $N=\infty$ are shown in table \ref{tab:1loop_ops}.

We include the characters counting 1-loop cohomology classes of various charges:
\begin{align} 
    \chi^{SU(2)}_{\text{1-loop}}(u,p,q)
     =& 1+p q u^2+ p^3 q^3 u^2  +p^5 q^5 u^2-p^5 q^5 u^5 +p^6 q^6 u^4 \cr
    &+ p^7q^7u^2-p^7q^7u^5 + \mathcal{O}(p^8 q^{8})\,, \\
 \chi^{SU(3)}_{\text{1-loop}}(u,p,q) =& 1 + p q u^2 + p^2 q^2 u^4 + p^2 q^2 u^2 + \mathcal{O}(p^3 q^{3})\,, \\
 \chi^{\infty}_{\text{1-loop}}(u,p,q) =& \frac{1}{\prod_{n=1}^\infty(1 - p^n q^n u^2)} = \chi^{SU(3)}_{\text{1-loop}}(u,p,q) + \mathcal{O}(p^3 q^{3})\,.\label{eq:inftychar}
\end{align}

If we replace the fugacity $u$ for $C_R$ with a fugacity $v$ for the nonperturbative cohomological degree $C$ (see table \ref{tab:charges}), the $SU(2)$ character becomes
\begin{align}
    \chi^{SU(2)}_{\text{1-loop}}(v,p,q)
     =& 1+p q+ \left(p^3 q^3 v^{-4}  -p^5 q^5 v^{-5}\right) +\left(p^5 q^5 v^{-8}-p^7q^7v^{-9}\right) +p^6 q^6 v^{-8} \cr
    &+ p^7q^7v^{-12} + \mathcal{O}(p^8 q^{8})\,.
\end{align}
We grouped together pairs of terms which could be lifted by 1-instanton effects.

It seems plausible that the term $p^7q^7v^{-12}$ coming from the cohomology class $C_6$ could be paired up with a term $-p^9q^9v^{-13}$ coming from a potential cohomology class $B_{5,2} \partial_2 C_0 \partial_1^2 C_0$ at charge $M_R=18$, which is above the values studied by us. More generally one can conjecture the contributions from cohomology classes $C_{2n}$ for $n>0$:
\begin{align}
    p^{2n+1}q^{2n+1} v^{-4n}
\end{align}
could be lifted by the contributions from ``non-multitrace" cohomology classes 
\[
B_{2n-1,2}\partial_2 C_0 \partial_1^2 C_0
\]
which take the form
\begin{align}
    -p^{2n+3}q^{2n+3} v^{-4n-1} \, .
\end{align}
The term $p^6q^6v^{-8}$ could potentially be cancelled by a contribution $-p^8q^8v^{-9}$ from a new ``non-multitrace" cohomology class at charge $M_R=18$, which is above the values we studied numerically.

\subsection{Comparison to Compactification on a Two-Torus}
We should elaborate on the possibility of matching operators between the UV and far IR theory. 
This discussion will be somewhat orthogonal to the rest of the section. 

In two-dimensional topological twists, it was found that the space of local operators computed in the UV could be fully matched to the space of local operators in the far IR, but only including the effect of gapped particles and solitons \cite{Gaiotto:2015aoa}. The two-dimensional $(2,2)$ supersymmetric $\bC P^1$ sigma model is closely related to pure four-dimensional ${\cal N}=1$ gauge theory. Indeed, it is expected to arise from a supersymmetric compactification on a two-torus of the four-dimensional theory \cite{Friedman:1997yq,Acharya:2001dz}. Solitons in the two-dimensional theory are related to domain walls in the four-dimensional theory. 

The supersymmetric torus compactification is compatible with the holomorphic twist: we can consider the twisted theory on $\bC^* \times \bC^*$ and reduce it to a topological theory on $\bR^2$. 
Local operators in the two-dimensional theory, though, could arise either from local operators in the four-dimensional theory or from holomorphic surface defects wrapping the internal two-torus. In particular, the local operators in the twisted $\bC P^1$ sigma model which create solitons are likely to arise in such manner. 

A better option, for our purpose, is to consider the twisted theory on the product of two holomorphic cigars and reduce it to the same topological theory on $(\bR^+)^2$, with specific boundary conditions. Then four-dimensional local operators map to ``corner'' local operators at the junction of the two boundaries. It would be nice to characterize the resulting boundary conditions in the twisted $\bC P^1$ sigma model and verify that the corner local operators computed via the web formalism of \cite{Gaiotto:2015aoa} match the $1$, $C_0$ local operators expected in four dimensions. 

\acknowledgments
We would like to thank Chris Beem, Pieter Bomans, Ramiro Cayuso, Kevin Costello, Richard Eager, Owen Gwilliam, Ji Hoon Lee, Surya Raghavendran, and Keyou Zeng for useful conversations. This research was supported in part by a grant from the Krembil Foundation. DG is supported by the NSERC Discovery Grant program and by the Perimeter Institute for Theoretical Physics. KB and DG are supported by the Simons Foundation through grant 994308 for the Simons Collaboration on Confinement and QCD Strings. JK is funded through the NSERC CGS-D program. JW is supported by the European Union's Horizon 2020 Framework: ERC grant 682608 and the ``Simons Collaboration on Special Holonomy in Geometry, Analysis and Physics''. Research at Perimeter Institute is supported in part by the Government of Canada through the Department of Innovation, Science and Economic Development Canada and by the Province of Ontario through the Ministry of Colleges and Universities.

\appendix

\section{Properties of \texorpdfstring{$\lambda$}{lambda}-Brackets}\label{appendix:lambdabrackets}
The $\lambda$-brackets describe a multilinear operation on the space of local operators modulo total derivatives.  They control how the BRST differential is affected by position dependent interactions with non-zero holomorphic momenta, denoted $\lambda$.\footnote{We discuss and derive properties of these brackets in our companion papers \cite{Budzik:2022mpd,Gaiotto:2024gii}. Mathematically, they should describe part of the information contained in a holomorphic factorization algebra \cite{CG1, CG2}. We also thank Ahsan Khan for sharing unpublished notes on $\lambda$-brackets at an early stage of this work.} A general $\lambda$-bracket $\{\bullet \,_{\lambda_1} \ldots\,_{\lambda_{n-1}} \bullet \}_{n}$ satisfies the properties 
\begin{align}
    \{\ldots_{\lambda_{i-1}} (\p + \lambda_i )O_{\lambda_i} \ldots  \}_n=0\,, \\
    \p \{\ldots_{\lambda_{n-1}} O \}_{n} = \{ \ldots _{\lambda_{n-1}} (\p-\sum_{i=1}^{n-1} \lambda_i)O \}_{n}\,.
\end{align}
Notice that the last entry of the brackets behaves in a slightly different manner from the others. The brackets are graded symmetric under permutations of the remaining arguments together with the $\lambda$ parameters. They are graded symmetric under permutation of all arguments if we define $\lambda_n \equiv -\sum_{i=1}^{n-1} \lambda_i - \partial$, with the latter symbol acting outside the bracket. 

The $\lambda$-brackets should be understood as generating series for a tower of brackets:
\begin{equation}
    \{\bullet \,_{\lambda_1} \ldots\,_{\lambda_{n-1}} \bullet \}_{n} \equiv \sum_{k_*} \left[ \prod_i\frac{\lambda_i^{k_i}}{k_i!}\right] \{\bullet \, , \ldots\, , \bullet \}_{n;k_*} \, .
\end{equation}
The brackets used in the BRST charges at the loop level and Maurer Cartan  equations are the $\lambda$-bracket evaluated at $\lambda_i=0$. The MC equations for holomorphic position-dependent couplings involve the full brackets. 

A particularly simple situation is one where all brackets with more than two entries vanish. Then the 2-ary $\lambda$ bracket $\{\bullet \,_{\lambda} \bullet \}$ satisfies axioms such as:
\begin{itemize}
	\item $\{\partial A\,{}_\lambda\,B\} = - \lambda \{A\,{}_\lambda\,B\}$.
	\item $\partial \{A\,{}_\lambda\,B\} = -\lambda \{  A\,{}_\lambda\,B\}+\{ A\,{}_\lambda\,\partial B\}$. This implies 
	$\{ A\,{}_\lambda\,\partial B\} = (\partial + \lambda) \{A\,{}_\lambda\,B\}$.
	\item $\{A\,{}_\lambda\,B\} = (-1)^{|A||B|} \{B\,{}_{-\lambda- \partial }\,A\}$.
	\item $\{A  \, {}_{\lambda_1} \{ B\, {}_{\lambda_2} \,C \}\} - (-1)^{(|A|+1)(|B|+1)}\{B  \, {}_{\lambda_2} \{ A\, {}_{\lambda_1} \,C \}\} +(-1)^{|A|} \{\{ A\, {}_{\lambda_1} \,B \} \, {}_{\lambda_1+\lambda_2} C   \}=0$
\end{itemize}
where $|A|$ and $|B|$ denotes the fermion parity of the operators.

In the case of one complex dimension with vanishing higher brackets, such a $\lambda$-bracket\footnote{$\lambda$ bracket in one complex dimension satisfies the same set of properties as above except a few sign differences. } generate the familiar Lie conformal algebra\footnote{There is also a great set of lecture notes by Victor G. Kac available at \url{https://web.archive.org/web/20220320093700/https://w3.impa.br/~heluani/files/lect.pdf}.} \cite{kac1998vertex}, which encodes the singular part of the OPE, i.e. the $\lambda$-bracket is a generating function for the non-negative VOA operations. The information contained in the negative VOA operations is captured by a \textit{regularized product}
which satisfy compatibility axioms with the $\lambda$-bracket. These axioms guarantee that together the $\lambda$-bracket and regularized product
encode the chiral algebra. 

\section{Bracket on the Tree Level}\label{app:treelevelbracket}
In this section, we give more details on the computations of the tree level bracket, i.e. $\lambda$-bracket of two arguments. Tree level brackets are simply given by performing a single Wick contraction of free fields. 
\begin{equation}
    \{b^A,c^B\}  = \delta^{AB} \,,
\end{equation}
where capital letters denote adjoint gauge indices. Our Lie algebra conventions are collected at the end of section \ref{sub:LieAlgCoho}. Therefore, given two operators $\mathcal{O}_1$ and $\mathcal{O}_2$, its tree level bracket is simply
\begin{equation}
    \{\mathcal{O}_1, \mathcal{O}_2\}=  \delta^{AB}\frac{\delta \mathcal{O}_1}{\delta c^A}\frac{\delta \mathcal{O}_2}{\delta b^B} +(-1)^{|\mathcal{O}_1|} \delta^{AB}\frac{\delta \mathcal{O}_1}{\delta b^A}\frac{\delta \mathcal{O}_2}{\delta c^B} \,,
\end{equation}
where $|\mathcal{O}|$ denotes the fermion degree. Care with signs is needed when taking functional derivative with respect to Grassman variables. Our convention is 
\begin{equation}
    \frac{\delta}{\delta c^A} \left( c^Bc^C\right) = \delta^{AB} c^C -\delta^{AC} c^B  \,.
\end{equation}

In the rest of the section, we will demonstrate the computation with more explicit examples.

\subsection{Bracket with Stress Tensor: Hamiltonian Symplectomorphisms}\label{appendix:hamiltoniansyp}
The stress tensor for the $bc$ system, i.e. holomorphic twist of the pure gauge theory, is $2 \Tr b \,\p_\alpha c$. We are interested in vector fields on $\mathbb{C}^2$ that preserves the symplectic form $\omega = \d z_1 \wedge \d z_2$. Since all closed forms are exact on $\mathbb{C}^2$, all symplectic vector fields are actually Hamiltonian
\begin{equation}
    X_f = \frac{\partial f}{\p z_1} \frac{\partial }{\p z_2}- \frac{\partial f}{\p z_2}\frac{\partial }{\p z_1} \,.
\end{equation}
Given any two functions $f,g$ on $\mathbb{C}^2$, the corresponding vector fields have the commutation relation 
\begin{equation}
    [X_f,X_g] = X_{\{f,g\}} \,,
\end{equation}
where the Poisson bracket is defined to be 
\begin{equation}
	\{f, g\}=\frac{\partial f}{\partial z_1} \frac{\partial g}{\partial z_2}-\frac{\partial g}{\partial z_1} \frac{\partial f}{\partial z_2} \,.
\end{equation}
Let's choose the basis $f_{m,n} = z_1^{m+1}z_2^{n+1}$ for any $m \in \mathbb{Z}_{\geq -1}$ and $n\in \mathbb{Z}_{\geq -1}$, which has the commutation relation
\begin{equation}
	\left\{f_{m, n}, f_{m^{\prime}, n^{\prime}}\right\}=\left[(m+1)\left(n^{\prime}+1\right)-\left(m^{\prime}+1\right)(n+1)\right] f_{m+m^{\prime}, n+n^{\prime}} \,.
 \label{eq:commutorofHamC2}
\end{equation}

Define the mode generator  
\begin{equation}
     g_{m,n} = z_1^{m+1}z_2^{n+1} \{\partial_\alpha S^\alpha, - \}\,, \label{eq:gmn}
\end{equation}
with $\p_\alpha S^\alpha = 2\Tr\partial_{\alpha}b\partial^{\alpha}c = -\partial_{\alpha}b_A\partial^{\alpha}c^A$. A straightforward computation gives
\begin{equation}
    [g_{m,n},g_{m',n'}] = \left((m+1)(n'+1)-(m'+1)(n+1)\right) g_{m+m',n+n'}\,, \quad m,n \geq -1.
\end{equation}
These are precisely the commutation relations \eqref{eq:commutorofHamC2} of the positive modes of the Hamiltonian vector field  $\mathrm{Ham}(\mathbb{C}^2)$ on $\mathbb{C}^2$. Acting on a test field, we have 
\begin{equation}
    g_{m,n}\mathcal{O} = - X_{f_{m,n}} \mathcal{O} \,,
\end{equation}
which, in particular, includes $SU(2)$ rotations and translations
\begin{equation}
    \begin{aligned}
        g_{-1,0} \mathcal{O} = \partial_1 \mathcal{O}, \quad g_{0,-1} \mathcal{O} = -\partial_2 \mathcal{O}\\
        -\frac12(g_{1,-1}+g_{-1,1})\mathcal{O} = \Big(x_1\partial_2 - x_2\partial_1\Big) \mathcal{O} \,.
    \end{aligned}
\end{equation}

\subsection{Brackets for Large \texorpdfstring{$N$}{N} Tree Level Cohomology}
In this section, we calculate the tree level $\lambda$-brackets between the three towers of operators introduced in section \ref{sub:LieAlgCoho} and section \ref{sec:infiniteN}.
We will work out the brackets explicitly in for $U(2)$. To  reduce clutter, we normalize $A_n = \frac{(2i)^n}{n} \Tr b^n$ and further define $B_{n,\alpha}$ and $C_n$ through the relation
\begin{equation}
    A_n[b+\eta^\alpha \partial_\alpha c] = A_n +\eta^\alpha B_{n-1,\alpha} + \eta^1 \eta^2 C_{n-2} \,.
\end{equation}
A straightforward computation of Wick contraction yields\footnote{We use the convention $\psi^\alpha\chi_\alpha = \psi_2\chi_1 - \psi_1 \chi_2 = -\psi_\alpha \chi^\alpha$ }
\allowdisplaybreaks
\begin{equation}
\begin{aligned}
    \{A_n \,{}_{\lambda}\, A_m\} & = 0\\
    \{A_n\,{}_{\lambda}\,B_{m\alpha} \} &= 2\Big(
    (n-1)\partial_{\alpha}+ (m+n-1)\lambda_\alpha
    \Big) A_{n+m-1}\\
    \{A_{n}\,{}_{\lambda}\,C_{m}\} & =  \frac{2(m+1)}{n+m-1} 
    \Big((n-1)\partial^\alpha +(m+1)\lambda^\alpha\Big) B_{n+m-1,\alpha} \\
	\{B_{n,\alpha}\,{}_{\lambda}\,A_m\} &= (-2n) \left(\partial_\alpha + \frac{n+m-1}{n}\lambda_\alpha\right) A_{n+m-1}\\
	\{B_{n\alpha}\,{}_{\lambda}\,B_{m\beta}\} & = (-2n) \left(\frac{m}{n+m-1} \left(\partial_\alpha + \frac{n+m-1+\delta_{\alpha\beta}}{n} \lambda_\alpha \right)B_{n+m-1,\beta} \right.\\
	&\left. \phantom{assdfasf} + \frac{n-1}{n+m-1} \left(\partial_\beta + \frac{n+m-1+\delta_{\alpha\beta}}{n-1+\delta_{\alpha\beta}} \lambda_\beta \right) B_{n+m-1,\alpha}\right)  \\
	\{B_{n,\alpha}\,{}_{\lambda}\,C_m\} & = (-2n) \frac{m+1}{n+m}\left(\partial_{\alpha} + \frac{n+m}{n} \lambda_\alpha\right) C_{n+m-1}\\
	\{C_n\,{}_{\lambda}\,A_m\} & = - 2n \frac{n+1}{n+m-1} \left(\partial^\alpha +\frac{m+n-1}{n}\lambda^\alpha\right) B_{n+m-1,\alpha}\\
	\{C_n\,{}_{\lambda}\, B_{m\alpha}\} & = 2n \frac{n+1}{n+m}\left( \partial_\alpha + \frac{n+m}{n}\lambda_\alpha\right) C_{n+m-1}\\ 
	\{C_n\,{}_{\lambda}\,C_m\} &= 0
\end{aligned}
\end{equation}
up to $Q_0$ exact terms. 

As discussed in section \ref{sec:holography}, calculations above could be reproduced by holographic consideration. We leave this as a future direction.

\section{Bracket at One Loop}\label{appendix:loopbracket}
In this section of the appendix, we explain the computation of the one loop brackets on operators, and perform some explicit examples, filling in some of the details in section \ref{subsection:1loopandbeyond}. According to the prescription in \eqref{eq:QnintermsofBracket}, we need to compute the triple bracket
\begin{equation}
    Q_1\mathcal{O} \equiv \frac12 \{\mathcal{I},\mathcal{I}, \mathcal{O}\} 
\end{equation}
with all $\lambda$ parameters set to zero. The nontrivial contribution comes from Wick contractions removing two fields in $\calO$ and two fields in each of the two interaction vertices, as shown below:
\begin{center}
\begin{tikzpicture}
		[
		baseline={(current bounding box.center)},
		line join=round
		]
		\coordinate (pd1) at (-0.866*\gS,-0.5*\gS);
		\coordinate (pd2) at (0.866*\gS,-0.5*\gS);
		\coordinate (pd3) at (0.*\gS,1.*\gS);
		\draw (pd1) node[GraphNode] {};
		\draw (pd2) node[GraphNode] {};
		\draw (pd3) node[GraphNode] {};
		\draw[GraphEdge] (pd1) -- (pd2);
		\draw[GraphEdge] (pd1) -- (pd3);
		\draw[GraphEdge] (pd2) -- (pd3);
		\draw[] (-0.866*\gS-.2,-0.5*\gS-.5) node{$\mathcal{I}(x)$};
		\draw[] (0.866*\gS+.2,-0.5*\gS-.5) node{$\mathcal{I}(y)$};
		\draw[] (0.*\gS+.2,1.*\gS+.3) node{$\calO$};
	\end{tikzpicture}\,.
\end{center}
We see immediately that if $\mathcal{O}$ consist of only a single field, the triple brackets vanish identically. $Q_1$ acting on a pair of fields can be nonvanishing. In the example of pure gauge theory, described in \ref{sec:twistedAsHolo}, the relevant operators we are interested in are $b_Ab_B$, $b_Ac^B$ and their derivatives, with open adjoint gauge indices $A,B$. On operators of more than two fields, the action of $Q_1$ is simply the sum of $Q_1$ on all pairs of fields,
\begin{equation}
    Q_1 \Big(f_1\dots f_{i-1}f_i f_{i+1}\dots f_{j-1} f_j f_{j+1} \dots \Big) = \sum_{i<j} (-1)^{\dots} Q_1(f_if_j)  \Big(f_1\dots f_{i-1}f_{i+1}\dots f_{j-1} f_{j+1} \dots \Big)
\end{equation}
where the signs $(-1)^{\dots}$ are due to the (anti)commutation relation to bring $f_i$ and $f_j$ to the front.

We will first compute the triple bracket by evaluating the one loop diagram given above. The one loop diagram has been calculated in \cite{Budzik:2022mpd}. As a result, the $Q_1$ action on $b_Ab_B$ is
\begin{equation}\label{eq:Qbb}
    Q_1(b_A b_B) = \p_\alpha b_A(x) \p^\alpha c^B(x)+  \p_\alpha b_B(x) \p^\alpha c^A(x) - 2 \delta_{AB} \p_\alpha b_C(x) \p^\alpha c^C(x)\,
\end{equation}
and on $b_Ac^B$ is
\begin{equation}
    Q_1(b_A c^B)  = (-1)\Big( \partial_\alpha c^A(x) \partial^\alpha c^B(x) -\delta^B_{A}\, \partial_\alpha c_D(x) \partial^\alpha c^D(x)\Big) \,.\label{eq:Q1onbc}
\end{equation}

On the other hand, in the example of $SU(2)$ pure gauge theory, we can also evaluate the $Q_1$ action on operators by bootstapping the relation \eqref{eq:Qfull2O}. Choose the initial ansatz 
\begin{equation}
Q_1(b_1 c_1) = C_2 \partial_1 c_1 \partial_2 c_1 + C_1 \partial_1 c_2 \partial_2 c_2 + \dots
\end{equation}
where the indices are adjoint gauge indices and $C_{1,2}$ are arbitrary coefficients. Comparing to \eqref{eq:Q1onbc}, we find $C_2 = 0$ and $C_1 = -2$ but we will leave it unfixed below. 
Surprisingly, up to three loop level, we find a unique solution of \eqref{eq:Qfull2O} parametrized by the arbitrary coefficients $C_{1,2}$. In particular, the action of $Q_1$ on the most general type of operator of the form $V(1,1,n_1+m_1,n_2+m_2)$ is given by:
\begin{align}
Q_1(\partial_1^{n_1}\partial_2^{n_2}b_A \partial_1^{m_1} \partial_2^{m_2} c_B) &=
-{C_2} \,\mathcal{F}(c_A, c_B)+C_1\frac{2\,\epsilon_{ADE}\,\epsilon_{BCE}}{m_1+m_2+2} \,\mathcal{G}(c_C,c_D)\,,\\
Q_1(\partial_1^{n_1}\partial_2^{n_2}b_A \partial_1^{m_1} \partial_2^{m_2} b_B) &=    -{C_2}\, \mathcal{H}(b_A,b_B,c_A,c_B) +C_1\frac{2\epsilon_{ADE}\,\epsilon_{BCE}}{m_1+m_2+2} \,\mathcal{K}(b_C,b_D,c_C,c_D)\,.
\end{align}
Where the explicit form of the functions are given by
\begin{align*}
    \mathcal{F}(c_A,c_B) &= \partial_1^{n_1+1}\partial_2^{n_2} c_A\partial_1^{m_1}\partial_2^{m_2+1} c_B-\partial_1^{n_1}\partial_2^{n_2+1} c_A\partial_1^{m_1+1}\partial_2^{m_2} c_B\,,\\
    \mathcal{H}(b_A,b_B,c_A,c_B) &= \partial_1^{n_1+1}\partial_2^{n_2} c_A\partial_1^{m_1}\partial_2^{m_2+1} b_B-\partial_1^{n_1}\partial_2^{n_2+1} c_A\partial_1^{m_1+1}\partial_2^{m_2} b_B\\
& \qquad -\partial_1^{n_1+1}\partial_2^{n_2} b_A\partial_1^{m_1}\partial_2^{m_2+1} c_B+\partial_1^{n_1}\partial_2^{n_2+1} b_A\partial_1^{m_1+1}\partial_2^{m_2} c_B\,,\\
     \mathcal{G}(c_C,c_D) &= \sum_{k = 0}^{m_1}\sum_{l = 0}^{m_2} \frac{1}{k+l+1}\left(\begin{array}{c}
		m_1\\k
	\end{array}\right) \left(\begin{array}{c}
	m_2\\l
\end{array}\right) \\
&\quad \Big[\partial_1^{m_1-k+1}\partial_{2}^{m_2-l}c_C\, \partial_1^{n_1+k}\partial_2^{n_2+l+1}c_D  - \partial_1^{m_1-k}\partial_{2}^{m_2-l+1}c_C\, \partial_1^{n_1+k+1}\partial_2^{n_2 + l}c_D\Big]\,,\\
\mathcal{K}(b_C,b_D,c_C,c_D) &= \sum_{k = 0}^{m_1}\sum_{l = 0}^{m_2} \frac{1}{k+l+1}\left(\begin{array}{c}
		m_1\\k
	\end{array}\right) \left(\begin{array}{c}
	m_2\\l
\end{array}\right) \\
&\quad  \Big[\partial_1^{m_1-k+1}\partial_{2}^{m_2-l}c_C \,\partial_1^{n_1+k}\partial_2^{n_2+ l+1}b_D- \partial_1^{m_1-k}\partial_{2}^{m_2-l+1}c_C \,\partial_1^{n_1+k+1}\partial_2^{n_2+l}b_D\\
& \quad  \ -\partial_1^{m_1-k+1}\partial_{2}^{m_2-l}b_C\, \partial_1^{n_1+k}\partial_2^{n_2+l+1}c_D+ \partial_1^{m_1-k}\partial_{2}^{m_2-l+1}b_C\, \partial_1^{n_1+k+1}\partial_2^{n_2+ l}c_D\Big]\,.\\
\end{align*}
%

One can check explicitly that the resulting $Q_1$ is not nilpotent. This implies the need for a two-loop correction, so that $Q_1^2 + \{ Q_0,Q_2\} = 0$ from \eqref{eq:Q2pert}. For example, 
\begin{align}
     Q_2 \left(b_Ab^Ab_Bb^B\right) &= \frac{20C_1^2}{3}  \epsilon_{ABC} b^A \partial_\alpha \partial_\beta b^B\partial^\alpha \partial^\beta c^C \\
    Q_2\left(b_Ab^A \partial_{\alpha}c_B\partial^{\alpha}c^B \right) &= -\frac{10 C_1^2}{9} \epsilon_{ABC} \partial_\alpha c^A \partial^\alpha\left[\partial_1^2 c^B\partial_2^2 c^C - \partial_1\partial_2 c^B \partial_1\partial_2 c^C \right]\,.
\end{align}

\section{Homotopy Transfer} \label{sec:HomotopyTransfer}

Consider a super vector space $X$ with a differential $Q$ (an odd nilpotent linear operator). 
Denote the cohomology $Q$ in $X$ as $H(X,Q)$. Suppose that we are given a decomposition of supervector spaces
\begin{equation}
    X = U \oplus V \oplus W 
\end{equation}
such that $Q U \subset U$ and $W = \text{im}(Q|_V)$, which implies $Q W=0$. 
This immediately implies that
\begin{equation} \label{eq:eqC}
    H(X,Q) = H(U,\tilde Q = Q|_U) \, .
\end{equation}
Indeed, the kernel of $Q$ equals the kernel of $\tilde Q$ plus $W$, and the image of $Q$ is the image of $\tilde Q$ plus $W$.

An explicit quasi-isomorphism between $(U, \tilde Q)$ and $(X, Q)$ can be presented in terms of a cochain {\it contraction} $(\pi, \iota, h)$ from $(X,Q)$ to $(U, \tilde Q)$.
Let $\pi: X \to U$ be the projection, $\iota: U \to X$ be the inclusion let $h: X \to X$ be such that the following identities hold
\begin{equation}
    \begin{aligned}
\pi Q &= \tilde Q \pi \cr
Q \iota &= \iota \tilde Q \cr
\pi \iota &= 1_U \cr
Q h + h Q &= 1_X - \iota \pi \cr
\pi h &= 0 \cr
h \iota &=0 \cr
h^2 &= 0 \,.
\end{aligned}\label{eq:definecontraction}
\end{equation}
Indeed, we can define $h$ as vanishing on the $U$ and $V$ summands and the inverse of $Q$ on $W$.
The first two relations say that $\pi,\iota$ are chain maps.
The remaining relations encode that $(\pi,\iota,h)$ is a cochain contraction.

A systematic calculation of the cohomology of any cochain complex $(X,Q)$ will effectively produce a contraction as above. 
Indeed, we can let $V \subset X$ be the complement of the kernel of $Q$. 
The cohomology $U$ is then presented as any complement of $W= \text{im}(Q|_V)$ in the kernel of $Q$.

\subsection{Deforming a Contraction}
Next, consider a situation where the differential $Q$ is a perturbation of a simpler differential $Q_0$, that is
\begin{equation}
    Q = Q_0 - \delta
\end{equation}
for some `small' differential $\delta$.
Also, assume that we have a contraction for $Q_0$:
\begin{equation}
    X = U_0 \oplus V_0 \oplus W_0 
\end{equation}
such that $Q_0 U_0 \subset U_0$ and $\text{im}(Q|_{V_0}) = W_0$. Denote by $\tilde Q_0$, $\pi_0$, $\iota_0$ and $h_0$ the corresponding maps supplying a contraction as above.

We can look for a deformation of the contraction for $Q_0$ to a contraction for $Q$. For example, define 
\begin{equation}
    \Delta \equiv \iota_0 \pi_0 + \{h_0, Q\} = 1_X - \{h_0, \delta\}\,.
\end{equation}
If we can invert $\Delta$, we can define new maps
\begin{equation}
\begin{aligned}
\tilde Q & \equiv \pi_0 Q \Delta^{-1} \iota_0 = \tilde Q_0 - \pi_0 \delta \frac{1_X}{1_X - h_0 \delta} \iota_0 \cr
\iota &\equiv \Delta^{-1} \iota_0 =  \frac{1_X}{1_X - h_0 \delta}\iota_0 \cr
\pi &\equiv \pi_0 \Delta^{-1}=  \pi_0 \frac{1_X}{1_X - \delta  h_0} \cr
h &\equiv h_0 \Delta^{-1} = h_0 \frac{1_X}{1_X - \delta  h_0}\,.
\end{aligned}\label{eq:Qtildeansatz}
\end{equation}
We claim that these maps define the desired deformed contraction. Some relations follow immediately from the expressions above: 
\begin{equation}
    \begin{aligned}
    \pi \iota &= 0 \,,\cr
\pi h &= 0 \,,\cr
h \iota &=0 \,,\cr
h^2 &= 0 \,.
\end{aligned}\label{eq:checkbasic}
\end{equation}
Using  
\begin{equation}
    \tilde Q = \pi_0 \Delta^{-1} Q \iota_0 = \pi Q \iota_0\,,
\end{equation}
and a bit of work, we can verify another set of relations:
\begin{align}
    \tilde Q \pi &= \pi Q \iota_0 \pi_0 \Delta^{-1}= \pi Q - \pi Q \{h_0, Q\} = 
     \pi Q - \pi \{h_0, Q\} Q  \cr
     &=  \pi Q - \pi \iota_0 \pi_0 Q + \pi \Delta Q =  \pi Q - \pi_0 Q + \pi_0 Q = \pi Q\,, \cr
     \iota \tilde Q &= \Delta^{-1} \iota_0 \pi_0 Q \iota = Q \iota -  \{h_0, Q\} Q \iota = 
     Q \iota - Q \{h_0, Q\} \iota \cr
     &=   Q \iota  - Q \iota_0 \pi_0 \iota + Q \Delta \iota =   Q \iota  - Q \iota_0 + Q \iota_0= Q \iota \,.
\end{align}
The final relation takes a bit more work. First, we can compute the commutator
\begin{align}
    \{ Q_0, h\} &= \frac{1_X}{1_X - h_0 \delta} \{Q_0,h_0\} \frac{1_X}{1_X - \delta h_0 } - h \{Q_0, \delta\} h \cr 
    &=\frac{1_X}{1_X - h_0 \delta}\frac{1_X}{1_X - \delta h_0 } - \iota \pi- h \delta^2 h \cr 
    &= 1_X + \frac{h_0 \delta}{1_X - h_0 \delta}+\frac{\delta h_0}{1_X - \delta h_0 } - \iota \pi \,.
\end{align}
From this it follows that
\begin{equation}
  \iota \pi+  \{ Q, h\} = 1_X\,,
\end{equation}
as desired.

\subsection{Perturbative Cohomology}
In a perturbative setting we may have a differential given as a power series in some formal parameter 
\begin{equation}
    Q = Q_0 + \hbar Q_1 + \hbar^2 Q_2 + \cdots
\end{equation}
Homotopy transfer offers a way to recursively {\it define} a ``perturbative cohomology'' in such setting, as long as the $Q_i$ have a sufficiently ``triangular'' form. 

Namely, we can start by taking the cohomology $U_0$ of $Q_0$ and treating the rest of the sum as $-\delta$, inverting the corresponding 
\begin{equation}
    \Delta \equiv \iota_0 \pi_0 + \{h_0, Q\} = 1_X + \hbar \{h_0, Q_1\}+ \hbar^2 \{h_0, Q_2\}+ \cdots
\end{equation}
perturbatively. The result is a new perturbative differential
\begin{equation}
    \tilde Q = \pi_0 (\hbar Q_1 + \cdots) \Delta^{-1} i_0 = \hbar Q^{(1)}_0 + \hbar^2 Q^{(1)}_1 + \cdots
\end{equation}
which is now a formal power series starting at order $\hbar$. 

We can then compute a contraction for $Q^{(1)}_0$ and repeat the procedure to get a new pertubative differential on the cohomology $U^{(1)}_0$ of $Q^{(1)}_0$, etcetera. As long as we have some kind of filtration 
controlling the form of the $Q_i$, the procedure will converge to some 
limiting $U^{(\infty)}_0$ which can be taken as the definition of perturbative cohomology.

\section{Maurer-Cartan Equations and Quantum Field Theories}\label{app:MC}
A student of QFT is usually familiar with the idea that any QFT can be formally/ perturbatively deformed by adding to the action a general linear combination of local operators (modulo total derivatives) multiplied by couplings:
\begin{equation}
    \sum_i g_i \int d^D x \, {\cal O}^i(x) \, .
\end{equation}
The parameterization of the space of deformations is scheme-dependent beyond the first order of perturbation theory. The deformation space can be described as a {\it formal} (i.e. defined in perturbation series) {\it pointed} (i.e. equipped with a base point) {\it super} (if we include deformations by Grassmann odd operators) {\it manifold}. 

Another familiar notion in QFT is that of a beta function. If the base theory is scale-invariant, an infinitesimal scale transformation will act on the couplings, thus defining a formal vector-field 
\begin{equation}
    \beta = \sum_i \beta_i(g) \partial_{g_i} = \beta_{1;i}^j g_j \partial_{g_i} + \frac12 \beta_{2;i}^{j k} g_j g_k \partial_{g_i} + \cdots \, .
\end{equation}
Even if the base theory is not scale invariant, an infinitesimal scale transformation will give an infinitesimal deformation of the theory. We will still get a formal vectorfield $\beta = \beta_{0;i} \partial_{g_i} + \cdots$.

If the base theory is described in a BRST formalism, we can extend the space of formal deformations to include all operators in the BRST complex. Accordingly, we have a larger collection of couplings $g_i$, some of which will have non-zero ghost number/cohomological degree. If we take the deformed theory with some couplings $g_i$ and act with the BRST differential on correlation function, a possible BRST anomaly will manifest itself as an odd variation of the couplings, i.e. an odd vectorfield vector field 
\begin{equation}
    {\cal Q} = \sum_i \eta_i(g) \partial_{g_i} = \eta_{1;i}^j g_j \partial_{g_i} + \frac12 \eta_{2;i}^{j k} g_j g_k \partial_{g_i} + \cdots \, .
\end{equation}
In other words, correlation functions will be annihilated by the action of $Q_{\mathrm{BRST}} + {\cal Q}$.
By definition, the odd vectorfield ${\cal Q}$ is nilpotent. We thus have what is called a {\it formal pointed dg-supermanifold}.

The condition for a deformation of the theory to be non-anomalous thus takes manifestly the form of a Maurer-Cartan equation,
with coefficients $\eta_{n;i}^{j_i \cdots j_n}$. The constraint ${\cal Q}^2=0$ tells us that the coefficients satisfy the axioms of an 
$L_\infty$ algebra. 

Notice that we did not assume that the QFT we are referring to should be topological or holomorphic. The appearance of $\infty$-structures is simply due to the formal deformation nature of the problem. 

We can also consider position-dependent deformations of the theory. This leads to $\lambda$-brackets and their non-holomorphic analogues. We refer to our companion paper \cite{Gaiotto:2024gii} for more details. 

\subsection{Homotopy Transfer for Odd Vector-Fields and \texorpdfstring{$L_\infty$}{L-infinity} Algebras}
The basic homotopy transfer formula from appendix \ref{sec:HomotopyTransfer} can be widely generalized. For example, consider an $L_\infty$ algebra, presented dually as the collection of coefficients in an odd nilpotent formal vectorfield ${\cal Q}$.

There is a ``sum over trees'' construction for $L_\infty$ algebras which is analogous to the contraction formulae above \cite{Kontsevich:2000yf}. The analogy becomes sharp in the language of the formal supermanifolds. Suppose that we have a contraction for $Q$, the linear part of ${\cal Q}$. This gives us dual linear maps
$\iota$, $\pi$, $h$ relating the $g_i$'s and some smaller set of $\tilde g_a$. 
We can define as before 
\begin{equation}
        \Delta = \iota \pi + \{ h, {\cal Q} \}\,,
\end{equation}
and 
\begin{equation}
\begin{aligned}
\tilde {\cal Q} & \equiv \pi {\cal Q} \Delta^{-1} \iota \,,\cr
{\cal I} &\equiv \Delta^{-1} \iota \,,\cr
\Pi &\equiv \pi \Delta^{-1} \,,\cr
{\cal H} &\equiv h \Delta^{-1}\,,
\end{aligned}\label{eq:calQtildeansatz}
\end{equation}
which give an odd vectorfield $\tilde {\cal Q}$ on the $\tilde g_a$ as well as morphisms relating it to ${\cal Q}$. 

Dually, these formulae reproduce the `sum over tree' formulae. As we expand out the expressions for the coefficients of $\tilde {\cal Q}$, we find sums of terms involving sequences of operations which alternate the action of $L_\infty$ brackets and of the coefficients in $h$.

\section{Rigid SUSY, Twists, and Indices}\label{sec:RigidSUSY}

In the bulk of the document, our goal is to understand the (perturbative) algebra of local operators in a SUSY QFT protected by the minimal amount of supersymmetry. These operators form a ``semi-chiral ring,'' and are computed by taking the $Q$-cohomology of any supercharge in the $\calN=1$ theory (the particular supercharge is immaterial). Essentially by definition of locality, the collection of local operators in any QFT should only depend on the details of the spacetime manifold through the contribution of gauge-invariant functions of the metric and other background fields, and their derivatives at the point, which can affect operator mixing. After a holomorphic twist, we do not expect the surviving background fields, such as the Beltrami differential encoding the complex structure of spacetime, to have local invariants. Hence the semi-chiral ring should be insensitive to the spacetime manifold that the theory is placed on.

There are many other quantities of interest in SUSY QFTs which \textit{are} sensitive to the underlying spacetime. Perhaps the best example of a BPS quantity which depends on the spacetime is the supersymmetric partition function $\calZ_{\calM}$. The partition function, with arbitrary background fields turned on, captures the correlation functions of local operators which couple to those background fields e.g. $\calZ_{\calM}[A]$ tells us about correlation functions of the current $j^\mu$ which couples to $A_\mu$. The partition function also describes the ``topological'' response of the theory to non-trivial background field configurations, including non-trivial topologies themselves. In the case that the flat space theory is superconformal (and thus has a $U(1)_R$ symmetry), the supersymmetric partition function on $S^3\times S^1$ is essentially the superconformal index, which in turn counts (with signs) the (dimensions of the $Q$-cohomologies of the) local operators in the semi-chiral ring. We shall review the precise correspondence between these quantities below.

In this appendix we will review and disambiguate the terminology employed here and in the supersymmetry literature. Specifically, we will:
\begin{enumerate}
    \item Give a brief review of the literature on rigid supersymmetry and define what we mean by the word ``twist.'' The goal here will be to disentangle the study of BPS quantities, roughly captured by passing to the $Q$-cohomology (and what we will call twisting), from the process of turning on a non-trivial background field. This is summarized in figure \ref{fig:twistFigure}.\footnote{To re-iterate, our definition of ``twist'' only means to pass to the cohomology of a nilpotent (not necessarily scalar) operator $Q$. In the literature, twist can sometimes mean the entire passage from $\calL_{\bbR^4} \to \calL_{\calM}^{\mathrm{tw}}$. Unfortunately, twist can also refer to turning on background gauge fields, which we will also do (but not call twisting).}
    \item Describe precisely how to do computations in a twisted theory by trading it for a simpler theory which is a deformation retract of the original one, and comment on the information of the untwisted theory that is captured by the twisted theory.
    \item Separate the notions of supersymmetric partition function, supersymmetric index, and superconformal index, and explain under what conditions and to what extent they are the same on $S^3\times S^1$.
\end{enumerate}

\begin{figure}[t]
\centering
\begin{tikzcd}
	{\mathcal{L}_{\mathbb{R}^{2 d_1 + d_2}}} && {\mathcal{L}^{\mathrm{tw}}_{\mathbb{C}^{d_1}\times \mathbb{R}^{d_2}}} \\
	\\
	{\mathcal{L}_{\mathcal{M}}} && {\mathcal{L}^{\mathrm{tw}}_{\mathcal{M}}}
	\arrow["{\begin{matrix}\text{Couple to}\\\text{off-shell}\\\text{SUGRA}\end{matrix}}"', from=1-1, to=3-1]
	\arrow["{Q\text{-cohomology}}", from=1-1, to=1-3]
	\arrow["{Q\text{-cohomology}}"', from=3-1, to=3-3]
	\arrow["{\begin{matrix}\text{Define}\\\text{on\,}\mathcal{M}\end{matrix}}", from=1-3, to=3-3]
\end{tikzcd}
\caption{Down-moving arrows always involve turning on some background fields in order to define the theory on a curved manifold: on the left, there is the extra constraint to preserve supersymmetry on $\calM$; on the right, many local aspects of the geometry are unimportant. Right-moving arrows always indicate passing to a $Q$-cohomology, i.e. restricting attention to BPS quantities of the theory.} \label{fig:twistFigure}
\end{figure}

As we will explain, there are two ways to obtain the twisted/cohomological theory on a non-trivial spacetime $\calL_{\calM}^{\mathrm{tw}}$.\footnote{Note: we write everything schematically with a Lagrangian $\calL$ for simplicity, but a Lagrangian is not essential for the rigid SUSY based arguments below.} Starting with the untwisted flat space theory, schematically denoted ${\mathcal{L}_{\mathbb{R}^{2 d_1 + d_2}}}$, one can:
\begin{enumerate}
    \item \textbf{Put it on $\calM$ then twist.} Start by trying to place the flat space theory on a desired curved manifold $\calM$. Since this generically breaks the supersymmetries of the original flat space theory, one must turn on particular background fields in order to preserve SUSY. This may not be possible at all and will generally require choices beyond the geometry of $\calM$, such as a complex structure for a 4d ${\cal N}=1$ theory equipped with a $U(1)_R$ symmetry, or a complex symplectic structure for a generic 4d ${\cal N}=1$ theory. If the theory can be placed on $\calM$ while preserving at least one nilpotent scalar supercharge, then one can study the $Q$-twisted theory on $\calM$.
    \item \textbf{Twist then put it on $\calM$.} Alternatively, we can start by passing from the flat space theory to the twisted theory $\calL^{\mathrm{tw}}_{\bbC^{d_1}\times \mathbb{R}^{d_2}}$, a holomorphic-topological QFT which we can try to place on different curved manifolds by using an atlas of coordinate patches related by diffeomorphisms which preserve the holomorphic-topological structure.\footnote{In some situations, e.g. with extended supersymmetry, there may be other non-geometric background fields which can be turned on and are needed to match the supergravity setup. The $S^2$ partition function of 2d $(2,2)$ theories or the $S^4$ partition function of 4d ${\cal N}=2$ theories are likely examples.} For example, the twist of a 4d ${\cal N}=1$ theory equipped with a $U(1)_R$ symmetry can be placed on an arbitrary complex manifold/coupled to an arbitrary Beltrami differential. In the absence of $R$-symmetry, it can be placed on any complex symplectic manifold. 
\end{enumerate}
We schematically sketch these two paths in figure \ref{fig:twistFigure}. In the main body of the text we take path 2. 

In the next section, we briefly review the literature on rigid supersymmetry that encompasses the left-most arrow of this diagram, since it is the only one which is not obvious from the text. Then we review what it means to compute and study the twisted/cohomological theory. Finally, we discuss the particular case that $\calM = S^3 \times \bbR$ and/or $\calM = S^3 \times S^1$, and the relationship between the supersymmetric partition function, the supersymmetric index, and the superconformal index.

\subsection{Rigid SUSY and Curved Spacetimes}\label{sec:rigidSUSY}
Let us very briefly review the first approach, and, in particular, explain how to place a SUSY theory on a curved spacetime $\calM$ while preserving supersymmetry. A systematic answer for this problem was first described by \cite{Festuccia:2011ws}, with solutions tabulated in \cite{Dumitrescu:2012ha, Dumitrescu:2012at}. Important follow-ups for index calculations are described in \cite{Closset:2013vra, Closset:2014uda} (see \cite{Kuzenko:2012vd} for a superspace approach and \cite{Dumitrescu:2016ltq} for a self-contained review).

As previously mentioned, if one naively tries to place a flat space theory $\calL_{\bbR^4}$ on some manifold $\calM$ by minimally coupling it to the metric $g_{\mu\nu}$ for $\calM$, then the theory on $\calM$ will not generically be supersymmetric.\footnote{Intuitively, perturbations to the flat space metric couple to the stress tensor $T^{\mu\nu}$, and $[Q,T^{\mu\nu}] \neq 0$, so generically $\calL_{\calM}$ is not supersymmetric. In a similar vein, (non-flat) background connections for global symmetries will also break SUSY.} In particular, there is only a supercharge on $\calM$ for each covariantly constant spinor $\zeta$
\begin{equation}\label{eq:covariantConstantSpinor}
    \nabla_\mu \zeta = 0\,.
\end{equation}
Note: even when SUSY is preserved, there's no reason to expect the SUSY algebra on $\calM$ to be the same as on $\bbR^4$ since the isometries of $\calM$ will be different from isometries of $\bbR^4$.\footnote{However, the SUSY algebra on $\calM$ should In\"onu-Wigner contract to the SUSY algebra for $\bbR^4$, where the contraction parameter is some characteristic length scale for $\calM$. Also see the comments around (1.1) of \cite{Festuccia:2011ws}.
}

The \textit{rigid supersymmetry} construction of \cite{Festuccia:2011ws} provides a systematic procedure for determining if and how one can place the flat space theory $\calL_{\bbR^4}$ on $\calM$, without breaking supersymmetry, for a much larger class of $\calM$ than specified by \eqref{eq:covariantConstantSpinor}. The idea is to couple the theory to background off-shell supergravity in a possibly non-minimal way, enriching the Killing spinor equation \eqref{eq:covariantConstantSpinor} with additional background field terms. Then, by tuning the values of the background fields in this supergravity multiplet, the Killing spinor equation may admit more solutions on $\calM$. The procedure is as follows:
\begin{enumerate}
    \item \textbf{Pick a supermultiplet $\calS_\mu$.} Every local Lorentz-invariant SUSY QFT has a real conserved symmetric stress tensor $T_{\mu\nu}$, which lives inside a stress tensor supermultiplet alongside the conserved supercurrent $S_{\mu\alpha}$ \cite{Dumitrescu:2011iu} (see also section \ref{subsec:stresstensor}). The other component fields $\{\mathcal{X}_i\}$ in the supermultiplet are determined by the precise supermultiplet $\calS_\mu$ we are using, but $T_{\mu\nu}$ and $S_{\mu\alpha}$ are universal.
    \item \textbf{Make $g_{\mu\nu}$ a superfield.} We promote the background metric $g_{\mu\nu}$ to a (component of a) background supergravity multiplet. The multiplet contains $g_{\mu\nu}$, the gravitino $\Psi_{\mu\alpha}$, and some additional auxiliary fields $\{\mathcal{Y}_i\}$ which can source the $\{\mathcal{X}_i\}$ operators in the supercurrent multiplet $\calS_\mu$. 
    \item \textbf{Couple $\calL_{\bbR^4}$ to supergravity background.} We couple $\calL_{\bbR^4}$ to the appropriate background off-shell supergravity multiplet (as dictated by $\calS_\mu$).
    \item \textbf{Tune the background so $\calL_{\calM}$ is supersymmetric.} To obtain a supersymmetric theory on $\calM$, we must tune the values for the background off-shell supergravity multiplet appropriately. We set $g_{\mu\nu}$ to look like $\calM$, and set the gravitinos (and any other fermionic background fields) to zero so we have a bosonic background, i.e. $\Psi = 0 = \tilde{\Psi}$. Since SUSY variations of the bosonic background fields are proportional to the gravitinos, they vanish. By solving for $\delta \Psi = 0$ and $\delta \tilde{\Psi} = 0$, we obtain the allowed values for the auxiliary bosonic background fields $\{\mathcal{Y}_i\}$ so that $\calL_{\calM}$ is supersymmetric.    
\end{enumerate}

That this last step of tuning the background off-shell supergravity fields works, and can be done at all, is non-obvious. The intuition is that one should actually think of the theory $\calL_{\bbR^4}$ as being coupled to standard dynamical off-shell supergravity. Off-shell because we \textit{do not} integrate out the auxiliary fields. Then, by rescaling the fields in the dynamical supergravity multiplet and taking the ``rigid limit'' Planck mass $M_P \to \infty$, the supergravity fluctuations decouple and we can ``freeze in'' the admissible background values. This also makes it clear that the resulting rigid SUSY algebra can be thought of as the subalgebra of local superdiffeomorphisms that preserve the particular background.

Some important notes about this procedure:
\begin{itemize}
    \item Since the gravitino variation always contains a covariant derivative of $\zeta$
    \begin{equation}\label{eq:generalizedKillingSpinor}
        0 = \delta_\zeta \Psi_{\mu\alpha} =\nabla_\mu \zeta_\alpha + F(g_{\mu\nu}, \mathcal{Y}_i,\zeta)\,,
    \end{equation}
    we obtain a generalization of the Killing spinor equation \eqref{eq:covariantConstantSpinor}, enriched by the background fields. There can be yet additional dependence of $F$ on background connections for global symmetries.
    \item We can use this construction in the reverse. By solving the Killing spinor equation \eqref{eq:generalizedKillingSpinor}, we can determine which $\calM$ admit supersymmetric backgrounds \cite{Dumitrescu:2012ha, Dumitrescu:2012at}.
    \item Similarly, the construction makes it clear that we do not need specific information about $\calL_{\bbR^4}$, other than the stress tensor multiplet $\calS_{\mu}$ it admits. In particular, we don't even need to have a Lagrangian description of the flat space theory.\footnote{With the caveat that background fields can enter the action non-linearly. A description of the coupling thus includes information on how the stress tensor multiplet varies as the background fields are turned on.}
\end{itemize}

To pick a guiding concrete example, if the flat space $\calN=1$ theory has an unbroken $U(1)_R$ symmetry, then the theory admits an $\calR$-multiplet and couples to background fields for ``new minimal supergravity'' \cite{Sohnius:1981tp,Sohnius:1982fw}
\begin{align}
    \calR_\mu 
        &= (j_\mu^{(R)}, S_{\alpha\mu}, \tilde{S}_{\mu\dot\alpha}, T_{\mu\nu}, C_{\mu\nu})\,,\\
    \calH_\mu 
        &= (A_\mu^{(R)}, \Psi_{\alpha\mu}, \tilde{\Psi}_{\mu\dot\alpha}, g_{\mu\nu}, B_{\mu\nu})\,,
\end{align}
where $C_{\mu\nu}$ is a conserved anti-symmetric two-form associated to ``string charges'' \cite{Dumitrescu:2011iu}. In Euclidean signature, left and right-handed spinors are independent, so their variations are set to zero independently $\delta_\zeta \Psi_{\alpha\mu} = 0$ and $\delta_{\tilde{\zeta}} \tilde{\Psi}_{\mu\dot\alpha}=0$. Thus we simply read off the new Killing spinor equations from the gravitino variations of new minimal supergravity:
\begin{align}\label{eq:newKillingSpinor1}
    (\nabla_\mu-i A_\mu^{(R)})\zeta 
        &= -\frac{i}{2} V^{\nu}\sigma_{\mu}\tilde{\sigma}_{\nu}\zeta\,,\\
    (\nabla_\mu+i A_\mu^{(R)})\tilde{\zeta} 
        &= +\frac{i}{2} V^{\nu}\tilde{\sigma}_{\mu}{\sigma}_{\nu}\tilde{\zeta}\,,\label{eq:newKillingSpinor2}
\end{align}
where $V^\mu = \frac{i}{2} \epsilon^{\mu\nu\rho\lambda}\partial_\nu B_{\rho\lambda}$ and is conserved $\nabla_\mu V^\mu=0$ \cite{Festuccia:2011ws, Closset:2013vra, Closset:2014uda}. In Euclidean signature, $A_\mu$ and $V^\mu$ are generally allowed to be complex, but one assumes $g_{\mu\nu}$ is real.

Solutions to the above Killing spinor equations in the off-shell new minimal supergravity background (\ref{eq:newKillingSpinor1}-\ref{eq:newKillingSpinor2}) exist if and only if $\calM$ is a Hermitian manifold, in which case the background field values $A_{\mu}^{(R)}$ and $V^\mu$ are (almost) fully determined from the complex structure ${J^\mu}_\nu$ and Hermitian metric on $\calM$ (see \cite{Dumitrescu:2012ha} for details). For example, $V^\mu = \frac{1}{2}\nabla_\nu {J^\nu}_\mu$. When $\calM$ is K\"ahler, ${J^\nu}_\mu$ is covariantly constant, so $V^\mu$ vanishes and the Killing spinor equations reduce to the familiar holomorphic twist of $\calN=1$ theories on K\"ahler manifolds by an $R$-symmetry background as described in \cite{Johansen:1994aw, Witten:1994ev, Vyas:2010pda}. We will return to this below. Other stronger facts are obtained if more supercharges are preserved.

The analysis above can be repeated without a $U(1)_R$ symmetry for theories admitting an FZ-multiplet. The FZ-multiplet couples to the auxilliary fields of ``old minimal supergravity'' \cite{Stelle:1978ye, Ferrara:1978em}: a metric $g_\mu$, gravitinos $\Psi_{\mu\alpha}$ and $\tilde{\Psi}_{\alpha\mu}$, a vector field $b^\mu$, and two scalars $M$ and $\tilde{M}$. There are also similar results for ``conformal supergravity'' and Lorentzian signatures \cite{Samtleben:2012gy, Klare:2012gn, Liu:2012bi, Cassani:2012ri, deMedeiros:2012sb, Hristov:2013spa}.

An important consequence of having a $U(1)_R$ symmetry in the new-minimal supergravity treatment, is that the $Q$ which is guaranteed to survive on a Hermitian manifold $\calM$ is a scalar under holomorphic coordinate trasformations \cite{Festuccia:2011ws, Dumitrescu:2012ha}. This is not the case without a $U(1)_R$ symmetry for the old-minimal supergravity treatment \cite{Dumitrescu:2012at}. A closely related (but strictly stronger) fact is that with and without $U(1)_R$ symmetry, an $\calN=1$ theory can be placed on a manifold locally isometric to $\calM_3 \times \bbR$, where $\calM_3$ is a maximally symmetric space, while preserving four supercharges; but only with a $U(1)_R$ symmetry they can be made time-independent (i.e. they commute with the generator of isometries along the $\bbR$ factor). 

One way to understand this scalarity of $Q$ is as follows \cite{Dumitrescu:2012ha}: the holonomy group of the Levi-Civita connection for a K\"ahler manifold $\calM$ is $U(2)$, and at the level of Lie algebras $\mathfrak{u}(2) = \mathfrak{u}(1) \times \mathfrak{su}(2)$ where these are subalgebras of the chiral $\mathfrak{su}(2)$ symmetries of flat spacetime. By turning on a $U(1)_R$ background, one can cancel the $U(1)$ component in the $U(2)$ holonomy group, leaving a scalar supercharge $\zeta$ to the Killing spinor equations
\begin{equation}
    (\nabla_\mu - i A_\mu^{(R)}) \zeta = 0\,
\end{equation}
as in \cite{Johansen:1994aw, Witten:1994ev, Vyas:2010pda}. On a more general Hermitian manifold $\calM$, the metric and complex structure are covariantly constant with respect to the Chern connection $\nabla_\mu^{(c)}$, not the Levi-Civita connection. In this general Hermitian case, the Killing spinor equations in the off-shell new minimal supergravity background (\ref{eq:newKillingSpinor1}-\ref{eq:newKillingSpinor2}) become the exceptionally simple
\begin{equation}
    (\nabla_\mu^{(c)} - i A_\mu^{(c)}) \zeta = 0\,
\end{equation}
when written in terms of the Chern connection (see \cite{Dumitrescu:2012ha} for details on the Chern-connection and gauge field $A_\mu^{(c)}$). Thus the entire story for the K\"ahler manifold and the holonomy of the Levi-Civita connection being cancelled by $A_\mu^{(R)}$ repeats for the $U(2)$ holonomy of the Chern connection and $A_\mu^{(c)}$.

\subsection{Twisted Theory on Curved Spacetime}
Oppositely to the story above, we can instead take a flat space theory $\calL_{\bbR^4}$ and \textit{twist} it by passing to the $Q$-cohomology. As previously mentioned, these $Q$-cohomologies capture the BPS local operators comprising the semi-chiral ring. In the main body of the paper, our journey ends here for $\calN=1$ SYM: \textit{we are computing the (perturbative) semi-chiral ring of local operators on flat space, and studying its structure and holomorphic descendants.}

There are now two (essentially equivalent) things we mean by the ``twisted theory'' $\calL^{\mathrm{tw}}_{\bbC^2}$. On one hand, we can study the cohomological theory written in terms of the original field variables on $\bbC^2$. On the other hand, we can study an effective repackaging of these $Q$-cohomologies into new fields, presenting the theory in a different way. Specifically, we find a different set of field variables that correspond to $Q$-cohomologies and when treating both in the BV formalism provides a deformation retract of the respective chain complexes. For example. In our case, the twisted sector of $\calN=1$ SYM can be presented in terms of the original field variables, or can be effectively described by a $bc$ system/holomorphic BF theory. 

Now, \textit{if} we are interested in trying to place the twisted flat-space theory $\calL^{\mathrm{tw}}_{\bbC^2}$ with Lorentz group $K$ on a non-trivial manifold $\calM$, and \textit{if} we also have a $U(1)_R$ symmetry (or other global symmetries), then we can talk about turning on background fields again to preserve the $Q$-cohomological/holomorphic structure on $\calM$. In particular, it may be possible to find a homomorphism $K \hookrightarrow K \times U(1)_R$, so that the twisting supercharge $Q$ is a scalar under the image of $K$: the new Lorentz group $K'$. In this case, there is a new stress tensor $T'$ (whose components integrate to new Lorentz group rotations) is different from $T$, satisfying $T' = \{Q,S\}$.\footnote{Here there is an unfortunate historical overloading of terminology which we have tried to avoid. In the literature, the homomorphism, ``new Lorentz group $K'$,'' and ``new stress tensor,'' are called the ``twisting homomorphism,'' ``twisted Lorentz group,'' and ``twisted stress tensor'' respectively. Referring to the fact that objects are affected by a nontrivial $U(1)_R$ background. We do not use such language because a priori these ``twists'' do not have to be related to taking a $Q$-cohomology.} In flat-space, this is a trivial redefinition of field data, but alters the way that the theory $\calL^{\mathrm{tw}}_{\bbC^2}$ couples to curved space. More specifically, a $Q$-exact stress tensor makes placing $\calL^{\mathrm{tw}}_{\bbC^2}$ on $\calM$ sensible at all. Since $Q$ is a $K'$ scalar, the theory can be put on any $\calM$ without breaking the $Q$-cohomology/holomorphic structure i.e. while preserving $Q$ and $Q^2=0$.

It is important to reiterate that our definition of twist in this paper is precisely passing to $Q$-cohomology, to study the BPS operators, i.e. the right-moving arrows in figure \ref{fig:twistFigure}. Not the act of turning on a background gauge field/twist for $U(1)_R$ symmetry, i.e. down-moving arrows in figure \ref{fig:twistFigure}. We stress this point because the flat-space twisted theory $\calL^{\mathrm{tw}}_{\bC^2}$, with or without a $U(1)_R$ symmetry, is a theory that exists, which captures the correlation functions of the semi-chiral subsector of the full flat-space theory $\calL_{\bbR^4}$. Turning on a background symmetry is required to put the theory on a non-trivial manifold.

Some extended remarks about twisting are in order.
We have emphasized that a main step in constructing the twist of a supersymmetric theory involves taking the cohomology with respect to a nilpotent supercharge $Q$.
More precisely, this step in the twisting procedure involves deforming the original BRST operator $Q_{\mathrm{BRST}}$ by the supercharge $Q$, see \cite{CostelloHol,ESW}. 
If one is only interested in what happens to the cohomology of local operators when passing to a twist, then there is a spectral sequence which does involve taking the $Q$-cohomology.

There is another step in the twisting procedure that we have not discussed.
To make sense of the deformed/twisted BRST operator $Q_{\mathrm{BRST}} + Q$ one uses $R$-symmetry to shift the ghost number grading so that this operator is homogenous. 
The output of twisting typically collapses the $\bbZ \times \bbZ/2$ (ghost $\times$ parity) bigrading of a supersymmetric field theory to a $\bbZ/2$ grading. 
In the original theory, the BRST operator has bidgree $(1,0)$ with respect to this bigrading and $Q$ has bidegree $(0,1)$.
One chooses a copy of $U(1)$ in the $R$-symmetry to lift this to a $\bbZ$ grading, which we interpret as the ghost grading in the twisted theory.
Of course, if the $R$-symmetry does not contain a $U(1)$ factor then this step is not possible.
We refer to \cite{CostelloHol,ESW} for more details.

\subsection{Partition Functions, Supersymmetric Index, and Superconformal Index}

In the rigid SUSY story, we emphasized the existence of a nilpotent time-independent scalar supercharge $Q$ on the Hermitian manifold $\calM$ when there is a $U(1)_R$ symmetry. This can be used to show, modulo anomalies, that $\calZ_{\calM}$ is independent of the Hermitian metric on spacetime, and is locally a holomorphic function of the complex structure moduli and line bundle moduli (in the presence of background fields) of $\calM$ \cite{Closset:2014uda}. 

Alternatively, in the $Q$-cohomology story, we emphasized the existence of a new Lorentz group on flat space under which $Q$ was a spacetime scalar, and such that a new stress tensor $T_{\mu\nu}'$ was $Q$-exact. Infinitesimal responses of the holomorphic theory to a curved background are computed by correlation functions of the new stress tensor $T_{\mu\nu}'$, and hence it is also clear that $\calZ_{\calM}$ (or really any correlation function of operators in the semi-chiral ring) only depends (locally) on the complex structure moduli since $Q$-exact deformations leave correlation functions unchanged. 

Indeed, using a linearized analysis around flat space, one can show that the change $\Delta \calL$ to the Lagrangian under infinitesimal variations of the complex structure and Hermitian metric lead to $Q$-exact deformations of the Lagrangian, except for terms proportional to the holomorphic complex structure moduli \cite{Closset:2013vra}. The conclusion is that $\calZ_{\calM}$ is locally a holomorphic function of the complex structure moduli. The fact that it is only \textit{locally} a holomorphic function emphasizes that we may still have anomalies leading to phase ambiguities of the partition function \cite{Cassani:2021fyv}.

Now let us focus on the special case of interest. The rigid SUSY analysis tells us that the only 4-manifolds admitting four supercharges must be locally isometric to $\calM_3 \times \bbR$ where $\calM_3$ is a space of constant curvature \cite{Festuccia:2011ws, Dumitrescu:2012ha}. In particular, manifolds with topology of $S^3 \times S^1$ are an example. Without $U(1)_R$ symmetry, a theory with an FZ-multiplet can be placed on a manifold locally isometric to $S^3\times \bbR$ while preserving 4 supercharges, but only with a $U(1)_R$ symmetry can those supercharges be made time-independent in the sense that they commute with the generator of translations along the $\bbR$ direction \cite{Festuccia:2011ws,Dumitrescu:2012ha}. Hence, we once again focus from hereout on theories which have a non-anomalous $U(1)_R$ symmetry.

Every complex manifold diffeomorphic to $S^3 \times S^1$ is a Hopf manifold: a quotient of $\bbC^2 - \{(0,0)\}$ by the free action of a discrete group $G$ \cite{kodaira1966complex, kodaira1966structure}. Of primary interest are those manifolds such that $G \cong \bbZ$, in which case we are left with a \textit{primary Hopf surface}; secondary Hopf surfaces can be obtained by finite group quotients from the primary Hopf surfaces. If the coordinates of $\bbC^2 - \{(0,0)\}$ are $z_1, z_2$, we obtain a primary Hopf surface by quotienting
\begin{equation}
    (z_1, z_2) \sim (p z_1+\lambda z_2^m, q z_2)
\end{equation}
where $m \in \mathbb{N}$  and $p,q,\lambda$ are complex parameters, such that $0 < |p| \leq |q| < 1$ and $(p-q^m)\lambda = 0$ \cite{Assel:2014paa}. We will take $\lambda=0$ from here.

Given all this, it is completely reasonable to ask for the supersymmetric partition function $\calZ_{\calM}(p,q,u)$ on $\calM = S^1 \times S^3$, where $p$ and $q$ label complex structure moduli of the Hopf surface, and $u$ labels line bundle moduli for background gauge fields. By supersymmetric partition function we mean the partition function with $(-1)^F$ inserted, so that fermions have periodic boundary conditions on the $S^1$. Antiperiodic boundary conditions would explicitly break supersymmetry.

We define the supersymmetric index to be the (refined) Witten index of the theory on $S^3\times \bbR$ in Hamiltonian quantization \cite{Sen:1985ph, Sen:1989bg, Romelsberger:2005eg, Kinney:2005ej, Romelsberger:2007ec, Dolan:2008qi}
\begin{equation}
    \calI(p,q,u) = \Tr_{\calH_{S^3}}\left((-1)^F p^{j_\ell + j_r - \frac{R}{2}}q^{j_\ell -j_r - \frac{R}{2}} \zeta^{J_{\mathrm{fl}}}\right)\,,
\end{equation}
where $2j_\ell$ and $2j_r$ are charges under the Cartan's $M_{\dot{+}\dot{-}}$ and $\bar{M}_{+-}$ of the spacetime $SU(2)$s, $R$ is $U(1)_R$ charge, and $J_{\mathrm{fl}}$ is some collection of $U(1)$ flavour charges associated to a corresponding set of fugacities $\zeta$. Since $S^3$ is compact, this is a well-defined (signed) count of the states of the theory on $S^3$. Based on the counting of quantum numbers, the index is well-defined along RG flows (so long as all pertinent symmetries are preserved along the flow). Note that this definition of the index as a refined Witten index makes sense if the theory is superconformal or not. However, if the theory is superconformal, then the counting of states in the supersymmetric index can be interpreted as a counting of $1/(4\,\calN)$-BPS local operators by the state-operator correspondence, which we will call the superconformal index. In theories with a superconformal symmetry, this turns the four-dimensional index into a counting problem of gauge invariant local operators in the limit of vanishing coupling.

The relationship between the supersymmetric index to the superconformal index is clear. The relationship between the supersymmetric partition function and the supersymmetric index is that when $\calM=S^3\times S^1$, then
\begin{equation}
    \calZ_{\calM}(p,q,u) = e^{-\mathcal{F}(p,q)} \calI(p,q,u)\,,
\end{equation}
where the prefactor $\mathcal{F}$ depends on the $a$ and $c$ central charges of the theory \cite{Cassani:2013dba}, and are interpreted as a supersymmetric Casimir energy \cite{Assel:2014paa, Assel:2015nca, Dumitrescu:2016ltq}.

\bibliographystyle{JHEP}

\bibliography{mono}

\end{document}